%% file: main.tex
\documentclass[11pt]{article}
\pdfoutput=1
\usepackage{jheppub}
\usepackage[dvipsnames]{xcolor}

\usepackage{microtype}
\usepackage{mathbbol}
\usepackage{physics}
\usepackage{accents}

\usepackage{cleveref}
\usepackage{nicematrix}

\newcommand{\sectorfig}[1]{\raisebox{-0.5\height}{\includegraphics{figures/#1}}}
\newcommand{\gsfig}[1]{\raisebox{-0.5\height}{\includegraphics[scale=1.0]{figures/#1}}}
\newcommand{\gsfigs}[1]{\raisebox{-0.5\height}{\includegraphics[scale=0.68]{figures/#1}}}
\usepackage[titles]{tocloft}

\usepackage{orcidlink}

\usepackage{enumitem}

\definecolor{gr}{gray}{0.7}
\definecolor{orcidlogocol}{named}{Maroon}

\newcommand{\im}{\mathrm{i}}
\newcommand{\mzero}{\color{gr}{\cdot}}
\newcommand{\mZeroBlock}{\color{gr}{\mathbb{0}}}

\newcommand{\midtilde}[1]{%
    \accentset{%
        \raisebox{-0.15ex}[\height][0pt]{\scalebox{0.85}{$\scriptstyle\sim$}}%
    }{#1}%
}

\newcommand{\transpose}{^{\!\top}}
\newcommand{\defas}{:=}
\newcommand{\Complex}{\mathbb{C}}

\newcommand{\cC}{\mathcal{C}}
\newcommand{\cF}{\mathcal{F}}
\newcommand{\cG}{\mathcal{G}}
\newcommand{\cI}{\mathcal{I}}
\newcommand{\cN}{\mathcal{N}}
\newcommand{\cO}{\mathcal{O}}
\newcommand{\cR}{\mathcal{R}}
\newcommand{\cU}{\mathcal{U}}

\newcommand{\brk}[1]{(#1)}
\newcommand{\lrbrk}[1]{\left(#1\right)}
\newcommand{\bigbrk}[1]{\bigl(#1\bigr)}
\newcommand{\Bigbrk}[1]{\Bigl(#1\Bigr)}
\newcommand{\biggbrk}[1]{\biggl(#1\biggr)}
\newcommand{\Biggbrk}[1]{\Biggl(#1\Biggr)}
\newcommand{\sbrk}[1]{[#1]}
\newcommand{\bigsbrk}[1]{\bigl[#1\bigr]}
\newcommand{\brc}[1]{\{#1\}}
\newcommand{\bigbrc}[1]{\bigl\{#1\bigr\}}
\newcommand{\Bigbrc}[1]{\Bigl\{#1\Bigr\}}
\newcommand{\supbrk}[1]{^{\brk{#1}}}
\renewcommand{\vev}[1]{\langle#1\rangle}
\newcommand{\bigvev}[1]{\big\langle#1\big\rangle}
\NewDocumentCommand{\regpair}{o m m}{%
    \IfValueTF{#1}
        {\vev{#2 \mid #3}_{\delta,(#1)}}
        {\vev{#2 \mid #3}_{\delta}}%
}
\renewcommand{\abs}[1]{\lvert#1\rvert}
\newcommand{\bigabs}[1]{\bigl\lvert#1\bigr\rvert}
\newcommand{\set}[1]{\{#1\}}

\title{Differential Equations for Feynman Integrals from Higher Residue Pairings}
\author[\!a,b,\orcidlink{0009-0004-4788-738X}]{Giacomo Brunello,}\emailAdd{giacomo.brunello@sns.it}
\author[\!c,\orcidlink{0000-0001-7067-0315}]{Vsevolod Chestnov,}\emailAdd{vsevolod.chestnov@maths.ox.ac.uk}
\author[d,\orcidlink{0000-0001-7147-501X}]{Ben Page}\emailAdd{ben.page@ugent.be}
\affiliation{$^a$Scuola Normale Superiore, Piazza dei Cavalieri 7, 56126, Pisa, Italy}
\affiliation{$^b$INFN Sezione di Pisa, Largo Pontecorvo 3, 56127 Pisa, Italy}
\affiliation{$^c$Mathematical Institute, University of Oxford, OX2 6GG, UK}
\affiliation{$^d$Department of Physics and Astronomy, Ghent University, 9000 Ghent, Belgium}
\abstract{
We develop a method for constructing differential equations for Feynman integrals directly from multivariate intersection numbers, without fibration.
The large-$\varepsilon$ expansion of the intersection pairing, with $\varepsilon$ the dimensional regulator, is given by Saito's higher residue pairings, which localise at the critical points of a potential and admit closed multivariate formulae.
In the Lee--Pomeransky representation, applying these formulae to Feynman integrals requires analytic regulators $\delta$ on the coordinate hyperplanes, and the physical limit $\delta \to 0$ is not manifest.
Assuming that the regulated critical points are analytic in $\delta$, we show that, as $\delta \to 0$, they organise into sectors labelled by sets of pinched propagators.
We derive closed formulae for the leading and first subleading higher residue pairings in this limit, written entirely in terms of unregulated sector data.
These suffice to reconstruct differential systems at most linear in $\varepsilon$, including canonical ones, bypassing integration-by-parts identities.
We apply the method to the massless planar four-point ladder families up to four loops and obtain their canonical differential equations, which are new at four loops.
}

\begin{document}

\addtocontents{toc}{\protect\thispagestyle{empty}}

\maketitle

\thispagestyle{empty}

\setcounter{page}{2}
\allowdisplaybreaks
\input{Sections/intro}
\input{Sections/representations}
\input{Sections/eps_expansion}
\input{Sections/deltaexpansion}
\input{Sections/applications}
\input{Sections/conclusions}
\acknowledgments
We would like to thank Pierpaolo Mastrolia for useful discussions.
The computations for this work were performed with computing and storage resources provided by the CloudVeneto initiative at the University of Padova and INFN, and the Mathematical Institute at the University of Oxford.
AI tools assisted in the editing of this work as well as with the final implementation
of the proof-of-concept pairing code.

G.B.'s research is supported by the Italian MIUR under contract 20223ANFHR (PRIN2022), by the ERC (NOTIMEFORCOSMO, 101126304), and by the INFN initiatives \textit{Amplitudes} and \textit{TPPC}.
The research of V.C. is funded by the European Research Council (ERC) Synergy Grant MaScAmp 101167287. 
The work of B.P. is funded by the European Union (ERC, MultiScaleAmp, Grant
Agreement No. 101078449).
Views and opinions expressed are however those of the author(s) only and do not necessarily reflect those of the European Union or the European Research Council Executive Agency. Neither the European Union nor the granting authority can be held responsible for them.
\appendix
\input{Sections/subleading}
\input{Sections/ladder_definitions}
\bibliographystyle{JHEP}
\bibliography{refs}
\end{document}

%% file: Sections/intro.tex
\section{Introduction}
The evaluation of multi-loop Feynman integrals represents a central problem in modern scattering amplitudes computations.
A standard approach is to reduce the integrals to a finite basis of master integrals and determine their dependence on the external kinematics through systems of differential equations (DEs)~\cite{Kotikov:1990kg,Kotikov:1991pm,Remiddi:1997ny,Gehrmann:1999as,Argeri:2007up}.
The construction of these differential systems conventionally relies on integration-by-parts (IBP) identities~\cite{Tkachov:1981wb,Chetyrkin:1981qh,Laporta:2000dsw}. Derivatives of the master integrals with respect to the kinematic variables generate integrals in the same family, which are then reduced back to the chosen basis using these identities.
Several efficient public tools for constructing and solving IBP systems are available~\cite{Lee:2013mka,Peraro:2019svx,Guan:2024byi,Lange:2025fba,Smirnov:2019qkx}.
Nevertheless, this procedure can become increasingly expensive as the number of loop integrations and external scales grows.
This motivates the search for approaches to constructing DEs that bypass the need for IBPs.

Intersection theory provides such a framework~\cite{%
    matsumoto1994, matsumoto1998, OST2003, doi:10.1142/S0129167X13500948,%
    goto2015, goto2015b, Yoshiaki-GOTO2015203, Mizera:2017rqa,matsubaraheo2019algorithm, Ohara98intersectionnumbers,%
https://doi.org/10.48550/arxiv.2006.07848,https://doi.org/10.48550/arxiv.2008.03176,https://doi.org/10.48550/arxiv.2104.12584}.
Feynman integrals belong to the mathematical class of twisted period integrals~\cite{Mastrolia:2018uzb}. 
In parametric representations, their integrands consist of a multivalued function, called the twist, and a differential form.
The latter represents a class in twisted de Rham cohomology and resides in a finite-dimensional vector space.
A target form can be decomposed onto a cohomology basis, with coefficients computed from
bilinear pairings with a dual basis, known as intersection
numbers~\cite{Mastrolia:2018uzb,Frellesvig:2019kgj,Frellesvig:2019uqt,Frellesvig:2020qot,%
Weinzierl:2020xyy}.
In particular, kinematic derivatives of the basis forms can be decomposed back onto the same basis in this way, providing a direct construction of the DEs~\cite{Mastrolia:2018uzb}.
This approach offers a geometric perspective on relations among Feynman integrals and their symmetries~\cite{Duhr:2024rxe,Duhr:2024xsy,Duhr:2025xyy,Bree:2026qww,Duhr:2026elp}.
Related intersection-theoretic methods have also been applied to Fourier integrals~\cite{Brunello:2023fef,Brunello:2024ibk,Brunello:2025cot}, correlation functions in quantum mechanics~\cite{Cacciatori:2022mbi}, lattice field theory~\cite{Weinzierl:2020nhw,Gasparotto:2022mmp,Gasparotto:2023roh}, and cosmological correlators~\cite{Arkani-Hamed:2023kig,De:2023xue,Benincasa:2024ptf}.

The practical construction of DEs in this framework requires efficient computation of intersection numbers.
For one-form intersection numbers, a closed computational algorithm is known~\cite{Mastrolia:2018uzb}.
Multivariate intersection pairings are usually evaluated using a fibration
strategy~\cite{Ohara98intersectionnumbers,Frellesvig:2019uqt,Frellesvig:2020qot}, which recursively reduces the
problem to a sequence of one-variable pairings.
This approach has proved effective in combination with polynomial division techniques~\cite{Fontana:2023amt,Brunello:2023rpq}, relative twisted cohomology~\cite{Caron-Huot:2021iev,Caron-Huot:2021xqj,Brunello:2023rpq}, companion matrices~\cite{Brunello:2024tqf}, and different parametric representations~\cite{Lu:2024dsb,Huang:2026xnq}.
In this approach, intermediate expressions depend simultaneously on the remaining integration variables and the external kinematics, and can grow rapidly with the number of loops and legs.
Direct multivariate formulae, which avoid this recursive structure, are therefore desirable.
At present, such formulae are available for special classes of forms, such as forms with logarithmic singularities~\cite{matsumoto1998,Mizera:2017rqa,Frellesvig:2020qot}.
Alternative strategies for the computation of multivariate intersection numbers
include approaches via higher-order partial differential
equations~\cite{Chestnov:2022xsy}, the secondary
equation~\cite{Chestnov:2022alh}, and a large-exponent expansion whose
coefficients are Saito's higher residue
pairings~\cite{Saito:1983,Mizera:2019vvs}.

This last approach is particularly natural for Feynman integrals, whose twist
exponents are governed by the dimensional regulator.
In dimensional regularisation, we set the spacetime dimension to $D=D_0-2 \varepsilon$.
Feynman integrals can then be written as integrals of $e^{\varepsilon W}$, with
the potential $W$ independent of $\varepsilon$.
The expansion of the intersection pairing around $\varepsilon\to\infty$, the
limit of large spacetime dimension, is a saddle-point expansion whose
coefficients are the higher residue pairings~\cite{Saito:1983,Mizera:2019vvs}.
These pairings localise at the critical points of $W$ and, order by order in
$\varepsilon^{-1}$, are given in closed form in terms of the differential forms,
the Hessian of $W$, and their derivatives~\cite{Mizera:2019vvs}.
For differential equations, such an expansion is sufficient.
The matrices of the differential systems are rational functions of
$\varepsilon$, and for suitable bases they are
polynomials~\cite{Lee:2014ioa,Henn:2014qga,e-collaboration:2025frv}.
In particular, for a canonical basis of master integrals~\cite{Henn:2013pwa}, the system is linear in $\varepsilon$ and its $\varepsilon$-dependence factorises from the kinematic one.
Such a system is then determined exactly by finitely many orders of the expansion of intersection numbers around $\varepsilon\to\infty$~\cite{Mizera:2019vvs}.
Higher residue pairings therefore provide direct multivariate formulae for the differential equations, without recourse to fibration.

For complete Feynman-integral families, applying these formulae directly requires an additional step. Standard parametric representations contain singularities along coordinate hyperplanes $z_i=0$ that are not regulated by the physical twist.
In the Lee--Pomeransky representation~\cite{Lee:2013hzt}, these hyperplanes are associated with pinched propagators.
In the Baikov representation~\cite{Baikov:1996iu,Frellesvig:2017aai,Frellesvig:2024ymq}, they correspond to cut propagators.
They can be regulated by introducing analytic regulators $z_i^{\varepsilon\delta_i}$, with the physical integrals recovered in the limit $\delta_i\to0$~\cite{Frellesvig:2019kgj,Frellesvig:2020qot}.
At finite values of the analytic regulators, the intersection pairing is defined in absolute twisted cohomology, and higher residue pairings apply.
As $\delta_i\to0$, however, some critical points approach the coordinate hyperplanes where the Hessian of $W$ diverges, so the limit of the pairings is not manifest.
Unregulated Feynman integrals are naturally described by relative twisted cohomology~\cite{Matsumoto:2018igm,Caron-Huot:2021iev,Caron-Huot:2021xqj,Brunello:2023rpq}, where the coordinate hyperplanes become components of the boundary.
Applying the higher residue pairings to physical Feynman-integral families therefore requires control over the vanishing-regulator limit, $\delta_i\to 0$, of the multivariate intersection pairings.
So far, this limit has been studied only within the fibration approach in Refs.~\cite{Fontana:2023amt,Brunello:2023rpq}.

In this work, we address this problem and determine the vanishing-regulator limit of the leading and first subleading higher residue pairings.
We assume that the regulated critical points are isolated and non-degenerate, remain finite, and admit a Taylor expansion in a common regulator $\delta_i=\delta$. 
As $\delta\to0$, the critical points organise into sectors, labelled by subsets of pinched propagators.
We further assume that, in each sector, their limits are isolated and non-degenerate critical points of the corresponding unregulated sector problems.
Before identifying graph symmetries, the number of master integrals is the sum of the numbers of critical points in each sector~\cite{Lee:2013hzt,Bitoun:2017nre,Frellesvig:2019uqt}.
The higher residue pairings decompose accordingly into sums over sectors.
We derive closed formulae for the sector contributions to the leading and first subleading higher residue pairings.
The resulting expressions are written entirely in terms of unregulated quantities associated with the individual sectors.
These two higher residue pairings are sufficient to reconstruct differential systems whose dependence on the dimensional regulator $\varepsilon$ is at most linear.
For such systems, our vanishing-regulator formulae therefore provide a direct multivariate construction of the differential equations from unregulated sector data, bypassing the generation of integration-by-parts identities.
Systems with a more general dependence on $\varepsilon$ require higher terms in the large-$\varepsilon$ expansion, whose analysis is left for future work.
The sums over the critical points of each sector are evaluated as traces of companion matrices, so the critical points are never determined explicitly.

We apply this framework to the massless planar ladder families up to four loops, deriving their DEs directly from multivariate intersection numbers.
In all cases, we recover canonical differential systems and validate the results against independent IBP computations.
Up to three loops, canonical bases for these families were known~\cite{Henn:2013pwa,Henn:2013tua}.
At four loops, we find $159$ non-vanishing sectors, with $246$ critical points in total.
After accounting for symmetry relations, these reduce to a basis of $88$ master integrals, for which we construct the complete canonical differential system.
To our knowledge, the complete canonical system for this four-loop family is new.

The paper is organised as follows.
In \cref{sec:representations}, we review the Lee--Pomeransky representation,
twisted period integrals, and differential equations for Feynman integrals.
In \cref{sec:eps_expansion}, we introduce intersection numbers and their
large-$\varepsilon$ expansion in terms of higher residue pairings, and relate them to differential systems linear in $\varepsilon$.
In \cref{sec:delta_expansion}, we analyse the vanishing-regulator limit and derive the sectorwise formulae for the leading and first subleading pairings.
In \cref{sec:applications}, we describe the algebraic implementation and apply the method to the massless planar ladder families up to four loops.
We conclude in~\cref{sec:conclusions} and provide further details of the
subleading pairing derivation and integral family definitions in
appendices~\labelcref{app:subleading_derivation,app:ladder_definitions}.

%% file: Sections/representations.tex
\section{Feynman Integrals as Twisted Period Integrals}
\label{sec:representations}
In this section, we review the Lee--Pomeransky representation of Feynman
integrals~\cite{Lee:2013hzt} and its analytic regularisation, which embeds all topologies of an integral family in the parameter space of the top-level graph.
In this representation, Feynman integrals can be regarded as twisted period integrals~\cite{Mastrolia:2018uzb}:
the multivalued part of the integrand defines the twist, while the remaining differential form
represents a class in twisted de~Rham cohomology.
We then introduce the differential equations satisfied by a basis of master integrals.
\subsection{Lee--Pomeransky Representation for Feynman Integrals}
\label{ssec:lee_pomeransky_representation}
We consider a family of Feynman integrals associated with a fixed top-level
graph $\Gamma$, with $n$ propagators and $L$ loops.
Lower topologies in the same family are obtained by pinching subsets of the
propagators of $\Gamma$, and will be denoted by
$\Gamma_{\set{i_1,\ldots,i_k}}$, where the set $\set{i_1,\ldots,i_k}$ labels
the pinched propagators. The top-level graph itself is $\Gamma = \Gamma_{\varnothing}$.
For propagator powers $\boldsymbol{\alpha}=(\alpha_1,\ldots,\alpha_n)$, $\alpha_i \not\in \mathbb{Z}_{\le 0}$,
the Lee--Pomeransky representation~\cite{Lee:2013hzt} is
\begin{equation}
\label{eq:leepom}
    I_{\alpha_1\ldots \alpha_n}
    =
    \frac{\Gamma\brk{D/2}}{\Gamma\bigbrk{\brk{L+1}D/2-|\boldsymbol{\alpha}|}}
    \int_{\mathbb{R}^n_+}
    \prod_{i=1}^n \frac{\dd z_i \, z_i^{\alpha_i-1}}{\Gamma\brk{\alpha_i}}\cG^{-D/2}
    \>,
\end{equation}
where $\cG= \cU+\cF$ is the Lee--Pomeransky polynomial, built from the Symanzik polynomials $\cU$ and $\cF$ associated with the top-level graph $\Gamma_{\varnothing}$.
The variables $z_i$ are in one-to-one correspondence with the
propagators of $\Gamma_{\varnothing}$. For simplicity, we write $\partial_i \defas \partial/\partial z_i$ and $|\boldsymbol{\alpha}|=\alpha_1+\cdots+\alpha_n$.

The twist factor $\cG^{-D/2}$ regulates the singular locus $\brc{\cG=0}$, but not the coordinate hyperplanes $\brc{z_i=0}$.
For non-positive integer $\alpha_i$, \cref{eq:leepom} is therefore singular: the factor $z_i^{\alpha_i-1}$ is not integrable at $z_i=0$, while $\Gamma\brk{\alpha_i}$ has a pole.
Such values are nevertheless required to describe the complete family: $\alpha_i=0$ corresponds to pinching propagator $i$, while negative integer values describe numerator insertions.
We therefore understand the corresponding integrals by analytic continuation in the $\alpha_i$.
To make this continuation explicit, and to treat all coordinate hyperplanes on the same footing, we introduce analytic regulators, the factors
$z_i^{\varepsilon\delta_i}$ with generic exponents
$\boldsymbol{\delta} = (\delta_1,\ldots,\delta_n)$.
The analytically regulated Lee--Pomeransky representation is
\begin{equation}
\label{eq:leepom_reg}
    I^{\brk{\boldsymbol{\delta}}}_{\alpha_1\ldots \alpha_n}
    =
    \frac{\Gamma\brk{D/2}}{\Gamma\brk{\brk{L+1}D/2-|\boldsymbol{\alpha}| - \varepsilon\, |\boldsymbol{\delta}|}}
    \int_{\mathbb{R}^n_+}
    \prod_{i=1}^n \frac{\dd z_i \, z_i^{\alpha_i-1+\varepsilon \delta_i}}{\Gamma\brk{\alpha_i+\varepsilon \delta_i}}\cG^{-D/2}
    \>,
\end{equation}
where $|\boldsymbol{\delta}| = \delta_1+\cdots+\delta_n$.
For generic non-zero $\delta_i$, the coordinate hyperplanes are regulated by the twist, and every integer $\alpha_i$ is allowed.
Unless stated otherwise, we use the \textit{democratic} choice of regulators,
\begin{equation}
\label{eq:democratic_choice}
    \delta_i = \delta
    \>,
    \qquad
    i = 1,\ldots,n
    \>,
\end{equation}
and denote the corresponding integrals by $I^{(\delta)}_{\alpha_1\ldots \alpha_n}$. 
In general, this specialisation may lead to degeneracy, as discussed in \cref{ssec:critical_sectors}. In such a case, the regulators $\delta_i$ must instead be kept generic. We leave such cases to future work and assume non-degeneracy.

The original integral family is recovered in the vanishing-regulator limit,
\begin{equation}
   I_{\alpha_1\ldots \alpha_n} = \lim_{\delta \to 0} I^{(\delta)}_{\alpha_1\ldots \alpha_n}
    \>.
    \label{eq:vanishing_regulator_limit}
\end{equation}
An integral associated with a pinch $\Gamma_S$ of the top-level graph $\Gamma$, i.e.\ with $\alpha_i = 0$ for $i \in S$,
can also be represented in this fashion, with integrand-level poles at $z_i = 0$.
This mechanism recovers the Lee--Pomeransky representation of such pinch integrals.
If we consider setting $\alpha_i = 0$ for $i \in S$, the integrand contains 
the factor $z_i^{-1+\varepsilon\delta}/\Gamma\brk{\varepsilon\delta}$ for each pinched edge.
The integral over $z_i$ localises on $z_i = 0$ as the regulator is removed.
This follows from the distributional identity\footnote{
    More precisely, in the sense of distributions~\cite{GelfandShilov1964}, one has the expansion
    $
    \frac{z_+^{-m - 1 + \eta}}{\Gamma\brk{-m + \eta}}
    =
    \delta\supbrk{m}_0 + \cO\brk{\eta}
    \>,
    $
    for $m \ge 0$, where $z_+^{-m - 1 + \eta}$ denotes the one-sided distributional power, 
    initially defined for $\Re\eta > m$ by integration over the positive ray $z > 0$ 
    and then analytically continued in $\eta$,
    while $\delta\supbrk{m}_0 \defas \partial_z^m \delta_0$ is the $m^{\text{th}}$ distributional 
    derivative of the Dirac delta distribution supported at $z = 0$.
}~\cite{Lee:2014tja, Lu:2024dsb}
\begin{equation}
\label{eq:lift_to_top}
    \int_0^\infty
    \frac{\dd z \, z^{-m - 1 + \varepsilon\delta}}{\Gamma\brk{-m + \varepsilon \delta}}
    \>
    f\brk{z}
    =
    \brk{-1}^m \> \partial_z^m f\brk{z} \bigr\rvert_{z = 0}
    + \cO\brk{\varepsilon \delta}
    \>,
    \qquad
    m \ge 0
    \>,
\end{equation}
applied with $m = 0$: the singular factor $z^{-1+\varepsilon\delta}$ produces
a pole $f\brk{0}/\brk{\varepsilon\delta}$, which is cancelled by the
$\Gamma$-function normalisation,
\begin{equation}
\label{eq:one_over_gamma_expansion}
    \frac{1}{\Gamma\brk{\varepsilon\delta}}
    =
    \varepsilon\delta\, \left(1 + \gamma_{\rm E}\,\varepsilon\delta + \cO\brk{\varepsilon^2 \delta^2}\right)
    \>.
\end{equation}
Given that the Lee--Pomeransky polynomial associated with $\Gamma_S$ is simply $\cG$ evaluated at $z_i = 0, i \in S$, 
integrating~\cref{eq:leepom_reg} with $\alpha_i = 0$, $i \in S$, over the pinched
variables gives the Lee--Pomeransky representation of the pinched graph as $\delta \to 0$.

Besides representing pinch integrals, \cref{eq:lift_to_top} also allows one to
represent Feynman integrals with irreducible scalar products (ISPs) of the top-level graph in the numerator.
To this end, we introduce an augmented Lee--Pomeransky polynomial $\cG^{\mathrm{aug}}$. Such a polynomial arises if one associates Feynman parameters not just with the propagators of an integral, but also 
with some of the ISPs. If we do this for all ISPs, then we say that the augmented representation is complete.
When the propagators form a complete and linearly independent set, the $\partial_{i} \log \cG^{\mathrm{aug}}$ can be identified with the images of the Baikov variables via the duality of Ref.~\cite{Chen:2023eqx}.
Let $A$ label the ISPs and $z_i$, $i \in A$, their auxiliary Feynman parameters.
An integral with the ISPs raised to positive powers $\boldsymbol{\beta} =
(\beta_i)_{i \in A}$ in the numerator can be represented in the same
Lee--Pomeransky family using~\cref{eq:lift_to_top}.
After removing their analytic regulators, the auxiliary integrations
become derivatives of $(\cG^{\mathrm{aug}})^{-D/2}$, of orders specified
by $\boldsymbol{\beta}$, evaluated at vanishing auxiliary parameters.
We define the corresponding parametric numerator factor as
\begin{equation}
    \label{eq:isp_numerator_factor}
    \cR_A^{\boldsymbol{\beta}}
    \defas
    \left.
    \frac{
        \prod_{i \in A}
        \brk{-\partial_{i}}^{\beta_i}
        \brk{\cG^{\mathrm{aug}}}^{-D/2}
    }{
        \cG^{-D/2}
    }
    \right\rvert_{z_i=0,\;i\in A}
    \>,
\end{equation}
where $\cG^{\mathrm{aug}}\bigr\rvert_{z_i=0,\;i\in A} = \cG$.
The argument of the denominator Gamma function in~\cref{eq:leepom_reg}
is increased by $|\boldsymbol{\beta}| = \sum_{i \in A} \beta_i$.
The resulting integral involves only the original $n$ Feynman parameters, with the
numerator insertion represented by $\cR_A^{\boldsymbol{\beta}}$.
Each derivative in~\cref{eq:isp_numerator_factor} lowers the power of $\cG^{\mathrm{aug}}$
by at most one, so $\cR_A^{\boldsymbol{\beta}}$ is a polynomial divided by
$\cG^{|\boldsymbol{\beta}|}$; in particular, it is rational, and hence single-valued, in $z$.
A negative propagator power $\alpha_i = -m$ is treated in the same way, without auxiliary parameters: integrating by parts $m$ times in $z_i$ reduces the pole $z_i^{-m - 1}$ to a simple one, as for a pinched edge, and produces the rational factor $\cG^{D/2} \brk{-\partial_i}^m \cG^{-D/2}$, which contains derivatives of $\cG$ with respect to $z_i$.
\subsection{Twisted Period Integrals and Critical Points}
\label{ssec:twisted_period_integrals}
Setting $D=D_0-2\varepsilon$, with $D_0$ an even integer,
the twist factorises as $\cG^{-D/2} = \cG^{-D_0/2}\,\cG^{\varepsilon}$.
For the democratic choice of regulators \cref{eq:democratic_choice}, the regulated integral \cref{eq:leepom_reg} can be rewritten as a twisted period integral~\cite{Mastrolia:2018uzb}:
\begin{equation}
\label{eq:leepom_tpi}
    I^{(\delta)}_{\alpha_1\ldots \alpha_n} = \int_{\cC} e^{\varepsilon W_\delta} \varphi
    \>,
    \qquad
    W_\delta 
     =
    \log\brk{\cG}+ \delta \sum_{i=1}^n\log \brk{z_i}
    \>,
\end{equation}
where $\cC = \mathbb{R}^n_+$ is the integration cycle and $W_\delta$ is a multivalued logarithmic potential.
The differential $n$-form $\varphi$ is defined as:
\begin{equation}
\label{eq:leepom_form}
\varphi = 
\hat{\varphi}\,\dd^n z,
\qquad
    \hat{\varphi} =
    \frac{\Gamma\brk{D/2}}{\Gamma\brk{\brk{L+1}D/2-|\boldsymbol{\alpha}|- n \varepsilon\delta}}
    \frac{1}{\cG^{D_0/2}}
    \prod_{i=1}^n \frac{ \, z_i^{\alpha_i-1}}{\Gamma\brk{\alpha_i+\varepsilon \delta}}
    \>.
\end{equation}
The dual family of twisted period integrals is obtained by flipping the sign of the logarithmic potential in \cref{eq:leepom_tpi},
\begin{equation}
\label{eq:leepom_tpi_dual}
    I^{(\delta)\, \vee }_{\alpha_1\ldots \alpha_n} = \int_{\cC^\vee} e^{-\varepsilon W_\delta} \varphi^{\vee}
    \>,
\end{equation}
where $\cC^\vee$ is a dual integration cycle and the dual differential
$n$-form $\varphi^\vee$ is constructed analogously to \cref{eq:leepom_form}.
The explicit choice of dual forms will be discussed in~\cref{sec:delta_expansion}.

The two twists define the twisted differentials
\begin{equation}
\label{eq:twisted_differential}
    \nabla_{\dd W_\delta}
    =
    \dd + \varepsilon\, \dd W_\delta\wedge\,,
    \qquad
    \nabla_{\dd W_\delta}^\vee
    =
    \dd - \varepsilon\, \dd W_\delta\wedge\,,
\end{equation}
where $\dd$ is the total differential in the integration variables.
Twisted total derivatives integrate to zero on $\cC$ and $\cC^\vee$. 
Consequently, differential $n$-forms related by $\varphi \sim \varphi + \nabla_{\dd W_\delta}\xi$, 
with $\xi$ a generic $(n-1)$-form, define the same integral. 
This is the equivalence relation of twisted cohomology, and it encodes the IBP identities of the family.
To discuss this in more detail, we define 
the relevant ambient space as the complement of the singular locus of the integrand,
\begin{equation}
\label{eq:Xdelta_def}
    T_\delta
    =T_\delta^\vee
    =
    \mathbb{C}^n
    \setminus
    \brk{
        \{\cG=0\}
        \cup
        \bigcup_{i=1}^n \brc{z_i=0}
    }
    \>.
\end{equation}
For $\delta\neq0$, all boundary components are regulated by the twist. 
Differential forms represent classes in \textit{absolute} twisted cohomology groups:
\begin{equation}
\label{eq:absolute_twisted_cohomology}
    [\varphi]
    \in
    H^n\bigbrk{T_\delta \,;\, \nabla_{\dd W_\delta}}
    \>,
    \qquad
    [\varphi^\vee]
    \in
    H^n\bigbrk{T_\delta^\vee \,;\, \nabla_{\dd W_\delta}^\vee}
    \>,
\end{equation}
defined as the spaces of closed modulo exact forms with respect to the twisted differentials in \cref{eq:twisted_differential}.

Throughout this work, we take representatives with at most simple poles on the coordinate
hyperplanes $\brc{z_i = 0}$. As we will see in \cref{sec:delta_expansion}, this allows us to
make a uniform analysis of the $\delta \rightarrow 0$ limit of intersection numbers.
Every integral of the family admits such a
representative: positive indices $\alpha_i$ produce no pole, while for negative
indices integrating by parts in $z_i$ reduces a higher-order pole in $z_i$ to a
simple pole, as described in~\cref{ssec:lee_pomeransky_representation}.
Higher-order coordinate poles are thereby traded for derivatives of $\cG$, so poles
along $\brc{\cG = 0}$ may be of any order.
For a form whose pinched edges $\Gamma$ have $\alpha_i \le 0$, with positive indices
otherwise, this gives
\begin{equation}
    \label{eq:form_pole_structure}
    \hat\varphi
    \propto
    \frac{\cR_A^{\boldsymbol{\beta}}}{\cG^{D_0/2}}
    \prod_{i \in \Gamma} \frac{1}{z_i}
    \prod_{i \notin \Gamma} z_i^{\alpha_i-1}
    \>,
\end{equation}
up to factors independent of the Feynman parameters, where $\cR_A^{\boldsymbol{\beta}}$
also includes the rational factors from negative propagator powers.
Each pinched edge thus contributes a single simple pole, and the remaining factor is
regular on the coordinate hyperplanes away from $\brc{\cG = 0}$.

The two cohomology groups in~\cref{eq:absolute_twisted_cohomology}
are finite-dimensional vector spaces of equal dimension $\nu$,
\begin{equation}
\label{eq:nu_def}
    \nu
    =
    \dim H^n\bigbrk{T_\delta \,;\, \nabla_{\dd W_\delta}}
    =
    \dim H^n\bigbrk{T_\delta^\vee \,;\, \nabla_{\dd W_\delta}^\vee}
    \>,
\end{equation}
which, for generic kinematics, equals the number of master integrals in the family~\cite{Bitoun:2017nre,Frellesvig:2019uqt}.
We fix bases of twisted cocycles and dual twisted cocycles. We will frequently denote cocycles as $e_a \in H^n\bigbrk{T_\delta \,;\, \nabla_{\dd W_\delta}}$ and dual cocycles as $h_b \in H^n\bigbrk{T_\delta^\vee \,;\, \nabla_{\dd W_\delta}^\vee}$. We will denote the bases themselves as
\begin{equation}
\label{eq:bases_def}
    \mathbf e = \{e_a\}_{a=1}^{\nu}
    \>,
    \qquad
    \mathbf h = \{h_b\}_{b=1}^{\nu}
    \>.
\end{equation}
Moreover, we will denote the associated twisted periods and dual periods as
\begin{equation}
    I^{(\delta)}_a = \int_{\cC} e^{\varepsilon W_\delta} e_a, \qquad \qquad I_b^{(\delta)\,\vee} = \int_{\cC^\vee} e^{-\varepsilon W_\delta} h_b.
    \label{eq:IntegralBasisDef}
\end{equation}
The twisted periods $I_a^{(\delta)}$ form a basis of master integrals.

The dimension $\nu$ is a topological invariant of the ambient space.
For generic exponents $\boldsymbol{\delta}$ in~\cref{eq:leepom_reg}, 
$\nu$ equals the signed Euler characteristic\footnote{
    For generic exponents, the twisted cohomology is concentrated in the top degree, and all other cohomologies vanish:
    $\dim H^k\bigbrk{T_\delta \,;\, \nabla_{\dd W_\delta}} = 0$ for $k \neq n$~\cite[Theorem 2.4 and Appendix A]{Agostini:2022cgv}.
    See also Refs.~\cite{Aomoto:1975,KitaNoumi:1982,Cho:1997} and Ref.~\cite[Section 2.8]{AomotoKitaKohnoIohara} for earlier results in hypergeometric settings.
}~\cite{Bitoun:2017nre,Agostini:2022cgv},
\begin{equation}
\label{eq:nu_euler}
    \nu
    =
    \brk{-1}^n \, \chi\brk{T_\delta}
    \>.
\end{equation}
For the democratic choice of regulators~\labelcref{eq:democratic_choice}, this holds provided that the specialisation preserves the genericity of the exponents, which we assume throughout.
The critical points of the potential $W_\delta$ form the set
\begin{equation}
\label{eq:def_critical_points_intro}
    \operatorname{Crit}\brk{W_\delta}
    \defas
    \Bigbrc{\>
        z \in T_\delta
        \,\,
    \>\Big|\>
    \,\,
        \dd W_\delta \equiv \sum_{i = 1}^n \Bigbrk{
            \frac{\partial_i \cG}{\cG} \,+\, \frac{\delta}{z_i}
        } \> \dd z_i
        = 0
    \>}
    \>.
\end{equation}
Under the genericity assumption, the critical points are isolated and non-degenerate, and their number is again the signed Euler characteristic~\cite{huh2013maximum}.
Combined with~\cref{eq:nu_euler}, this gives the geometric interpretation
\begin{equation}
\label{eq:nu_crit}
    \nu
    =
    (-1)^n\,\chi\brk{T_\delta}
    =
    \bigabs{\operatorname{Crit}\brk{W_\delta}}
    \>.
\end{equation}
Master integrals can often also be counted from the critical points of the Lee--Pomeransky polynomial itself~\cite{Lee:2013hzt}. The number of master integrals is the sum, over all subsets $S$ of pinched propagators, of the Milnor numbers of the proper critical points of $\cG_S \defas \cG\bigr\rvert_{z_i = 0,\; i \in S}$, namely the critical points at which $\cG_S \neq 0$.
These are unregulated critical points, and their relation to $\operatorname{Crit}\brk{W_\delta}$ as $\delta \to 0$ will be discussed in \cref{sec:delta_expansion}, where the sum over $S$ becomes a sum over sectors.
\subsection{Differential Equations for Feynman Integrals}
\label{ssec:de_from_intersection}
The regulated Feynman integral $I_a^{(\delta)}$ defined in
\cref{eq:IntegralBasisDef} is a function of the external kinematics, which we
denote generically as $x$. We now describe how differential equations for
Feynman integrals arise in the language of twisted cohomology.

A Feynman integral $I_a^{(\delta)}$ depends on $x$ through the potential and the differential forms. 
Under the integral, kinematic derivatives act on twisted forms through the
covariant derivative $\nabla_x$,
\begin{equation}
\label{eq:GM_connection}
    \dd_x I_a^{(\delta)} 
    = 
    \dd_x \int_{\cC} e^{\varepsilon W_\delta}\, e_a
    =
    \int_{\cC} e^{\varepsilon W_\delta}\, \nabla_x\, e_a
    \>, 
    \qquad 
    \nabla_x 
    = 
    \dd_x + \varepsilon \, \dd_x W_\delta\wedge
    \>.
\end{equation}
Here $\dd_x$ denotes the differential with respect to the kinematic variables only.
Reducing $\nabla_x e_a$ back to the basis $\mathbf e$ gives 
the differential system satisfied by the vector of master integrals 
$\mathbf I^{(\delta)} = \brk{I_1^{(\delta)},\ldots,I_\nu^{(\delta)}}\transpose$,
\begin{equation}
\label{eq:DE_system}
    \dd_x \mathbf{I}^{(\delta)} = \mathbf\Omega_{\delta}(x, \varepsilon) \,\cdot\, \mathbf{I}^{(\delta)}
    \>,
\end{equation}
where $\mathbf{\Omega}_{\delta}(x, \varepsilon)$ is a matrix of one-forms in the external kinematics, known as the (regulated) connection matrix.
Naturally, the differential system for the unregulated Feynman integrals can be recovered
by taking the limit $\delta\to0$ of the regulated system,
\begin{equation}
    \dd_x \mathbf{I} 
    = 
    \mathbf{\Omega}(x, \varepsilon) \,\cdot\, \mathbf{I}
    \>, 
    \qquad  
    \mathbf{\Omega} = \lim_{\delta\to 0}\mathbf\Omega_{\delta} \>,
    \label{eq:unregulatedDiffSystem}
\end{equation}
which defines the unregulated connection matrix $\mathbf{\Omega}$.
The matrix $\mathbf{\Omega}(x, \varepsilon)$ depends on the external kinematics $x$ alongside the dimensional 
regulator $\varepsilon$. For a basis of forms that are rational in both the kinematics $x$ and 
the regulator $\varepsilon$, $\mathbf{\Omega}$ is also rational.
A particularly important form of differential system is one that is ``canonical''~\cite{Henn:2013pwa}. In cases where 
the underlying Feynman integrals are polylogarithmic, the connection matrix of a canonical differential 
system can be written as
\begin{equation}
    \mathbf{\Omega} = \varepsilon \sum_k \mathbf A_k \> \dd\log\brk{w_k\brk{x}}
    \>,
    \label{eq:epFactorDE}
\end{equation}
where the $\mathbf A_k$ are constant matrices, and the letters $w_k\brk{x}$ are algebraic functions of the external kinematics.

The goal of this work is to construct $\mathbf{\Omega}$ from intersection
numbers. In~\cref{sec:eps_expansion}, we express the regulated connection matrix
$\mathbf\Omega_\delta$ through intersection numbers and their
large-$\varepsilon$ expansion. In \cref{sec:delta_expansion}, we discuss how to
compute the $\delta \to 0$ limit in~\cref{eq:unregulatedDiffSystem}.

%% file: Sections/eps_expansion.tex
\section{%
    \texorpdfstring
    {Large-$\varepsilon$ Expansion of Regulated Intersection Pairings}
    {Large-epsilon Expansion of Regulated Intersection Pairings}
}
\label{sec:eps_expansion}
In this section, we introduce the regulated intersection pairing and recall
details of its asymptotic expansion in the limit $\varepsilon\to\infty$, keeping
the analytic regulator $\delta$ fixed.
The coefficients of this expansion are the higher residue
pairings~\cite{Saito:1983}, which allow us to compute intersection
numbers order by order.
Intersection numbers, IBP
reduction coefficients, and connection matrices are rational
in $\varepsilon$ and can therefore be reconstructed from
sufficiently many terms in this expansion.
We focus on bases for which the connection matrix is polynomial
in $\varepsilon$ and restrict to degree at most one.
We show how the leading and first subleading pairings fully determine
such systems, including the $\varepsilon$-factorised
form of~\cref{eq:epFactorDE}.

\subsection{Intersection Numbers}
\label{ssec:intersection_numbers}
The two twisted cohomology groups introduced in \cref{eq:absolute_twisted_cohomology} admit a non-degenerate bilinear
pairing known as the \textit{intersection number},
\begin{equation}
\label{eq:intersection_pairing_def}
    \regpair{\cdot}{\cdot}
    \quad : \quad
    H_c^n\bigbrk{T_\delta \,;\, \nabla_{\dd W_\delta}}
    \times
    H^n\bigbrk{T_\delta^\vee \,;\, \nabla_{\dd W_\delta}^\vee}
    \,\,
    \longrightarrow
    \,\,
    \mathbb C
    \>,
\end{equation}
where $H_c^n$ denotes compactly supported twisted cohomology.
The pairing associates to a twisted cocycle $\varphi$ and a dual twisted cocycle $\varphi^\vee$
the complex number $\regpair{\varphi}{\varphi^\vee}$.
It is defined as
\begin{equation}
\label{eq:pairing_integral_def}
    \regpair{\varphi}{\varphi^\vee}
    \defas
    \lrbrk{\frac{\varepsilon}{2\pi\im}}^{\!n}
    \int_{T_\delta} \iota_c\brk{\varphi}\wedge\varphi^\vee
    \>,
\end{equation}
where $\iota_c\brk{\varphi}$ denotes a compactly supported representative of the cohomology class of $\varphi$. 
The normalisation in \cref{eq:pairing_integral_def} is 
chosen so that the asymptotic expansion of the intersection
number around $\varepsilon\to\infty$ starts at order $\varepsilon^0$.
The intersection pairings between the bases $\mathbf e$ and $\mathbf h$ of~\cref{eq:bases_def}
define the intersection matrix, or $\mathbf{C}$-matrix,
\begin{equation}
\label{eq:C_matrix_def}
    \mathbf C_{\delta,ab}
    =
    \regpair{e_a}{h_b}
    \>.
\end{equation}
Since the pairing is non-degenerate, the matrix $\mathbf C_{\delta}$ is invertible.

The connection
matrix $\mathbf\Omega_{\delta}$ of \cref{eq:DE_system} can be expressed in terms of intersection numbers~\cite{Mastrolia:2018uzb,Frellesvig:2019uqt,Fontana:2023amt,Brunello:2023rpq},
\begin{equation}
\label{eq:Omega_from_C}
    \mathbf\Omega_{\delta} 
    = 
    \mathbf D_{\delta} \,\cdot\, \left(\mathbf C_{\delta}\right)^{-1}
    \>,
    \qquad \qquad
    \mathbf D_{\delta,ab} 
    = 
    \regpair{ \nabla_x e_a}{ h_b} 
    \>.
\end{equation}
We introduce the coordinate components of
the twisted differentials,
\begin{equation}
\label{eq:twisted_differential_components}
    \nabla_i
    =
    \partial_i + \varepsilon\,\partial_i W_\delta
    \>,
    \qquad
    \nabla_i^\vee
    =
    \partial_i - \varepsilon\,\partial_i W_\delta
    \>.
\end{equation}
As shown in Ref.~\cite{Mizera:2019vvs},
intersection numbers admit an equivalent representation in terms of multivariate residues localised on
the critical points of the potential,
\begin{equation}
\label{eq:grothendieck_residue}
    \regpair{\varphi}{\varphi^\vee}
    =
    \varepsilon^n\,
    \operatorname{Res}_{\dd W_\delta = 0}
    \lrbrk{\hat{\varphi}^\vee\,\nabla_1^{-1}\nabla_2^{-1}\cdots \nabla_n^{-1}\hat{\varphi}\;\dd^n z }
    \>,
\end{equation}
where the formal inverses $\nabla_i^{-1}$ act on holomorphic functions in a neighbourhood of $\operatorname{Crit}\brk{W_\delta}$.
The operators commute, so their order in
\cref{eq:grothendieck_residue} is immaterial.
The symbol $\operatorname{Res}_{\dd W_\delta=0}$ denotes the sum of the local Grothendieck residues at the critical points $\dd W_\delta = 0$, given by
\begin{equation}
\label{eq:global_residue_def}
    \operatorname{Res}_{\dd W_\delta = 0}\brk{\eta}
    =
    \frac{1}{(2\pi\im)^n}
    \oint_{|\partial_1 W_\delta| = \rho_1 } \oint_{|\partial_2 W_\delta| = \rho_2 } 
    \cdots 
    \oint_{|\partial_n W_\delta| = \rho_n }\eta
    \>,
\end{equation}
with $\eta$
a meromorphic $n$-form, holomorphic  away from the hypersurfaces $\partial_i W_\delta = 0$,
and $\rho_i$ sufficiently small positive constants.
Exchanging the roles of the two forms gives the equivalent dual representation~\cite{Mizera:2019vvs}:
\begin{equation}
\label{eq:grothendieck_residue_dual}
    \regpair{\varphi}{\varphi^\vee}
    =
    \brk{-\varepsilon}^n\,
    \operatorname{Res}_{\dd W_\delta = 0}
    \lrbrk{\hat{\varphi}\,\brk{\nabla_1^\vee}^{-1}\brk{\nabla_2^\vee}^{-1}\cdots \brk{\nabla_n^\vee}^{-1}\hat{\varphi}^\vee\;\dd^n z }
    \>.
\end{equation}
The formal inverses $(\nabla_i^\vee)^{-1}$ are obtained
from $\nabla_i^{-1}$ by replacing
$\varepsilon\to-\varepsilon$.
Expanding either representation in $\varepsilon^{-1}$
gives the higher residue pairings, which we consider next.
\subsection{Higher Residue Pairings}
\label{sec:hrp}
The intersection pairing is a rational function of $\varepsilon$ and so admits an asymptotic expansion in the limit $\varepsilon\to\infty$,
\begin{equation}
\label{eq:pairing_eps_expansion}
    \regpair{\varphi}{\varphi^\vee}
    =
    \sum_{k=0}^{\infty}
    \varepsilon^{-k}\,
    \regpair[k]{\varphi}{\varphi^\vee}
    \>.
\end{equation}
The coefficients $\regpair[k]{\varphi}{\varphi^\vee}$ are known as \textit{higher residue pairings}~\cite{Saito:1983,Mizera:2019vvs}.
They obey the symmetry property
\begin{equation}
\label{eq:pairing_symmetry}
    \regpair[k]{\varphi}{\varphi^\vee}
    =
    (-1)^k\,
    \regpair[k]{\varphi^\vee}{\varphi}
    \>,
\end{equation}
where the pairing on the right-hand side has the two twisted cohomology groups exchanged.
In particular, the odd-order pairings are antisymmetric.

For an $\varepsilon$-dependent representative, the form is first expanded in $\varepsilon^{-1}$, and the
pairings are then evaluated term by term.
The higher residue pairings are obtained by expanding the formal inverses in~\cref{eq:grothendieck_residue} at large $\varepsilon$,
\begin{equation}
\label{eq:nabla_inverse}
    \nabla_i^{-1}\,\hat{\varphi}
    =
    \frac{1}{\varepsilon}\,
    \frac{\hat{\varphi}}{\partial_i W_\delta}
    -
    \frac{1}{\varepsilon^2}\,
    \frac{1}{\partial_i W_\delta}\,
    \partial_i\lrbrk{\frac{\hat{\varphi}}{\partial_i W_\delta}}
    +
    \cO\brk{\varepsilon^{-3}}
    \>,
\end{equation}
which follows by imposing $\nabla_i\,\nabla_i^{-1}\hat{\varphi} = \hat{\varphi}$ order by order in $1/\varepsilon$~\cite{Mizera:2019vvs}.
The leading residue pairing is
\begin{equation}
    \regpair[0]{\varphi}{\varphi^\vee}
    =
    \operatorname{Res}_{\dd W_\delta = 0}
    \lrbrk{
        \hat\varphi\,\hat\varphi^\vee\; \frac{\dd^n z} {\partial_1 W_\delta \cdots \partial_n W_\delta}
    }
    \>.
\end{equation}
The residue can be evaluated by changing variables from $z_i$ to $\partial_i W_\delta$. The Jacobian is $\det\bigbrk{\, \Phi_\delta \,}$,
where $\Phi_\delta$ is the Hessian of the regulated potential $W_\delta$,
\begin{equation}
\label{eq:def_regulated_Hessian}
    \brk{\Phi_\delta}_{ij}
    =
    \partial_i\partial_j W_\delta
    =
    \Phi_{ij}
    -
    \delta\,\frac{\delta_{ij}}{z_i^2}
    \>,
    \qquad
    \Phi_{ij}
    \defas
    \partial_i\partial_j \log \cG
    \>,
\end{equation}
and $\Phi$ is the Hessian of the potential at $\delta = 0$.
For non-degenerate critical points, $\Phi_\delta$ is invertible, and hence the Jacobian is non-zero.
Performing the residue, one thus finds that it localises at the critical points~\cite{Mizera:2017rqa,Mizera:2019vvs},
\begin{equation}
\label{eq:lo_hrp}
    \regpair[0]{\varphi}{\varphi^\vee}
    =
    \sum_{z\brk{\delta} \in \operatorname{Crit}\brk{W_\delta}}
    \left.
        \frac{\hat\varphi\,\hat\varphi^\vee}{\det\bigbrk{\, \Phi_\delta \,}}
    \right\rvert_{z=z\brk{\delta}}
    \>,
\end{equation}
where $\hat\varphi$ and $\hat\varphi^\vee$ are assumed to be regular on the critical locus.
By expanding to higher order in $\varepsilon^{-1}$, one can find analogous results for higher order pairings.
In this work, we will limit ourselves to studying the $\delta \to 0$ limit up to the first subleading pairing, which
can be written as~\cite{Mizera:2019vvs}
\begin{equation}
\label{eq:nlo_hrp}
    \regpair[1]{\varphi}{\varphi^\vee}
    =
    \frac{1}{2}
    \sum_{z\brk{\delta} \in \operatorname{Crit}\brk{W_\delta}}
    \>
    \sum_{i ,\, j = 1}^n
    \brk{\Phi_\delta^{-1}}_{ij}\,
    \partial_j
    \left.
        \frac{
            \hat\varphi\,\partial_i\hat\varphi^\vee
            -
            \hat\varphi^\vee\,\partial_i\hat\varphi
        }{\det\bigbrk{\, \Phi_\delta \,}}
    \right\rvert_{z = z\brk{\delta}}
    \>,
\end{equation}
which is manifestly antisymmetric, in agreement with \cref{eq:pairing_symmetry}.

\subsection{%
    \texorpdfstring
    {Large-$\varepsilon$ Expansion of Differential Equation Matrices}
    {Large-epsilon Expansion of Differential Equation Matrices}
}
\label{ssec:expansion_de_system}
We now study the large-$\varepsilon$ expansion of the intersection matrices $\mathbf C_{\delta}$, $\mathbf D_{\delta}$, and $\mathbf\Omega_{\delta}$ introduced in \cref{eq:C_matrix_def,eq:Omega_from_C}.
We choose bases of twisted cocycles $\mathbf e$ and $\mathbf h$ as in~\cref{eq:bases_def} such that the elements of $\mathbf e$ admit Laurent expansions in $\varepsilon^{-1}$ starting at order
$\varepsilon^0$, while the elements of
$\mathbf h$ are independent of $\varepsilon$. 
The $\Gamma$-function normalisation in~\cref{eq:leepom_form} does not directly have this property: its large-$\varepsilon$ asymptotics is governed by Stirling's formula, and terms of the form $\varepsilon^\varepsilon$ can arise.
For the democratic choice of regulator, we introduce the kinematic-independent rescaling
\begin{equation}
\label{eq:rational_normalisation}
    \varphi_{\alpha_1\ldots \alpha_n}
    =
    \frac{\Gamma\brk{c-\brk{L+1}\varepsilon}}{\varepsilon^{p}\,\Gamma\brk{\brk{L+1}D/2-|\boldsymbol{\alpha}|- n \varepsilon\,\delta}}
    \frac{1}{\cG^{D_0/2}}
    \prod_{i=1}^n \frac{\dd z_i \, z_i^{\alpha_i-1}}{\Gamma\brk{\alpha_i+\varepsilon \delta}}
    \>, 
\end{equation}
where $c$ and $p$ are integers and $c$ is fixed across the family.
For integer propagator powers, the $\Gamma$-function prefactor is rational in $\varepsilon$ and $\delta$, up to a kinematic-independent factor common to all basis elements, which cancels from $\mathbf\Omega_\delta$.
The powers $p$ are chosen so that each basis form starts at order $\varepsilon^0$. This guarantees that rational integrals have rational expansions.

With this normalisation, each element of the basis of differential forms $\mathbf e$ has a Laurent expansion that starts at order $\varepsilon^0$:
\begin{equation}
\label{eq:e_eps_expansion}
    e_a
    =
    \sum_{r=0}^{\infty}\varepsilon^{-r}\,e_a^{(-r)}
    \>.
\end{equation}
In the applications, the dual basis elements are chosen proportional to the corresponding
$e_a$, with the proportionality factors removing their $\varepsilon$-dependence.

The expansions~\labelcref{eq:pairing_eps_expansion,eq:e_eps_expansion} induce corresponding expansions of the intersection matrices.
Since $\nabla_x$ contains an additional power of $\varepsilon$, the leading power of
$\mathbf D_{\delta}$ is shifted by one unit with respect to that of $\mathbf C_{\delta}$:
\begin{equation}
\label{eq:C_expansion}
   \ \ \  \mathbf C_{\delta}
    =
    \mathbf C_{\delta}^{(0)}
    +
    \varepsilon^{-1}\mathbf C_{\delta}^{(-1)}
    +
    \varepsilon^{-2}\mathbf C_{\delta}^{(-2)}
    + \cdots
    \>,
\end{equation}
\begin{equation}
\label{eq:D_expansion}
    \mathbf D_{\delta}
    =
    \varepsilon\,\mathbf D_{\delta}^{(1)}
    +
    \mathbf D_{\delta}^{(0)}
    +
    \varepsilon^{-1}\mathbf D_{\delta}^{(-1)}
    + \cdots
    \>,
\end{equation}
and therefore:
\begin{equation}
\label{eq:Omega_expansion}
    \mathbf\Omega_{\delta}
    =
    \varepsilon\,\mathbf\Omega_{\delta}^{(1)}
    +
    \mathbf\Omega_{\delta}^{(0)}
    +
    \varepsilon^{-1}\mathbf\Omega_{\delta}^{(-1)}
    + \cdots\>.
\end{equation}
For brevity, in the following formulae, we suppress the indices $a$ and $b$ labelling the basis elements, so that $\regpair{e}{h}$ denotes the matrix with entries $\regpair{e_a}{h_b}$.
In terms of higher residue pairings, the coefficients of $\mathbf C_{\delta}$ read
\begin{equation}
\label{eq:C_theta_formula}
    \mathbf C_{\delta}^{(s)}
    =
     \sum_{r=0}^{-s}
    \regpair[-r-s]{e^{(-r)}}{h}
    \>,
    \qquad
    s \leq 0
    \>,
\end{equation}
while $\mathbf D_{\delta}$ admits the expansion
\begin{align}
\label{eq:D_theta_formula}
    \mathbf D_{\delta}^{(s)}
    ={}&
    \sum_{r=0}^{1-s}
    \regpair[-r-s+1]{\dd_x W_\delta\wedge e^{(-r)}}{h}+
    \sum_{r=0}^{-s}
    \regpair[-r-s]{\dd_x e^{(-r)}}{h}
    \>,
    \qquad
    s \leq 1
    \>.
\end{align}
The second sum in~\cref{eq:D_theta_formula} is empty for $s=1$.
Finally, the unregulated connection matrix defined in~\cref{eq:unregulatedDiffSystem} has the analogous expansion
\begin{equation}
\label{eq:Omega_unregulated_expansion}
    \mathbf\Omega
    =
    \varepsilon\,\mathbf\Omega^{(1)}
    +
    \mathbf\Omega^{(0)}
    +
    \varepsilon^{-1}\mathbf\Omega^{(-1)}
    + \cdots
    \>.
\end{equation}
Next we show how the connection matrix for systems linear in $\varepsilon$ is determined by the leading and first subleading residue pairings.

\subsection{%
    \texorpdfstring
        {Differential Systems Linear in $\varepsilon$}
        {Differential Systems Linear in epsilon}
}
\label{ssec:eps_linear_systems}
We restrict our discussion to differential systems whose connection matrix is polynomial in $\varepsilon$ of degree at most one, so that only $\mathbf\Omega_{\delta}^{(1)}$ and $\mathbf\Omega_{\delta}^{(0)}$ are non-vanishing.
To construct these matrices, we assume that, for generic kinematics and finite $\delta$, the leading coefficient $\mathbf{C}_\delta^{\brk{0}}$ of the regulated intersection matrix is invertible.
This assumption holds for the bases considered in this work.
Expanding \cref{eq:Omega_from_C} order by order in $\varepsilon$ gives
\begin{align}
    \label{eq:Omegalin}
    \mathbf\Omega_{\delta}^{(1)}
    &  =
    \mathbf D_{\delta}^{(1)}\cdot \left(\mathbf C_{\delta}^{(0)}\right)^{-1} 
    \> ,
    \\
    \mathbf\Omega_{\delta}^{(0)}
    &  =
    \left( \mathbf D_{\delta}^{(0)}-\mathbf D_{\delta}^{(1)}\cdot \left(\mathbf C_{\delta}^{(0)}\right)^{-1} \cdot  \mathbf C_{\delta}^{(-1)}\right) \cdot \left(\mathbf C_{\delta}^{(0)}\right)^{-1} 
    \>.
\end{align}
The coefficients of $\mathbf C_{\delta}$ and $\mathbf D_{\delta}$ entering these expressions
follow from \cref{eq:C_theta_formula,eq:D_theta_formula}:
\begin{align}
    \label{eq:CDlin}
    \mathbf C_{\delta}^{(0)}
    & =
    \regpair[0]{e^{(0)}}{h}\, ,
    \\
    \mathbf C_{\delta}^{(-1)}
    & =
    \regpair[1]{e^{(0)}}{h}
    +
    \regpair[0]{e^{(-1)}}{h}\, ,
    \\
    \mathbf D_{\delta}^{(1)}
    & =
    \regpair[0]{\dd_x W_\delta\wedge e^{(0)}}{h}\, ,
    \\
    \mathbf D_{\delta}^{(0)}
    & =
    \regpair[1]{\dd_x W_\delta\wedge e^{(0)}}{h}
    +
    \regpair[0]{\dd_x W_\delta\wedge e^{(-1)}}{h}
    +
    \regpair[0]{\dd_x e^{(0)}}{h}\, .
\end{align}
Therefore, only the leading and
first subleading higher residue pairings~\labelcref{eq:lo_hrp,eq:nlo_hrp} enter
$\varepsilon$-linear differential systems.

For the canonical bases considered in~\cref{sec:applications}, the unregulated connection matrix is $\varepsilon$-factorised, although the regulated matrix need not have this property.
Consequently,
\begin{equation}
    \lim_{\delta\to0}\mathbf\Omega_{\delta}^{(0)}=\mathbf\Omega^{(0)}=\mathbf 0
    \>.
\end{equation}
This cancellation of contributions to $\mathbf\Omega^{(0)}$ will provide a non-trivial check of the calculations in~\cref{sec:applications}.
We now turn to the $\delta \to 0$ limit of the residue pairings needed to compute the unregulated connection matrix $\mathbf\Omega$.

%% file: Sections/deltaexpansion.tex
\section{Vanishing-Regulator Limit of Intersection Pairings}
\label{sec:delta_expansion}
The leading and higher residue pairings of two forms $\varphi_a$ and
$\varphi_b$, as introduced in \cref{sec:hrp}, depend on both the kinematic
variables $x$ and the analytic regulator $\delta$.
For practical calculations, however, we are interested in their
vanishing-regulator limit $\delta \to 0$, which has been previously studied
within the fibration-based framework in
Refs.~\cite{Fontana:2023amt,Brunello:2023rpq}.
As is well known (see e.g.~\cite{Mizera:2019vvs,Fontana:2023amt}), intersection
pairings can be singular or vanish in the limit $\delta \to 0$.
As we will see in~\cref{ssec:lo_pairing}, the pairings
$\regpair[k]{\varphi_a}{\varphi_b^\vee}$ vanish as $\delta \to 0$ at least as
$\delta^{N_{\Gamma_b}}$, where $N_{\Gamma_b}$ is the number of edges in
$\Gamma_b$, the set of pinched edges associated with $\varphi_b$.
Since the connection matrix~\labelcref{eq:Omega_from_C} does not depend on the
normalisation of the dual forms, we avoid this suppression by pairing instead with the
\textit{rescaled dual forms} of Ref.~\cite{Fontana:2023amt},
\begin{equation}
\label{eq:basis_normalisation}
    \vartheta_b
    \defas
    \delta^{-N_{\Gamma_b}}\;\varphi_b^{\vee}
    \>.
\end{equation}
The aim of this section is to derive closed formulae for the limits
\begin{equation}
\label{eq:pairing_limits}
    \vev{\varphi_a \mid \vartheta_b}_{(k)}
    \defas
    \lim_{\delta \to 0} \regpair[k]{\varphi_a}{\vartheta_b}
    \>,
    \qquad
    k = 0, 1
    \>,
\end{equation}
of the leading and first subleading residue pairings, entirely in terms of unregulated sector data.

Our strategy is pedestrian.
The residue formulae~\labelcref{eq:lo_hrp,eq:nlo_hrp} express the pairings as sums over the
regulated critical points $z\brk{\delta}$ of rational functions of the forms, the
Hessian $\Phi_\delta$ and their derivatives.
We simply expand each of these ingredients in $\delta$ at $z\brk{\delta}$.
As we will see, the critical points organise into sectors $\Gamma$ as $\delta \to 0$,
labelled by the edges whose coordinates vanish in the limit.
The dependence of the leading pairing on $\delta$ can then be collected into a
single factor, the residue measure, built from the poles of the two forms on the
coordinate hyperplanes and the determinant of the Hessian.
Its behaviour as $\delta \to 0$ decides which sectors contribute: only those
with $\Gamma_a \subseteq \Gamma \subseteq \Gamma_b$, where $\Gamma_a$ and
$\Gamma_b$ are the pinch sets of the two forms.
The subleading pairing carries the same residue measure, multiplied by a factor
that stays finite in the limit, so the same sectors contribute.
In both cases, the limit is expressed entirely through unregulated sector data.

\subsection{Behaviour of Critical Points}
\label{ssec:critical_sectors}
We begin by considering the expansion of the critical points of the regulated
potential $W_\delta$ around $\delta = 0$.
To this end, we introduce the logarithmic derivatives of the Lee--Pomeransky polynomial
\begin{equation}
\label{eq:def_rho_function}
    \omega_i\brk{z}
    \defas
    \partial_i \log\cG
    \>,
\end{equation}
which allows us to write the critical point equations of
\cref{eq:def_critical_points_intro} as
\begin{equation}
\label{eq:critical_points_componentwise}
    z_i \> \omega_i\brk{z} + \delta = 0
    \>,
    \qquad
    \cG\brk{z} \neq 0
    \>,
    \qquad
    z_i \neq 0
    \>.
\end{equation}
Our task is to solve \cref{eq:critical_points_componentwise} as a series expansion in the analytic regulator.
In order to proceed, we assume that
the critical points of the democratically regulated $W_\delta$ are isolated and
non-degenerate, and that their number equals the number $\nu$ of master
integrals,
cf.~\cref{eq:nu_crit}~\cite{Lee:2013hzt,Frellesvig:2019uqt,Crisanti:2026rbc,Wu:2026xea}.
In general, the critical points are algebraic functions of $\delta$ and, by the
Newton--Puiseux theorem, are only guaranteed to admit Puiseux expansions in
fractional, possibly negative, powers $\delta^{k/p}$.
We assume, more restrictively, that each critical point admits a Taylor
expansion around $\delta=0$,
\begin{equation}
\label{eq:crit_taylor}
    z_i\brk{\delta}
    =
    \sum_{k \ge 0} z_{i, k} \, \delta^{k}
    \>,
    \quad
    \text{for $i \in \brc{1, \ldots, n}$}
    \>.
\end{equation}
Experience shows that these assumptions hold for a wide range of Feynman-integral families, including those studied in~\mbox{\cref{sec:applications}}.
Each critical point therefore stays finite as $\delta \to 0$, and we denote its limit by
\begin{equation}
\label{eq:def_zstar}
    z^\star \defas z\brk{0}
    \>.
\end{equation}
It can be shown\footnote{%
    The critical point equations imply $\partial_\delta \log \cG\bigbrk{z\brk{\delta}} = - \delta \sum_{i = 1}^n \partial_\delta \log z_i\brk{\delta}$.
    An algebraic function $f\brk{\delta} \not\equiv 0$ behaves as $c\, \delta^{m}$ with $c \neq 0$ as $\delta \to 0$, so $\partial_\delta \log f$ has at most a simple pole with residue $m$.
    The right-hand side is therefore regular at $\delta = 0$, so $\cG\bigbrk{z\brk{\delta}}$ has $m = 0$ and a non-zero limit, which equals $\cG\brk{z^\star}$ since $z\brk{\delta}$ stays finite by~\labelcref{eq:crit_taylor}.
}
that expansion~\labelcref{eq:crit_taylor}, together with the regulated
critical point equations~\labelcref{eq:critical_points_componentwise}, implies that $\cG\brk{z^\star}$ is non-zero.
We therefore see that the logarithmic derivatives~$\omega_i\bigbrk{z\brk{\delta}}$ also admit Taylor expansions.

An important observation about Taylor series solutions to the critical point
equations~\labelcref{eq:critical_points_componentwise} is that they develop a
combinatorial structure.
That is, at leading order, \cref{eq:critical_points_componentwise} implies that either $z^\star_i$ or $\omega_i\brk{z^\star}$ must vanish.
The critical points therefore fall into sectors
labelled by $\Gamma \subseteq \brc{1, \ldots, n}$, the indices for which $z^\star_i = 0$.
We refer to the corresponding edges as \textit{pinched} and write $N_\Gamma = \abs{\Gamma}$ for their total number.
The complementary set $\overline{\Gamma} \defas \brc{1, \ldots, n} \setminus \Gamma$ consists of the \textit{cut} edges.
The indices of pinched and cut edges are written as $i \in \Gamma$ and $\bar{i} \in \overline{\Gamma}$, respectively.
We stress that, throughout this work, sectors are labelled by their pinched edges rather than by the complementary set of unpinched propagators commonly used in the Feynman-integral literature.

In order to later treat the critical points computationally, we formulate
\cref{eq:critical_points_componentwise} as a set of polynomial equations in
$z^\star$.
Specifically, a $z^\star$ associated with a sector $\Gamma$ is a solution of\footnote
{%
    Here we have used that there are no solutions where $z_i\brk{\delta}$ is
    exactly zero.
    This follows since setting $z_i\brk{\delta} = 0$ in the first equality of \cref{eq:critical_points_componentwise} gives the constraint $\delta = 0$.
    Hence, when considering the limit $\delta \rightarrow 0$ (where $\delta \ne 0$) there are no such solutions.
}
\begin{equation}
\label{eq:critical_eqs_full}
    z^\star_{i} = 0 \quad \text{for $i\in\Gamma$},
    \qquad
    \text{$\partial_{\bar{i}} \cG\brk{z^\star} = 0$} \quad \text{for $\bar{i} \in \overline{\Gamma}$},
    \qquad
    \cG\brk{z^\star} \neq 0
    \>,
\end{equation}
where we have used $\cG\brk{z^\star} \neq 0$ to replace the conditions $\omega_{\bar{i}}\brk{z^\star} = 0$ by the equivalent polynomial conditions $\partial_{\bar{i}} \cG\brk{z^\star} = 0$.
These are the unregulated critical point equations of Ref.~\cite{Lee:2013hzt},
imposed sector by sector. (See Ref.~\cite{Crisanti:2026cuf} for recent
applications to discontinuities of Feynman integrals.)
The solutions of \cref{eq:critical_eqs_full} form a set that we call
the sector critical locus, $\operatorname{Crit}_\Gamma\brk{W}$.
In general, the sector critical locus may not be composed of a set of isolated
points (see e.g.~\cite{Lee:2013hzt,Page:2025gso,Crisanti:2026rbc,Wu:2026xea}).
Nevertheless, in this work, we assume that the points of $\operatorname{Crit}_\Gamma\brk{W}$
are isolated and non-degenerate.
Under this assumption, the sector critical locus has finite cardinality $\nu_\Gamma \defas
\abs{\operatorname{Crit}_\Gamma\brk{W}}$.
Since the number of regulated critical points equals the dimension $\nu$ of the cohomology group~\labelcref{eq:absolute_twisted_cohomology},
we obtain~\cite{Lee:2013hzt,Frellesvig:2019uqt}
\begin{equation}
    \nu = \sum_\Gamma \nu_\Gamma
    \>.
\end{equation}
Importantly, the grouping of the critical points into sectors also organises the
  residue sums of~\cref{eq:lo_hrp,eq:nlo_hrp}.
That is, the sum over critical points in the residue formulae splits into a
sum over sectors $\Gamma$ and, within each, a sum over
$\operatorname{Crit}_\Gamma\brk{W}$.

Altogether we see that, for  a cut edge, the coordinate tends to a non-zero constant,
\begin{equation}
\label{eq:cut_edge_limit}
    z_{\bar{i}}\brk{\delta}
    \xrightarrow{\;\delta\to0\;}
    z^\star_{\bar{i}} \neq 0
    \>,
    \quad
    \text{for $\bar{i} \in \overline{\Gamma}$}
    \>.
\end{equation}
For the pinched coordinates, however, the zeroth-order term $z^\star_i$ of the
expansion~\labelcref{eq:crit_taylor} vanishes and is therefore not the leading
order.
This behaviour enters the residue pairings, for example, as the forms carry
pinch factors $\delta/z_i$ (see~\cref{ssec:pinch_decomposition}).
For a pinched edge, such a ratio gives a finite and non-zero limit as
\begin{equation}
\label{eq:def_rho}
    \frac{\delta}{z_i\brk{\delta}}
    \xrightarrow{\;\delta\to0\;}
    -\,\omega^\star_i
    \>,
    \qquad
    \omega^\star_i \defas \omega_i\brk{z^\star}
    \neq 0
    \>,
    \quad
    \text{for $i \in \Gamma$}
    \>.
\end{equation}
The pinched logarithmic derivatives $\omega^\star_i$ and the cut coordinates
$z^\star_{\bar{i}}$ are the sector data entering the pairing formulae
of~\cref{ssec:lo_pairing,ssec:subleading_order_pairing}.

\subsection{%
    \texorpdfstring
        {Leading Residue Pairing as $\delta \to 0$ and the Residue Measure}
        {Leading Residue Pairing as delta -> 0 and the Residue Measure}
}
\label{ssec:lo_pairing}
\label{ssec:pinch_decomposition}
We now address the $\delta \to 0$ limit of the leading
pairing~\labelcref{eq:lo_hrp} of a form $\varphi_a$ with a rescaled dual form
$\vartheta_b$.
At each regulated critical point, its summand is the product of the two form
coefficients divided by the determinant of the regulated Hessian.
As we will see, the structure of the limit is governed by the pinch sets of the
two forms, with contributions coming from sectors interpolating between them.
We take the limit in three steps. First, we factorise each form into a pinch
factor, which carries its explicit dependence on $\delta$, and a regular factor.
Second, we collect the pinch factors and the determinant into a single universal
factor of our construction, the residue measure, which generalises the
determinant denominator of~\cref{eq:lo_hrp}. Third, we expand the ingredients of
the residue measure at a critical point of a sector $\Gamma$ and take its limit,
which yields the leading pairing.

Recall from~\cref{eq:form_pole_structure} that each form $\varphi_a$ has a
set $\Gamma_a$ of pinched edges, those with $\alpha_i \le 0$, each of which
contributes a simple pole at $z_i = 0$ and we write $N_{\Gamma_a} \defas \abs{\Gamma_a}$.
Analogously to the sectors from~\cref{ssec:critical_sectors}, we refer to $\Gamma_a$ as the sector associated with the form $\varphi_a$.
Combining~\cref{eq:form_pole_structure} with the
expansion~\labelcref{eq:one_over_gamma_expansion} of the $\Gamma$-function
normalisation, we find that, at leading order in $\delta$, the differential form
$\varphi_a$ constructed from the Lee--Pomeransky
representation~\labelcref{eq:leepom_form} factorises as
\begin{equation}
\label{eq:phi_g_decomp}
    \hat\varphi_a
    =
    \Delta_a\,\psi_a
    +
    \cO\brk{\delta^{N_{\Gamma_a}+1}}
    \>,
\end{equation}
where
\begin{equation}
\label{eq:Delta_psi_g}
    \Delta_a
    \defas
    \prod_{i\in\Gamma_a}\frac{\delta}{z_i}
    \>,
    \qquad
    \psi_a
    \defas
    \frac{\cN_a(z)}{\cG(z)^{D_0/2 + q_a}}
    \prod_{\bar{i} \in \overline{\Gamma}_a} \frac{1}{\Gamma\brk{\alpha_{\bar{i}}}}
    \>.
\end{equation}
The pinch factor $\Delta_a$ on the left encodes the pinch structure of the form and carries its explicit leading $\delta$-dependence, while $\psi_a$ is independent of $\delta$ at fixed $z$.
On the right, $q_a$ is a non-negative integer that specifies the common denominator $\cG^{q_a}$ generated via~\cref{eq:lift_to_top,eq:isp_numerator_factor}, and the polynomial $\cN_a$ collects the resulting numerator together with the factors $z_i^{\alpha_i-1}$ for $i\notin\Gamma_a$.

Inserting the factorisation~\labelcref{eq:phi_g_decomp} and the rescaled
dual forms~\labelcref{eq:basis_normalisation} into the leading residue
formula~\labelcref{eq:lo_hrp}, we write it as
\begin{equation}
\label{eq:lo_hrp_measure}
    \regpair[0]{\varphi_a}{\vartheta_b}
    =
    \sum_{z\brk{\delta} \in \operatorname{Crit}\brk{W_\delta}}
    \mu_{\delta, ab} \, \psi_a \psi_b
    \Bigr\rvert_{z = z\brk{\delta}}
    \bigbrk{1 + \cO\brk{\delta}}
    \>,
\end{equation}
where we define the \textit{residue measure}
\begin{equation}
\label{eq:def_measure}
    \mu_{\delta, ab}\brk{z}
    \defas
    \frac{\delta^{-N_{\Gamma_b}} \, \Delta_a \Delta_b}{\det\bigbrk{\, \Phi_\delta \,}}
    \>.
\end{equation}
It depends on the forms only through their pinch sets $\Gamma_a$ and $\Gamma_b$
and collects the pinch and Hessian structure of the residue formula, so that the
leading pairing is a sum over critical points of the regular factors $\psi_a
\psi_b$, weighted by $\mu_{\delta, ab}$.
The non-trivial work, therefore, is in finding the limit of $\mu_{\delta, ab}$
as $\delta \to 0$.
As we will see, this limit is indeed finite, though this arises through
non-trivial cancellations and the choice of the normalisation of the dual forms.

Our approach is simply to evaluate the pinch factors and the determinant at a
regulated critical point $z\brk{\delta}$ of a sector $\Gamma$.
We begin with the pinch factors.
The factors $\delta / z_i$ entering $\Delta_x$, $x = a, b$,
in~\cref{eq:Delta_psi_g} behave differently depending on whether $i \in
\Gamma$:
\begin{alignat}{3}
\label{eq:edge_cases}
    i &\in \Gamma_x\cap\Gamma:
    &\qquad
    z_i\brk{\delta} &= \cO\brk{\delta}
    \>,
    &\qquad
    \frac{\delta}{z_i\brk{\delta}} &\xrightarrow{\; \delta \to 0 \;} -\omega^\star_i \neq 0
    \>,
    \nonumber\\
    \bar{i} &\in \Gamma_x\setminus\Gamma:
    &\qquad
    z_{\bar{i}}\brk{\delta} &\xrightarrow{\; \delta \to 0 \;} z^\star_{\bar{i}} \neq 0
    \>,
    &\qquad
    \frac{\delta}{z_{\bar{i}}\brk{\delta}} &= \cO\brk{\delta}
    \>.
\end{alignat}
Thus, an edge which is pinched in both the form and the sector contributes $-\omega^\star_i$, while an edge pinched only in the form contributes one power of $\delta$.
The pinch factors therefore behave as
\begin{equation}
\label{eq:pinch_factor_limit}
    \Delta_x \Bigr\rvert_{z = z\brk{\delta}}
    =
    \delta^{N_{\Gamma_x \setminus \Gamma}}
    \>
    \Biggbrk{
        \brk{-1}^{N_{\Gamma_x \cap \Gamma}}
        \biggbrk{\,
            \prod_{i \in \Gamma_x \cap \Gamma}
                \omega_i
            \prod_{\bar{i} \in \Gamma_x \setminus \Gamma}
                \frac{1}{z_{\bar{i}}}
        \,}\bigg\rvert_{z = z^\star}
        + \cO\brk{\delta}
    }
    \>,
    \qquad
    x = a, b
    \>.
\end{equation}

We next turn to the determinant of the regulated Hessian.
At a regulated critical point $z\brk{\delta}$, the regulated Hessian
matrix~\labelcref{eq:def_regulated_Hessian} develops a simple pole in $\delta$
in its pinched block, since $z_i\brk{\delta} = \cO\brk{\delta}$ for $i \in
\Gamma$ makes the diagonal regulator term $-\delta / z_i^2$
of order $\delta^{-1}$.
Ordering the pinched indices $i \in \Gamma$ before the cut indices $\bar{i} \in \overline{\Gamma}$, we write
\begin{align}
\label{eq:hessian_exp}
    \Phi_{\delta}
    \equiv
    \begin{bNiceMatrix}
        \Phi_{\delta ,\, \Gamma} &
        \Phi_{\Gamma \overline{\Gamma}}
        \\
        \Phi_{\overline{\Gamma} \Gamma} &
        \Phi_{\delta ,\, \overline{\Gamma}}
    \end{bNiceMatrix}
    \Biggr\rvert_{z = z\brk{\delta}}
    =
    \Biggl(
    -\frac{1}{\delta}
    \>
    \begin{bNiceMatrix}
        D_{\Gamma} & \mZeroBlock
        \\
        \mZeroBlock & \mZeroBlock
    \end{bNiceMatrix}
    +
    \begin{bNiceMatrix}
        \Phi_\Gamma & \Phi_{\Gamma \overline{\Gamma}}
        \\
        \Phi_{\overline{\Gamma} \Gamma} & \Phi_{\overline{\Gamma}}
    \end{bNiceMatrix}
    -
    {\delta}
    \>
    \begin{bNiceMatrix}
        \mZeroBlock & \mZeroBlock
        \\
        \mZeroBlock & D_{\overline{\Gamma}}
    \end{bNiceMatrix}
    \Biggr)\Biggr\rvert_{z = z\brk{\delta}}
    \>,
\end{align}
where the blocks of the unregulated Hessian matrix of~\cref{eq:def_regulated_Hessian} are
\begin{equation*}
    \Phi_{\Gamma}
    \defas
    \begin{bNiceMatrix}
        \partial_i \partial_j \log \cG
    \end{bNiceMatrix}_{i, j \in \Gamma}
    \>,
\end{equation*}
and the analogous expressions for $\Phi_{\Gamma \overline{\Gamma}}$, $\Phi_{\overline{\Gamma} \Gamma}$ and $\Phi_{\overline{\Gamma}}$, while
the diagonal matrices are
\begin{equation}
\label{eq:def_Dgamma}
    D_{\Gamma}
    \defas
    \operatorname{diag}_{i \in \Gamma} \Bigbrk{\,
        \omega_i^2
    \,}
    \>,
    \qquad \qquad
    D_{\overline{\Gamma}}
    \defas
    \operatorname{diag}_{\bar{i} \in \overline{\Gamma}} \Bigbrk{\,
        \frac{1}{z_{\bar{i}}^2}
    \,}
    \>.
\end{equation}
In \cref{eq:hessian_exp} all blocks are evaluated at $z\brk{\delta}$ and
have finite and non-zero entries as $\delta \to 0$,
see~\cref{eq:def_rho,eq:cut_edge_limit}.

Given the block structure of \cref{eq:hessian_exp}, the determinant of
$\Phi_\delta$ at $z = z\brk{\delta}$ can be expressed through the Schur
complement $S_\delta$ of its cut block,
\begin{align}
    S_\delta
    &\defas
    - \frac1\delta D_{\Gamma}
    + \Phi_{\Gamma}
    - \Phi_{\Gamma \overline{\Gamma}} \cdot \bigbrk{\Phi_{\overline{\Gamma}} - \delta D_{\overline{\Gamma}}}^{-1} \cdot \Phi_{\overline{\Gamma} \Gamma}
    \>.
    \label{eq:regulated_Schur_complement}
\end{align}
By the Schur determinant formula, the determinant is the product of
$\det(S_\delta)$ and $\det\bigbrk{\, \Phi_{\delta ,\, \overline{\Gamma}} \,}$. Each factor can be
easily expanded to leading order in $\delta$ and we find
\begin{align}
\label{eq:hessian_ltd}
    \det\bigbrk{\, \Phi_\delta \,}
    \Bigr\rvert_{z = z\brk{\delta}}
    &=
    \Bigbrk{-\frac{1}{\delta}}^{N_\Gamma}
    \>
    \left(
        \biggbrk{\,
            \prod_{i \in \Gamma} \omega_{i}^2
            \>
            \det\bigbrk{\, \Phi_{\overline{\Gamma}} \,}
        \,}\bigg\rvert_{z = z^\star}
        +\cO\brk{\delta}
    \right)
    \>.
\end{align}
Thus, the determinant scales as $\delta^{-N_{\Gamma}}$ with a leading
coefficient given by the logarithmic derivatives~\labelcref{eq:def_rho} and the
determinant of the cut block of the unregulated Hessian at $z^\star$, which is
non-vanishing because the sector critical points are non-degenerate.

With these expressions in hand, we are now in a position to consider at which
order in $\delta$ the residue measure contributes.
When combining the expressions for the pinch factors and the determinant, we see
that the residue measure at a critical point of the sector $\Gamma$ is of order
$\delta^{\lambda_{ab}\brk{\Gamma}}$, where
\begin{equation}
\label{eq:lambda_def}
    \lambda_{ab}\brk{\Gamma}
    =
    N_\Gamma
    +
    {N_{\Gamma_a \setminus \Gamma}}
    +
    {N_{\Gamma_b \setminus \Gamma}}
    -
    N_{\Gamma_b}
    \>.
\end{equation}
Here $N_\Gamma$ comes from $1 / \det\bigbrk{\, \Phi_\delta \,}$, while
$N_{\Gamma_a \setminus \Gamma}$ and $N_{\Gamma_b \setminus \Gamma}$ count the
edges pinched in $\varphi_a$ and $\varphi_b^\vee$, respectively, but cut in the
sector, each of which contributes a factor $\delta$
by~\labelcref{eq:edge_cases}, and $-N_{\Gamma_b}$ comes from the
normalisation~\labelcref{eq:basis_normalisation}.
Using the inclusion--exclusion identity for the sets $\Gamma_a \setminus \Gamma$
and $\Gamma_b \setminus \Gamma$, we write the exponent as
\begin{equation}
\label{eq:lambda_final}
    \lambda_{ab}\brk{\Gamma}
    =
    N_{\Gamma_a \setminus \Gamma_b}
    +
    N_{\brk{\Gamma_a \cap \Gamma_b} \setminus \Gamma}
    +
    N_{\Gamma \setminus \brk{\Gamma_a \cup \Gamma_b}}
    \>.
\end{equation}
Here, the last two terms are non-negative and vanish simultaneously when
\begin{equation}
\label{eq:contributing_sectors}
    \Gamma_a \cap \Gamma_b
    \> \subseteq \>
    \Gamma
    \> \subseteq \>
    \Gamma_a \cup \Gamma_b
    \>.
\end{equation}
The minimum of $\lambda_{ab}$ is therefore $N_{\Gamma_a \setminus \Gamma_b}$,
attained by the sectors~\labelcref{eq:contributing_sectors}. Unless
$\Gamma_a \subseteq \Gamma_b$, the residue measure therefore vanishes as $\delta
\to 0$ at every critical point.
For $\Gamma_a \subseteq
\Gamma_b$, the contributing sectors~\labelcref{eq:contributing_sectors} reduce
to
\begin{align}
\label{eq:contributing_sectors_hierarchical}
    \Gamma_a \subseteq \Gamma \subseteq \Gamma_b
    \>.
\end{align}
In the Feynman-integral language, the topology associated with a contributing
sector must be a supertopology of that associated with $\varphi_b$ and a
subtopology of that associated with $\varphi_a$.

At the critical points of these sectors, the
exponent~\labelcref{eq:lambda_final} vanishes and the residue measure
has a finite and non-zero limit.
Inserting the leading coefficients of~\cref{eq:pinch_factor_limit,eq:hessian_ltd} into~\cref{eq:def_measure}, we find
\begin{equation}
\label{eq:measure_limit}
    \mu_{ab}\brk{z^\star}
    \defas
    \lim_{\delta \to 0}
    \mu_{\delta, ab} \Bigr\rvert_{z = z\brk{\delta}}
    =
    \brk{-1}^{N_{\Gamma_a}}
    \Biggbrk{\,
        \prod_{i \in \Gamma \setminus \Gamma_a} \omega_i^{-1}
        \;
        \prod_{\bar{i} \in \Gamma_b \setminus \Gamma} z_{\bar{i}}^{-1}
        \;
        \frac{1}{\det\bigbrk{\, \Phi_{\overline{\Gamma}} \,}}
    \,}\Bigg\rvert_{z = z^\star}
    \>,
\end{equation}
where $z\brk{\delta}$ is the regulated critical point that approaches $z^\star \in \operatorname{Crit}_\Gamma\brk{W}$.
The two products reflect the two possibilities for an edge in $\Gamma_b \setminus \Gamma_a$ to behave in the sector $\Gamma$.
In the pinched case, $i \in \Gamma$, it contributes $\omega_i^{-1}$; in the cut case, $\bar{i} \in \overline{\Gamma}$, it contributes $z^{-1}_{\bar{i}}$.
Summing $\mu_{ab} \, \psi_a \psi_b$ over the critical points of the sectors~\labelcref{eq:contributing_sectors_hierarchical}, we obtain the leading pairing
\begin{equation}
\label{eq:lo_pairing}
    \vev{\varphi_a \mid \vartheta_b}_{(0)}
    =
    \sum_{\Gamma_a \subseteq \Gamma \subseteq \Gamma_b}
    \;
    \sum_{z^\star \in \operatorname{Crit}_\Gamma\brk{W}}
    \mu_{ab} \, \psi_a \psi_b
    \Bigr\rvert_{z = z^\star}
    \>.
\end{equation}
All the ingredients of~\cref{eq:lo_pairing} are unregulated quantities evaluated
at the critical points of the sector equations~\labelcref{eq:critical_eqs_full}:
the analytic regulator has disappeared entirely.

Since the leading pairing vanishes unless $\Gamma_a \subseteq \Gamma_b$, the
matrix of leading pairings of a basis of forms with its rescaled duals is block
triangular once the basis is ordered by pinch sets, whereas at non-vanishing
$\delta$ it does not in general have this structure.
Its diagonal blocks pair forms with $\Gamma_a = \Gamma_b$, for which only the
sector $\Gamma = \Gamma_a$ contributes and~\cref{eq:lo_pairing} reduces, up to a
sign, to the leading residue pairing~\labelcref{eq:lo_hrp} of $\psi_a$ and
$\psi_b$ on the maximal cut of their common topology.

\subsection{%
    \texorpdfstring
        {Subleading Residue Pairing as $\delta \to 0$}
        {Subleading Residue Pairing as delta -> 0}
}
\label{ssec:subleading_order_pairing}
\label{sssec:subleading_formula}
\label{sssec:vanishing_factors}
Let us now consider the $\delta \to 0$ limit of the first subleading residue
pairing $\regpair[1]{\varphi_a}{\vartheta_b}$. Analogously to the leading
pairing, we expand~\cref{eq:nlo_hrp} to leading order in $\delta$, with the dual
forms normalised as in~\cref{eq:basis_normalisation}.
The result has the same structure as the leading pairing
of~\cref{eq:lo_pairing}: it is a sum over sectors, and each contribution is
expressed through unregulated sector data.
Taking the limit is, however, more intricate than for the leading pairing,
since the subleading pairing also involves the inverse Hessian and derivatives
of the forms and of the determinant, which can have poles in $\delta$.
The key steps parallel~\cref{ssec:lo_pairing}: we factor out the residue measure
and show that the remaining trace has a finite limit. We carry out these steps
below, deferring some lengthy argumentation to~\cref{app:subleading_derivation}.

We begin by factoring the residue measure~\labelcref{eq:def_measure} out of each summand and write
\begin{equation}
    \label{eq:subleading_trace_form}
    \regpair[1]{\varphi_a}{\vartheta_b}
    =
    \frac12
    \sum_{z\brk{\delta}\in\operatorname{Crit}\brk{W_\delta}}
    \mu_{\delta, ab}
    \>
    T_{\delta, ab}
    \Biggr\rvert_{z=z\brk{\delta}}
    \>,
    \qquad
    T_{\delta, ab}
    \defas
    \Tr\Bigbrk{\, \Phi_\delta^{-1}\cdot M_{ab} \,}
    \>,
\end{equation}
where
\begin{equation}
    \label{eq:def_Mab}
    M_{ab}^{ij}
    \defas
    \mu_{\delta, ab}^{-1}
    \>
    \partial_j\Biggbrk{
        \frac{
            \hat\varphi_a \> \partial_i\hat\vartheta_b
            -
            \hat\vartheta_b \> \partial_i\hat\varphi_a
        }{
            \det\bigbrk{\, \Phi_\delta \,}
        }
    }
    \>.
\end{equation}
Each critical point thus contributes its summand of the leading
pairing~\labelcref{eq:lo_hrp_measure}, with the product $\psi_a \psi_b$ of the
regular factors replaced by $T_{\delta, ab}/2$.
Since $\Phi_\delta^{-1}$ is symmetric, only the symmetric part of $M_{ab}$ contributes to $T_{\delta, ab}$.
We therefore write $M_{ab} \simeq M'_{ab}$ when their difference is antisymmetric.
Using the factorised representation~\labelcref{eq:phi_g_decomp}, we
obtain\footnote{
    We note that here we have dropped the remainder of~\cref{eq:phi_g_decomp}.
    It carries the same pinch factors, so by bilinearity it only adds terms of the same form multiplied by $\delta$, which vanish in the limit.%
}
\begin{align}
    \label{eq:symmetric_Mab}
        M_{ab}^{ij}
        &\simeq
            \frac{\psi_a}{\Delta_b}\>\partial_i\partial_j\brk{\Delta_b\psi_b}
            -\frac{\psi_b}{\Delta_a}\>\partial_i\partial_j\brk{\Delta_a\psi_a}
        \\
        &\quad-\Bigbrk{
            \psi_a\partial_i\psi_b-\psi_b\partial_i\psi_a
            +\psi_a\psi_b\>\partial_i\log\Bigbrk{\frac{\Delta_b}{\Delta_a}}
        }
            \>\partial_j\log\det\bigbrk{\, \Phi_\delta \,}
    \>.
    \nonumber
\end{align}
This representation is bilinear in the regular factors and their derivatives,
whose evaluation at $z(\delta)$ remains finite as $\delta \to 0$.
In contrast, derivatives of the pinch factors and pinched derivatives of
$\log\det\bigbrk{\, \Phi_\delta \,}$ can produce poles in $\delta$ after
evaluation at $z\brk{\delta}$.
As we will see, these poles cancel in the contraction against the inverse
Hessian in $T_{\delta, ab}$ and $T_{\delta, ab}$ is finite in the limit $\delta
\to 0$.

To control this cancellation, let us split $T_{\delta, ab}$ along the block
structure of the Hessian of~\cref{eq:hessian_exp}.
Fix a critical point $z^\star$ of some sector $\Gamma$ and let $z\brk{\delta}$
be the regulated critical point that approaches it.
In the block ordering of~\cref{eq:hessian_exp}, block inversion expresses
$\Phi_\delta^{-1}$ in terms of the inverse of the regulated Schur
complement~\labelcref{eq:regulated_Schur_complement}, and together with
$\Phi_{\delta ,\,
  \overline{\Gamma}}^{-1} = \Phi_{\overline{\Gamma}}^{-1} + \cO\brk{\delta}$,
one finds the expansion
\begin{equation}
\label{eq:invhessian_expanded}
    \Phi_{\delta}^{-1}
    =
    \begin{bNiceMatrix}
        \mZeroBlock & \mZeroBlock
        \\
        \mZeroBlock & \Phi_{\overline{\Gamma}}^{-1}
    \end{bNiceMatrix}
    +
    \delta
    \begin{bNiceMatrix}
        -D_{\Gamma}^{-1}
        &
        D_{\Gamma}^{-1} \cdot \Phi_{\Gamma \overline{\Gamma}} \cdot \Phi_{\overline{\Gamma}}^{-1}
        \\
        \Phi_{\overline{\Gamma}}^{-1} \cdot \Phi_{\overline{\Gamma}\Gamma} \cdot D_{\Gamma}^{-1}
        &
        \ast
    \end{bNiceMatrix}
    +
    \delta^2
    \begin{bNiceMatrix}
        -D_{\Gamma}^{-1} \cdot S_\Gamma \cdot D_{\Gamma}^{-1}
        &
        \ast
        \\
        \ast
        &
        \ast
    \end{bNiceMatrix}
    +
    \cO\brk{\delta^3}
    \>,
\end{equation}
where both sides are evaluated at $z = z\brk{\delta}$, $\ast$ denotes terms that will not contribute to the subleading residue pairing
and $S_\Gamma$ denotes the unregulated Schur complement
\begin{equation}
\label{eq:unregulated_Schur_complement}
    S_\Gamma
    \defas
    \Phi_{\Gamma}
    - \Phi_{\Gamma \overline{\Gamma}} \cdot \Phi^{-1}_{\overline{\Gamma}} \cdot \Phi_{\overline{\Gamma} \Gamma}
    \>.
\end{equation}
We split $M_{ab}$ in the same way as
\begin{equation}
    \label{eq:Mab_blocks}
    M_{ab}
    =
    \renewcommand{\arraystretch}{1.2}
    \begin{bNiceMatrix}
        M_{ab}^{\Gamma}
        &
        M_{ab}^{\Gamma\overline{\Gamma}}
        \\
        M_{ab}^{\overline{\Gamma}\Gamma}
        &
        M_{ab}^{\overline{\Gamma}}
    \end{bNiceMatrix}
    \>.
\end{equation}
Since $T_{\delta, ab} = \Tr\bigbrk{\Phi_\delta^{-1} \cdot M_{ab}}$ and
$\Phi_\delta^{-1}$ is symmetric, this decomposition splits $T_{\delta, ab}$
exactly into a cut, a mixed and a pinched part,
\begin{subequations}
\label{eq:TraceDecomp}
\begin{align}
    \label{eq:T_split}
    T_{\delta, ab}
    &=
    T_{\delta, ab}^{\overline{\Gamma}}
    +
    T_{\delta, ab}^{\Gamma\overline{\Gamma}}
    +
    T_{\delta, ab}^{\Gamma}
    \>,
    \\[4pt]
    \label{eq:T_cut_def}
    T_{\delta, ab}^{\overline{\Gamma}}
    &\defas
    \Tr\Bigbrk{\,
        \bigsbrk{\Phi_\delta^{-1}}_{\overline{\Gamma}\,\overline{\Gamma}} \cdot M_{ab}^{\overline{\Gamma}}
    \,}
    \>,
    \\[4pt]
    \label{eq:T_mixed_def}
    T_{\delta, ab}^{\Gamma\overline{\Gamma}}
    &\defas
    \Tr\Bigbrk{\,
        \bigsbrk{\Phi_\delta^{-1}}_{\Gamma\overline{\Gamma}}
        \cdot
        \bigbrk{
            M_{ab}^{\overline{\Gamma}\Gamma}
            +
            \brk{M_{ab}^{\Gamma\overline{\Gamma}}}\transpose
        }
    \,}
    \>,
    \\[4pt]
    \label{eq:T_pinch_def}
    T_{\delta, ab}^{\Gamma}
    &\defas
    \Tr\Bigbrk{\,
        \bigsbrk{\Phi_\delta^{-1}}_{\Gamma\Gamma} \cdot M_{ab}^{\Gamma}
    \,}
    \>,
\end{align}
\end{subequations}
where $\sbrk{\Phi_\delta^{-1}}_{XY}$ denotes the $XY$ block of the inverse Hessian.

The split is useful because each part has a definite behaviour as $\delta \to 0$.
By~\cref{eq:invhessian_expanded}, the cut block of the inverse Hessian is
finite at $z\brk{\delta}$, while its mixed and pinched blocks are of order
$\delta$, and the off-diagonal pinched entries even of order $\delta^2$.
The entries of $M_{ab}$ behave in the opposite way: they are finite in the cut
directions but acquire a simple pole for each pinched index, since each
derivative $\partial_i$ along a pinched edge of a pinch factor or of
$\log\det\bigbrk{\, \Phi_\delta \,}$ produces a factor $1/z_i =
\cO\brk{\delta^{-1}}$.
Every pole of $M_{ab}$ is therefore compensated by a power of $\delta$ from the
inverse Hessian, except on the diagonal of the pinched block, where a double
pole meets a single power of $\delta$.
These double poles cancel so that each part of $T_{\delta, ab}$ has a finite
limit at every critical point. Details of the determinant derivative and double
pole cancellation can be found in~\cref{app:subleading_derivation}.

Since $T_{\delta, ab}$ stays finite as $\delta \to 0$, the residue measure alone
decides which critical points contribute
to~\cref{eq:subleading_trace_form}.
As for the leading pairing, the subleading pairing therefore vanishes as $\delta
\to 0$ unless $\Gamma_a \subseteq \Gamma_b$, and then only the sectors $\Gamma_a
\subseteq \Gamma \subseteq \Gamma_b$
of~\cref{eq:contributing_sectors_hierarchical} contribute, at which
$\mu_{\delta, ab}$ tends to the limit $\mu_{ab}$ of~\cref{eq:measure_limit}.
It remains to find the limits of the three parts of $T_{\delta, ab}$ at the critical points of these sectors.
We denote them by $T_{ab}^{\overline{\Gamma}}$, $T_{ab}^{\Gamma\overline{\Gamma}}$ and $T_{ab}^{\Gamma}$.
Their derivation from the block expansions of $M_{ab}$ is laborious, and we
leave details to appendix~\labelcref{app:subleading_derivation}.
We find
\begin{align}
    \label{eq:T_cut_block}
    T_{ab}^{\overline{\Gamma}}
    &=
        \Tr\Bigbrk{\, \Phi_{\overline{\Gamma}}^{-1} \cdot \midtilde{M}_{ab} \,}
    \\
    &\quad +
        \psi_a
        \sum_{\substack{
            \bar{i} \in \Gamma_b \setminus \Gamma
            \\
            \bar{j} \in \overline{\Gamma}
        }}
        \frac{\bigbrk{\Phi_{\overline{\Gamma}}^{-1}}_{\bar{i}\bar{j}}}{z_{\bar{i}}}
        \biggbrk{
            \frac{\delta_{\bar{j} \Gamma_b} \psi_b}{z_{\bar{j}}}
            \brk{1 + \delta_{\bar{i}\bar{j}}}
            - 2 \> \partial_{\bar{j}} \psi_b
            + \psi_b \> \partial_{\bar{j}} \log \det\bigbrk{\, \Phi_{\overline{\Gamma}} \,}
        }
        \nonumber
        \>,
\end{align}
\begin{equation}
    \label{eq:T_mixed_block}
    T_{ab}^{\Gamma\overline{\Gamma}}
    =
        \psi_b
        \sum_{\substack{
            i \in \Gamma \setminus \Gamma_a
            \\
            \bar{j} \in \overline{\Gamma}
        }}
        \frac{1}{\omega_i}
        \bigbrk{\Phi_{\Gamma\overline{\Gamma}} \cdot \Phi_{\overline{\Gamma}}^{-1}}_{i\bar{j}}
        \>
        \Bigbrk{
            2 \> \partial_{\bar{j}} \psi_a
            - \psi_a \> \partial_{\bar{j}} \log \det\bigbrk{\, \Phi_{\overline{\Gamma}} \,}
        }
        \>,
\end{equation}
\begin{equation}
    \label{eq:T_pinch_block}
    T_{ab}^{\Gamma}
    =
        \psi_a \psi_b
        \sum_{i ,\, j \in \Gamma \setminus \Gamma_a}
        \brk{1 + \delta_{ij}}
        \frac{S_{\Gamma ,\, ij}}{\omega_i \, \omega_j}
        -
        \psi_b
        \sum_{i \in \Gamma \setminus \Gamma_a}
        \frac{1}{\omega_i}
        \Bigbrk{
            2 \> \partial_i \psi_a
            - \psi_a \> \partial_i \log \det\bigbrk{\, \Phi_{\overline{\Gamma}} \,}
        }
        \>,
\end{equation}
where the indicator $\delta_{\bar{j}\Gamma_b}$ equals one if $\bar{j} \in
\Gamma_b$ and zero otherwise and the matrix $\midtilde{M}_{ab}$ is given by
\begin{align}
    \label{eq:def_Mtilde}
    \midtilde{M}_{ab}^{\bar{i}\bar{j}}
    \defas
    \psi_a \> \partial_{\bar{i}} \partial_{\bar{j}} \psi_b
    - \psi_b \> \partial_{\bar{i}} \partial_{\bar{j}} \psi_a
    - \brk{
        \psi_a \> \partial_{\bar{i}} \psi_b
        - \psi_b \> \partial_{\bar{i}} \psi_a
    }
    \> \partial_{\bar{j}} \log \det\bigbrk{\, \Phi_{\overline{\Gamma}} \,}
    \>.
\end{align}
The indicator terms of $T_{ab}^{\overline{\Gamma}}$ come from the edges in $\Gamma_b \setminus \Gamma$ and vanish for $\Gamma = \Gamma_b$, while $T_{ab}^{\Gamma}$ and $T_{ab}^{\Gamma\overline{\Gamma}}$ come from the edges in $\Gamma \setminus \Gamma_a$ and vanish for $\Gamma = \Gamma_a$.

Inserting these limits and the limit~\labelcref{eq:measure_limit} of the residue measure into~\cref{eq:subleading_trace_form}, we find
\begin{align}
\label{eq:subleading_pairing_final}
    \vev{\varphi_a \mid \vartheta_b}_{(1)}
    &=
    \frac12
    \sum_{\Gamma_a \subseteq \Gamma \subseteq \Gamma_b}
    \>
    \sum_{z^\star \in \operatorname{Crit}_\Gamma\brk{W}}
    \>
    \mu_{ab}
    \>
    \bigbrk{T_{ab}^{\overline{\Gamma}} + T_{ab}^{\Gamma\overline{\Gamma}} + T_{ab}^{\Gamma}}
    \Biggr\rvert_{z = z^\star}
    \>.
\end{align}
Compared with~\cref{eq:subleading_trace_form}, the sum runs only over the
critical points of the contributing sectors, and the residue measure and the
three parts of $T_{\delta, ab}$ are replaced by their limits, all of which are
evaluated on the unregulated problem.

It is interesting to note that the pairing simplifies considerably on the
diagonal blocks of the pairing matrix.
When the two forms have the same pinch set, $\Gamma_a = \Gamma_b$, only the
sector $\Gamma = \Gamma_a$ contributes. Since the sets $\Gamma \setminus
\Gamma_a$ and $\Gamma_b \setminus \Gamma$ are then empty, the sums
in~\cref{eq:T_cut_block,eq:T_mixed_block,eq:T_pinch_block}
vanish, so that only the trace term
of~\cref{eq:T_cut_block} survives, giving
\begin{equation}
\label{eq:subleading_maxcut}
    \vev{\varphi_a \mid \vartheta_b}_{(1)}
    =
    \frac{(-1)^{N_{\Gamma_a}}}{2}
    \sum_{z^\star \in \operatorname{Crit}_{\Gamma_a}\brk{W}}
    \frac{1}{\det\bigbrk{\, \Phi_{\overline{\Gamma}_a} \,}}
    \>
    \Tr\Bigbrk{\, \Phi_{\overline{\Gamma}_a}^{-1} \cdot \midtilde{M}_{ab} \,}
    \Biggr\rvert_{z = z^\star}
    \>,
\end{equation}
which is, up to the sign, the first subleading higher residue pairing~\labelcref{eq:nlo_hrp} of $\psi_a$ and $\psi_b$ on the maximal cut of their common topology.
For $\Gamma_a = \Gamma_b = \varnothing$ it is the pairing for the unregulated
potential $W = \log\cG$, and for $a = b$ it vanishes since $\midtilde{M}_{aa} =
0$, in agreement with~\cref{eq:pairing_symmetry}.
As for the leading pairing, the diagonal blocks of the subleading pairing matrix are thus residue pairings on maximal cuts.

In summary, the subleading pairing is a sum over the same sectors, with the
same residue measure as the leading pairing, and $\psi_a \psi_b$ replaced by a
finite combination of unregulated quantities.
The analytic regulator has therefore disappeared from both pairings, which are
computed one critical point at a time on the unregulated problem.
In~\cref{sec:applications}, we use these formulae to compute the intersection
matrices, and from them canonical differential equations, for the planar
four-point ladder families up to four loops.

%% file: Sections/applications.tex
\section{Applications}
\label{sec:applications}
\begin{figure}[t]
\centering
\includegraphics[scale=0.85]{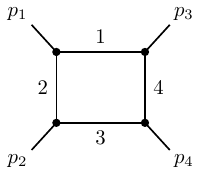}\hfill
\includegraphics[scale=0.85]{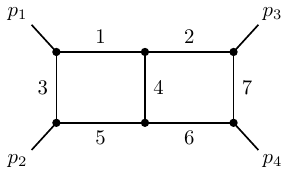}\hfill
\includegraphics[scale=0.85]{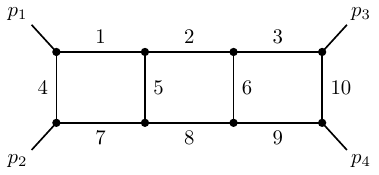}

\vspace{0.9cm}

\includegraphics[scale=0.85]{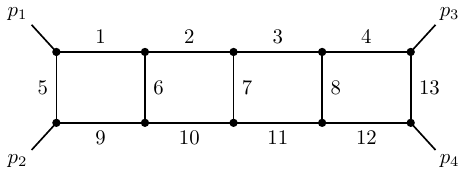}
\caption{\label{fig:ladders}Multi-loop ladder topologies. In our conventions all external legs are incoming.}
\end{figure}
In this section, we apply the formalism of \cref{sec:delta_expansion} to compute
canonical differential equations for the massless planar four-point ladder
families of~\cref{fig:ladders}, from one to four loops.
This series of examples allows us not only to elaborate on features of our
calculational approach but also to showcase its computational efficiency.

The kinematics of the ladder families depend on the massless momenta
$\{p_i\}_{i=1}^4$ through the Mandelstam invariants
\begin{equation}
\label{eq:mandelstam}
    s=(p_1+p_2)^2 \,, \qquad t=(p_1+p_3)^2 .
\end{equation}
Up to an overall scale, the families depend only on the dimensionless ratio
\begin{equation}
\label{eq:y_variable}
    y=\frac{t}{s} \,.
\end{equation}
We set $s=-1$, so that $t=-y$.
We work in $D=4-2\varepsilon$ and use the democratic regulator $\delta_i=\delta$ of~\cref{eq:democratic_choice}.

\subsection[Implementation]{Implementation}
\label{ssec:sector_sums}

Before turning to the examples, we describe how we implement the
construction of differential equations from higher residue pairings.
For each family, the computation follows the same steps, which we summarise
here:
\begin{enumerate}[label=(\roman*)]
    \item For each sector $\Gamma$, we determine the critical points
          of~\cref{eq:critical_eqs_full} and count them: summed over sectors, their number
          gives the number of master integrals before identifying graph symmetries.
    \item We choose a basis of forms $e_a$, one set for each sector, normalised
        as in~\cref{eq:rational_normalisation} so that it starts at order $\varepsilon^0$.
        The dual forms $\vartheta_b$ are chosen independent of $\varepsilon$ and rescaled
        as in~\cref{eq:basis_normalisation}.
    \item We evaluate the intersection matrices of~\cref{eq:CDlin} in the limit
          $\delta\to0$, through the leading and subleading pairings
          of~\cref{eq:lo_pairing,eq:subleading_pairing_final}.
    \item We assemble the differential system from~\cref{eq:Omegalin} in the limit $\delta\to0$.
\end{enumerate}
Most of these steps apply the formulae
of~\cref{sec:eps_expansion,sec:delta_expansion} directly.
Two points need more care: handling the critical points and summing over them,
in steps~(i) and~(iii), and taking the limit $\delta\to0$ of the matrix formulae
of~\cref{ssec:eps_linear_systems}, in steps~(ii)--(iv).

In order to evaluate the leading and subleading pairings in
\cref{eq:lo_pairing,eq:subleading_pairing_final} in practice, we must be able to evaluate a rational function
on all the critical points of a sector, over which we then sum.
Solving the sector-by-sector critical point equations~\labelcref{eq:critical_eqs_full} explicitly is, in general,
a non-trivial task, which we avoid by using techniques from computational algebraic geometry.
Specifically, we encode the solutions of \cref{eq:critical_eqs_full} as the vanishing set of the 
\textit{sector ideal}
\begin{equation}
\label{eq:critical_points_expanded}
    \cI_\Gamma
    =
    \bigvev{ z_i,\,\partial_{\bar{i}} \cG \,\,\,\big|\,\,\, i\in\Gamma,\, \bar{i} \in \overline{\Gamma} }
    :
    \vev{\cG}^\infty
    \>,
\end{equation}
where $:\vev{\cG}^\infty$ denotes saturation with respect to the principal ideal
$\vev{\cG}$, which enforces the condition $\cG \neq 0$ on the solution set.
The saturation in \cref{eq:critical_points_expanded} can be evaluated using
computational algebraic geometry packages such as \texttt{Singular}~\cite{DGPS}
or \texttt{msolve}~\cite{msolve}, which return an explicit generating set of
$\cI_\Gamma$.
Given a generating set of $\cI_\Gamma$, one could in principle solve the equations.
However, solving them explicitly would in general
introduce algebraic functions, such as square roots, which eventually cancel.
To avoid this intermediate complication, we work in the companion matrix
formalism~\cite{CoxLittleOShea2005,Brunello:2024tqf,Chestnov:2025svg}, which we
briefly review here.

Since we restrict to cases where the solution set of $\cI_\Gamma$ consists of
isolated and non-degenerate critical points, the quotient ring
\begin{equation}
\label{eq:quotient_ring}
    Q_\Gamma
    \defas \Complex\sbrk{z_1,\ldots,z_n}/\cI_\Gamma
\end{equation}
is a finite-dimensional vector space, of dimension $\nu_\Gamma$, equal to the
number of critical points counted in step~(i).
A monomial basis $\brc{m_1,\ldots,m_{\nu_\Gamma}}$ of $Q_\Gamma$
can be computed from a Gr\"obner basis of $\cI_\Gamma$.
To every $f \in Q_\Gamma$ we associate its \textit{companion matrix} $M\sbrk{f}$, defined by
the multiplication map in the quotient ring,
\begin{equation}
\label{eq:companion_matrix}
    f\, m_i
    =
    \sum_{j=1}^{\nu_\Gamma} M\sbrk{f}_{ij}\; m_j
    \qquad \operatorname{mod}\;\; \cI_\Gamma
    \>,
\end{equation}
where $\operatorname{mod}\;\; \cI_\Gamma$ denotes reduction by the
aforementioned Gr\"obner basis.
The eigenvalues of $M\sbrk{f}$ are the evaluations of $f$ on the points of
$\operatorname{Crit}_\Gamma\brk{W}$. Therefore the sums over critical points appearing
in~\cref{eq:lo_pairing,eq:subleading_pairing_final} can be computed as traces,
\begin{equation}
\label{eq:trace_formula}
    \sum_{z^\star\in\operatorname{Crit}_\Gamma\brk{W}} f\brk{z^\star}
    =
    \Tr\bigbrk{\, M\sbrk{f} \,}
    \>.
\end{equation}
Rational functions are handled by representing numerator and denominator separately, the
latter being invertible in $Q_\Gamma$ whenever it does not vanish on the critical
locus.
In this approach, the computation of the differential system is thus reduced to multivariate
polynomial division and linear algebra over the field of rational functions in the kinematic
variables, which is amenable to finite-field
reconstruction~\cite{vonManteuffel:2014ixa,Peraro:2016wsq}.

It remains to assemble the pairings into the differential system. The
relations~\labelcref{eq:Omegalin} hold at finite $\delta$, whereas our formulae
for the pairings give only their limits as $\delta\to0$, so we need the
intersection matrices themselves to have finite limits.
This is what the rescaling of the dual forms in step~(ii) achieves, with $h_b =
\vartheta_b$ in~\cref{eq:CDlin}.
Since the connection matrix~\labelcref{eq:Omega_from_C} does not depend on the
normalisation of the dual forms, this leaves $\mathbf\Omega_\delta$ unchanged,
while the intersection matrices acquire finite limits,
\begin{equation}
\label{eq:C_matrix_delta_limit}
    \mathbf{C}
    \defas
    \lim_{\delta \to 0} \mathbf{C}_\delta
    =
    \mathbf{C}^{\brk{0}}
    +
    \varepsilon^{-1}\, \mathbf{C}^{\brk{-1}}
    +
    \cdots
    \>,
    \qquad
    \mathbf{D}
    \defas
    \lim_{\delta \to 0} \mathbf{D}_\delta
    =
    \varepsilon\, \mathbf{D}^{\brk{1}}
    +
    \mathbf{D}^{\brk{0}}
    +
    \cdots
    \>.
\end{equation}
These limiting matrices are what step~(iii) evaluates, by applying our results for the $\delta \to
0$ limits of the intersection pairing to \cref{eq:CDlin}.
Taking $\delta\to0$ in~\cref{eq:Omegalin} then gives the coefficients $\mathbf\Omega^{(1)}$ and $\mathbf\Omega^{(0)}$ of
the unregulated connection matrix~\labelcref{eq:Omega_unregulated_expansion} in terms of the limiting matrices,
\begin{align}
\label{eq:Omegalin_unregulated}
    \mathbf\Omega^{(1)}
    &=
    \mathbf D^{(1)} \cdot \bigbrk{\mathbf C^{(0)}}^{-1}
    \>,
    \\
    \mathbf\Omega^{(0)}
    &=
    \Bigbrk{\mathbf D^{(0)} - \mathbf D^{(1)} \cdot \bigbrk{\mathbf C^{(0)}}^{-1} \cdot \mathbf C^{(-1)}} \cdot \bigbrk{\mathbf C^{(0)}}^{-1}
    \>.
\end{align}
Note that this construction requires $\mathbf{C}^{\brk{0}}$ to be invertible for generic
kinematics, which we assume from now on.
Indeed, as $\mathbf{C}^{\brk{0}}$ is block triangular (see~\cref{ssec:lo_pairing}), the
assumption amounts to the invertibility of its diagonal blocks and hence
constrains the basis sector by sector.
We use bases that are expected to be canonical: up to three loops they follow
Refs.~\cite{Henn:2013pwa,Henn:2013tua}, and at four loops they are constructed
as described in~\cref{ssec:four_loop_ladder}. For such bases the differential
system is $\varepsilon$-factorised in the $\delta\to0$ limit, so that
$\mathbf\Omega^{(0)}$ should vanish.
That it does is a non-trivial check of the entire construction.

All of the above is implemented in a proof-of-concept code, which relies on
\texttt{FiniteFlow}~\cite{Peraro:2019svx} for finite-field reconstruction.
We have checked the resulting canonical differential equations for the two-,
three- and four-loop ladders against independent computations based on IBP
reduction, with \texttt{Blade}~\cite{Guan:2024byi} and, at two and three loops,
also with \texttt{FiniteFlow}~\cite{Peraro:2019svx}.
In all examples below, the leading connection matrix takes the $\dd\log$ form
over the two-letter alphabet $\brc{y,\,1+y}$,
\begin{equation}
\label{eq:dlog_form}
    \mathbf\Omega^{(1)}
    =
    \mathbf A_{0} \, \dd\log \brk{y}
    +
    \mathbf A_{1} \, \dd\log \brk{1+y}
    \>,
\end{equation}
with constant matrices $\mathbf A_0$ and $\mathbf A_1$.
For each ladder, we provide the canonical basis and the matrices $\mathbf C^{(0)}$, $\mathbf C^{(-1)}$,
$\mathbf D^{(1)}$, $\mathbf D^{(0)}$, $\mathbf\Omega^{(1)}$ and
$\mathbf\Omega^{(0)}$ in the ancillary files \texttt{$L$\_loop\_ladder\_matrices.m}, with
$L=1,\dots,4$ the number of loops.
For the four-loop ladder, we also provide the canonical differential equation in
the basis reduced by graph symmetries in \texttt{4\_loop\_ladder\_de\_can.m}.

\subsection{One-Loop Box: A Pedagogical Example}
\label{ssec:one_loop_box}

We begin with the one-loop box, which was treated with analytically-regulated
higher residue pairings in Ref.~\cite{Mizera:2019vvs}.
The family has three non-vanishing sectors, each containing a single critical point.
The sums over critical points can therefore be evaluated directly, without
companion matrices, which makes the box a convenient first illustration of the
procedure of~\cref{ssec:sector_sums}.
In the following, we go through the steps, computing one leading and one subleading pairing explicitly as sums of sector contributions, each a rational function evaluated at the corresponding critical point.

The box integral family shown in~\cref{fig:ladders} is defined by
\begin{equation}
    I_{\alpha_1 \alpha_2 \alpha_3 \alpha_4} = \int \frac{\dd^D \ell}{\im\pi^{D/2}}\frac{1}{D_1^{\alpha_1}D_2^{\alpha_2}D_3^{\alpha_3}D_4^{\alpha_4}}\, ,
\end{equation}
where the denominators are
\begin{align}
    D_1 = -\ell^2 \,, & & D_2 = -(\ell+p_1)^2 \,, & & D_3 = -(\ell+p_1+p_2)^2 \,, & & D_4 = -(\ell-p_3)^2 \, .
\end{align}
The regulated Lee--Pomeransky representation of the family, as defined in~\cref{eq:leepom_reg}, reads
\begin{equation}
    I_{\alpha_1 \alpha_2 \alpha_3 \alpha_4}^{\brk{\delta}} = \frac{\Gamma\brk{2-\varepsilon}}{\Gamma\brk{4-2\varepsilon-\vert\boldsymbol{\alpha}\vert-4\varepsilon\delta}} \int \cG^{\varepsilon-2}\prod_{i=1}^4\frac{\dd z_i \, z_i^{\alpha_i-1+\varepsilon\delta}}{\Gamma\brk{\alpha_i+\varepsilon\delta}} \, ,
\end{equation}
with $\cG$ given, for $s=-1$ and $t=-y$, by
\begin{align}
\label{eq:box_G}
    \cG &= z_1+z_2+z_3+z_4+z_1z_3+y\,z_2z_4 \, .
\end{align}
We rewrite this family as a twisted period integral family, normalising the forms as
in~\cref{eq:rational_normalisation}, with $L=1$, $c=0$ and $p=1$:
\begin{equation}
\frac{\Gamma\brk{-2\varepsilon}}{\varepsilon\, \Gamma\brk{2-\varepsilon}}   I_{\alpha_1 \alpha_2 \alpha_3 \alpha_4}^{\brk{\delta}}= \int_{\mathbb{R}_+^4}e^{\varepsilon W_\delta} \varphi_{\alpha_1 \alpha_2 \alpha_3 \alpha_4} \, , \qquad   W_\delta
    = \log \brk{\cG}+\delta\sum_{i=1}^4 \log \brk{z_i}
    \, ,
    \label{eq:box_integral_definition}
\end{equation}
where
\begin{align}
\label{eq:BoxFormDefinition}
    \varphi_{\alpha_1 \alpha_2 \alpha_3 \alpha_4} &= \frac{\Gamma\brk{-2\varepsilon}}{\varepsilon\, \Gamma\brk{4-2\varepsilon-\vert\boldsymbol{\alpha}\vert-4\varepsilon\delta}} \frac{1}{\cG^2}\prod_{i=1}^4 \frac{\dd z_i \, z_i^{\alpha_i-1}}{\Gamma\brk{\alpha_i+\varepsilon\delta}} \, .
\end{align}
The $\Gamma$-function prefactor is chosen so that the large-$\varepsilon$ expansion of the
forms starts at order $\varepsilon^0$ and is rational order by order in $\delta$.

\begin{figure}[t]
\centering
\begin{tabular}{@{}ccc@{}}
\gsfig{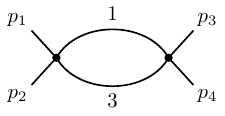} & \gsfig{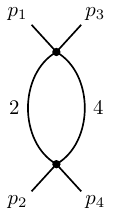} & \gsfig{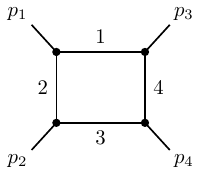} \\[14pt]
$\Gamma_{\set{2,4}}$ & $\Gamma_{\set{1,3}}$ & $\Gamma_{\varnothing}$
\end{tabular}
\caption{\label{fig:one_loop_sectors}%
The sectors of the one-loop box with associated master integrals.  Propagators are numbered as in \cref{fig:ladders}.}
\end{figure}

Our first step is to locate the critical points sector by sector.
We obtain the sector ideals with non-empty critical loci from
\cref{eq:critical_points_expanded} computationally. As there is only one
master per sector, the ideals are linearly generated and given by
\begin{align}
 \cI_{\set{2,4}}& = \langle z_2,\,z_4,\,z_1-z_3,\, z_3+1\rangle \, ,
    \nonumber \\
    \cI_{\set{1,3}} &= \langle z_1,\,z_3,\,z_2-z_4,\, y \,z_4+1\rangle \, ,
    \nonumber \\
    \cI_{\varnothing} &= \langle z_1-z_3,\,z_2-z_4,\, y\, z_4+1,\, z_3+1\rangle \,,
\end{align}
corresponding, respectively, to the sector topologies of
\cref{fig:one_loop_sectors}.
Here, $\Gamma_{\varnothing}$ is the top topology of the one-loop box, while $\Gamma_{\set{2,4}}$ and
$\Gamma_{\set{1,3}}$ are the one-loop bubbles obtained by pinching the edges $\set{2,4}$ and
$\set{1,3}$, respectively.
For generic kinematics, $y\neq0,-1$, each ideal has a single solution,
\begin{equation}
\label{eq:box_critical_points}
    z^\star_{\set{2,4}}=(-1,0,-1,0)
    \>,\qquad
    z^\star_{\set{1,3}}=\left(0,-\frac{1}{y},0,-\frac{1}{y}\right)
    \>,\qquad
    z^\star_{\varnothing}=\left(-1,-\frac{1}{y},-1,-\frac{1}{y}\right)
    \>,
\end{equation}
where the coordinates are ordered as $(z_1,z_2,z_3,z_4)$.
Hence, the corresponding dimensions are
\begin{equation}
\label{eq:box_nu}
    \nu_{\set{2,4}}=1 \,, \qquad \nu_{\set{1,3}}=1 \,, \qquad \nu_{\varnothing}=1 \,,
\end{equation}
so that this integral family has three master integrals.

Next, we choose for each sector $\Gamma$ a set $e_\Gamma$ of $\nu_\Gamma$ master forms.
For the one-loop box, each set contains a single element,
\begin{equation}
\label{eq:box_basis}
 e_{\set{2,4}}= \bigbrc{\varphi_{1\, 0 \, 2 \, 0 }}\, ,
 \qquad
 e_{\set{1,3}}=  \bigbrc{y \, \varphi_{0\, 1 \, 0 \, 2 }} \, ,
 \qquad
 e_{\varnothing}= \bigbrc{y \, \varepsilon \, \varphi_{1\, 1 \, 1 \, 1 }} \, ,
\end{equation}
whose elements we denote by $e_{\set{2,4},1}$, $e_{\set{1,3},1}$ and $e_{\varnothing,1}$, respectively.
Following~\cref{eq:basis_normalisation}, we take as dual basis the corresponding
$\delta$-rescaled forms,
\begin{equation}
\label{eq:box_dual_basis}
    \vartheta_{\set{2,4},1} = \delta^{-2}\,e_{\set{2,4},1}^\vee
    \>,
    \qquad
    \vartheta_{\set{1,3},1} = \delta^{-2}\,e_{\set{1,3},1}^\vee
    \>,
    \qquad
    \vartheta_{\varnothing,1} = e_{\varnothing,1}^\vee
    \>,
\end{equation}
where $e_{\Gamma,k}^\vee$ is the dual form associated with $e_{\Gamma,k}$, obtained by stripping off its $\varepsilon$-dependence. 
The power of $\delta^{-1}$ equals the number of pinched edges of the sector: two for the bubble
sectors $\set{2,4}$ and $\set{1,3}$, and none for the top sector $\varnothing$.
The pairings below require the pinch factors and regular factors of these forms.
At leading order in the $\delta$-expansion, the decomposition~\labelcref{eq:phi_g_decomp} of the
master forms reads
\begin{align}
\label{eq:box_delta_psi}
    \Delta_{\set{2,4},1} &= \frac{\delta^2}{z_2 z_4}
    \, , &
    \Delta_{\set{1,3},1} &= \frac{\delta^2}{z_1 z_3}
    \, , &
    \Delta_{\varnothing,1} &= 1
    \, ,
    \\
\label{eq:box_psi}
    \psi_{\set{2,4},1} &= -\frac{z_3}{2 \,\cG^2}
    \, , &
    \psi_{\set{1,3},1} &= -\frac{y \, z_4}{2\, \cG^2}
    \, , &
    \psi_{\varnothing,1} &= \frac{y}{\cG^2}
    \, .
\end{align}

The third step is the evaluation of the pairings. For simplicity, we consider
the evaluation of an entry of $\mathbf{C}$ to subleading order.
In order to see how the sector sums work, we choose the off-diagonal entry of
$\mathbf C^{(0)}$ associated with pairing the two forms $e_{\varnothing,1}$ and 
$\vartheta_{\set{1,3},1}$.
The pinch sets of these two forms are
\begin{equation}
    \Gamma_{a}= \varnothing \, ,
    \qquad
    \Gamma_{b}= \set{1,3}
    \, .
\end{equation}
From~\cref{eq:lo_pairing}, only the sectors with $\Gamma_{a}\subseteq \Gamma \subseteq \Gamma_b$
can contribute. Among these, the only sectors with associated critical points are
\begin{equation}
\label{eq:1loop_sectors}
   \Gamma\in \set{\varnothing,\, \set{1,3}}
   \>.
\end{equation}
The edges of $\Gamma_b\setminus\Gamma_a = \set{1,3}$ are pinched in the sector $\set{1,3}$,
where they contribute $1/\omega_i$, and cut in the top sector, where they contribute $1/z_i$.
The pairing therefore reads
\begin{align}
    \vev{e_{\varnothing,1}\mid \vartheta_{\set{1,3},1}}_{(0)}
    & =
    \sum_{z^\star\in\operatorname{Crit}_{\set{1,3}}\brk{W}}\frac{1}{\omega_1 \omega_3}\,
    \frac{\psi_{\varnothing,1}\,\psi_{\set{1,3},1}}{\det\bigbrk{\, \Phi_{\brc{2,4}} \,}}
    \bigg|_{z^\star}
    +
    \sum_{z^\star\in\operatorname{Crit}_{\varnothing}\brk{W}}\frac{1}{z_1 z_3}\,
    \frac{\psi_{\varnothing,1}\,\psi_{\set{1,3},1}}{\det\bigbrk{\, \Phi_{\brc{1,2,3,4}} \,}}
    \bigg|_{z^\star}
    \nonumber\\
    &= -\frac{1}{2y}+\frac{1}{2y}=0
    \>,
\end{align}
where we used $\omega_1=\omega_3=-y$ and
$\det\bigbrk{\, \Phi_{\brc{2,4}} \,}=-y^4$ at $z^\star_{\set{1,3}}$, while
$\det\bigbrk{\, \Phi_{\brc{1,2,3,4}} \,}=y^6/(1+y)^4$ at $z^\star_{\varnothing}$.
The bubble and the top sector contribute with opposite signs, so this entry vanishes.

We now turn to the subleading counterpart of this pairing, which contributes to
the corresponding entry of $\mathbf C^{(-1)}$.
The pinch sets of the two forms are the same as in the leading pairing
calculation, and therefore so are the contributing sectors
(see \cref{eq:1loop_sectors}).
Using~\cref{eq:subleading_pairing_final}, we can write
\begin{align}
    \vev{e_{\varnothing,1}\mid \vartheta_{\set{1,3},1}}_{(1)}
    &=
    \left.
    \frac{1}{2}\sum_{z^\star\in\operatorname{Crit}_{\set{1,3}}\brk{W}}\frac{1}{\omega_1\omega_3}
    \,
    \frac{1}{\det\bigbrk{\, \Phi_{\set{2,4}} \,}}
    \bigbrk{T^{\overline{\Gamma}}_{ab}+T^{\Gamma}_{ab}+T^{\Gamma\overline{\Gamma}}_{ab}}
    \right|_{z^\star}
    \nonumber \\
    &\quad+
    \left.
    \frac{1}{2}\sum_{z^\star\in\operatorname{Crit}_{\varnothing}\brk{W}}\frac{1}{z_1z_3}
    \,
    \frac{1}{\det\bigbrk{\, \Phi_{\set{1,2,3,4}} \,}}
    \,
    T^{\overline{\Gamma}}_{ab}
    \right|_{z^\star}
    \>,
\end{align}
with $a = \brk{\varnothing,1}$ and $b = \brk{\set{1,3},1}$, and with the regular factors $\psi_a=\psi_{\varnothing,1}$ and $\psi_b=\psi_{\set{1,3},1}$ of~\cref{eq:box_psi}.
In the bubble sector, $\Gamma=\set{1,3}$, the cut directions are $\overline{\Gamma}=\set{2,4}$.
Since $\Gamma=\Gamma_b$, the indicator terms of~\cref{eq:T_cut_block} vanish, and
\begin{equation}
    T^{\overline{\Gamma}}_{ab}\Bigr\rvert_{\Gamma=\set{1,3}}
    =
    \Tr\Bigbrk{\, \Phi_{\set{2,4}}^{-1} \cdot \midtilde{M}_{ab} \,}
    \>.
\end{equation}
In contrast, the pinched edges $\Gamma\setminus\Gamma_a=\set{1,3}$ enter
through $T^{\Gamma}_{ab}$ and $T^{\Gamma\overline{\Gamma}}_{ab}$ of~\cref{eq:T_pinch_block,eq:T_mixed_block},
\begin{equation}
\begin{split}
    \bigbrk{T^{\Gamma}_{ab}+T^{\Gamma\overline{\Gamma}}_{ab}}\Bigr\rvert_{\Gamma=\set{1,3}}
    &=
    \psi_a \psi_b
    \sum_{i,j\in\set{1,3}}
    \brk{1+\delta_{ij}}
    \frac{S_{\set{1,3} ,\, ij}}{\omega_i\,\omega_j}
    \\
    &\quad-
    \psi_b
    \sum_{i\in\set{1,3}}\frac{1}{\omega_i}
    \Bigbrk{
        2 \> \partial_i \psi_a
        - \psi_a \> \partial_i \log\det\bigbrk{\, \Phi_{\set{2,4}} \,}
    }
    \\
    &\quad+
    \psi_b
    \sum_{\substack{i\in\set{1,3} \\ \bar{j} \in \set{2,4}}}
    \frac{1}{\omega_i}
    \bigbrk{\Phi_{\Gamma\overline{\Gamma}} \cdot \Phi_{\overline{\Gamma}}^{-1}}_{i\bar j}
    \Bigbrk{
        2 \> \partial_{\bar j} \psi_a
        - \psi_a \> \partial_{\bar j} \log\det\bigbrk{\, \Phi_{\set{2,4}} \,}
    }
    \>,
\end{split}
\end{equation}
where $S_{\set{1,3}}$ is the Schur
complement~\labelcref{eq:unregulated_Schur_complement}.

In the top sector, $\Gamma=\varnothing$, there are no pinched directions, and only the cut
contribution $T_{ab}^{\overline{\Gamma}}$ survives, with all four variables cut.
Since $\Gamma_b\setminus\Gamma = \set{1,3}$, the edges $1$ and $3$ contribute to the indicator terms:
\begin{equation}
\begin{split}
    T^{\overline{\Gamma}}_{ab}\Bigr\rvert_{\Gamma=\varnothing}
    &=
    \Tr\Bigbrk{\, \Phi_{\set{1,2,3,4}}^{-1} \cdot \midtilde{M}_{ab} \,}
    +
    \psi_a
    \sum_{\substack{
        \bar{i} \in \set{1,3}
        \\
        \bar{j} \in \set{1,2,3,4}
    }}
    \frac{\bigbrk{\Phi_{\set{1,2,3,4}}^{-1}}_{\bar i\bar j}}{z_{\bar i}}
    \biggl(
        \frac{\delta_{\bar j\set{1,3}} \, \psi_b}{z_{\bar j}}\bigbrk{1+\delta_{\bar i\bar j}}
    \\
    &\hspace{12em}
        - 2 \> \partial_{\bar j} \psi_b
        + \psi_b \> \partial_{\bar j}\log\det\bigbrk{\, \Phi_{\set{1,2,3,4}} \,}
    \biggr)
    \>.
\end{split}
\end{equation}
A direct evaluation at the two critical points gives
\begin{align}
    T^{\overline{\Gamma}}_{ab}\Bigr\rvert_{\Gamma=\set{1,3}}&=0
    \>,&
    \bigbrk{T^{\Gamma}_{ab}+T^{\Gamma\overline{\Gamma}}_{ab}}\Bigr\rvert_{\Gamma=\set{1,3}}
    &=-y^4\brk{1+y}
    \>,&
    T^{\overline{\Gamma}}_{ab}\Bigr\rvert_{\Gamma=\varnothing}
    &=-\frac{y^4}{\brk{1+y}^3}
    \>.
\end{align}
Multiplying these by one half of the residue measure~\labelcref{eq:measure_limit} at the corresponding critical point, we obtain the subleading pairing
\begin{equation}
    \vev{e_{\varnothing,1}\mid \vartheta_{\set{1,3},1}}_{(1)}
    =
    \frac{1+y}{2y^2}
    -
    \frac{1+y}{2y^2}
    =0
    \>.
\end{equation}
As at leading order, the bubble and the top sector cancel, so this entry vanishes at both orders.

The remaining entries of $\mathbf{C}^{(0)}$ and $\mathbf{C}^{(-1)}$ follow in the same way.
Since only sectors with $\Gamma_a\subseteq\Gamma\subseteq\Gamma_b$ contribute, at either order an off-diagonal pairing can be non-zero only if $\Gamma_a\subsetneq\Gamma_b$.
This leaves the pairings of $e_{\varnothing,1}$ with $\vartheta_{\set{1,3},1}$ and with $\vartheta_{\set{2,4},1}$.
The first was computed above, and the second vanishes by the same cancellation between a bubble and the top sector.
Each diagonal entry of $\mathbf C^{(0)}$ receives a contribution from a single sector, $\Gamma=\Gamma_a=\Gamma_b$, while the diagonal subleading pairings vanish by the antisymmetry~\labelcref{eq:pairing_symmetry}.
Finally, at leading order in $\delta$ the coefficients $e_a^{(-1)}$ vanish for this basis, so they give no additional contribution to $\mathbf C^{(-1)}$ in~\cref{eq:CDlin}.
Ordering the basis as $\brk{e_{\set{2,4},1},\,e_{\set{1,3},1},\,e_{\varnothing,1}}$, and the dual basis accordingly, the intersection matrices are
\begin{equation}
    \mathbf C^{(0)} = \begin{bmatrix}
     -\frac{1}{4} & \mzero & \mzero \\
    \mzero & -\frac{1}{4} & \mzero \\
     \mzero & \mzero & 1 \\
    \end{bmatrix}
    \, , 
    \qquad
    \mathbf C^{(-1)} = \mathbf 0
    \, .
\end{equation}
The entries of $\mathbf D^{(1)}$ are leading pairings of $\dd_y W\wedge e_a$ with $\vartheta_b$, where $\dd_yW=(z_2z_4/\cG)\,\dd y$.
This factor takes different values at the critical points of the bubble and the top sector, so the contributions that cancelled in $\mathbf C^{(0)}$ no longer do.
The subleading matrix $\mathbf D^{(0)}$, which also receives the terms from the $y$-dependence of the forms themselves, vanishes. Altogether, we find
\begin{equation}
    \mathbf D^{(1)} = \begin{bmatrix}
     \mzero & \mzero & \mzero \\
 \mzero & \frac{1}{4 y} & \mzero \\
 -\frac{1}{2 (y+1)} & \frac{1}{2y ( y+1)} &
   -\frac{1}{y(y+1)} \\
    \end{bmatrix}\,\dd y
    \, , 
    \qquad
    \mathbf D^{(0)} = \mathbf 0
    \, .
\end{equation}

Finally, we assemble the differential equation. Inserting these matrices
into~\cref{eq:Omegalin_unregulated}, we obtain the leading and subleading
differential-equation matrices $\mathbf\Omega^{(1)}$ and $\mathbf\Omega^{(0)}$.
Since $\mathbf C^{(-1)}$ and $\mathbf D^{(0)}$ vanish, so does
$\mathbf\Omega^{(0)}$, as required for a canonical basis. The leading matrix
$\mathbf\Omega^{(1)}$ is of the form~\cref{eq:dlog_form} with
\begin{equation}
    \mathbf A_{0} = \begin{bmatrix}
    \mzero & \mzero & \mzero \\
    \mzero & -1 & \mzero \\
    \mzero & -2 & -1
    \end{bmatrix}
    \, ,
    \qquad
    \mathbf A_{1} = \begin{bmatrix}
    \mzero & \mzero & \mzero \\
    \mzero & \mzero & \mzero \\
    2 & 2 & 1
    \end{bmatrix}
    \, .
\end{equation}
As expected, the differential equation is $\varepsilon$-factorised and agrees with the result of Ref.~\cite{Mizera:2019vvs}.

\subsection{Two-Loop Ladder}
\label{ssec:two_loop_ladder}
\begin{figure}[t]
\centering
\setlength{\tabcolsep}{0pt}
\begin{tabular}{@{}*{4}{>{\centering\arraybackslash}p{0.25\textwidth}}@{}}
\gsfigs{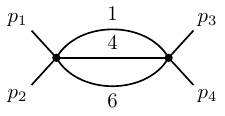} & \gsfigs{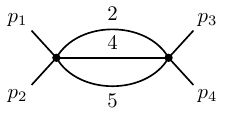} & \gsfigs{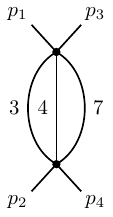} & \gsfigs{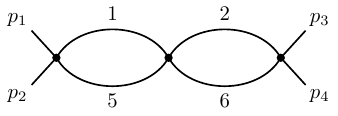} \\[-2pt]
$\Gamma_{\set{2,3,5,7}}$ & $\Gamma_{\set{1,3,6,7}}$ & $\Gamma_{\set{1,2,5,6}}$ & $\Gamma_{\set{3,4,7}}$ \\[12pt]
\gsfigs{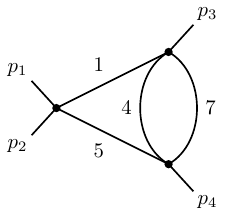} & \gsfigs{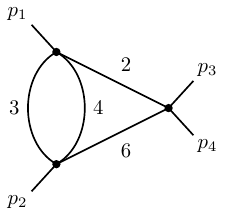} & \gsfigs{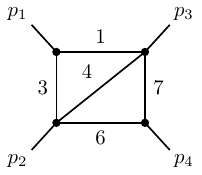} & \gsfigs{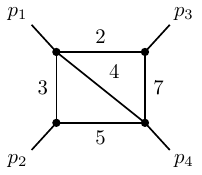} \\[-2pt]
$\Gamma_{\set{2,3,6}}$ & $\Gamma_{\set{1,5,7}}$ & $\Gamma_{\set{2,5}}$ & $\Gamma_{\set{1,6}}$ \\[12pt]
\gsfigs{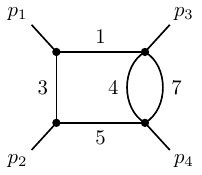} & \gsfigs{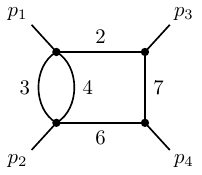} & \gsfigs{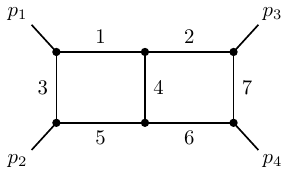} & \\[-2pt]
$\Gamma_{\set{2,6}}$ & $\Gamma_{\set{1,5}}$ & $\Gamma_{\varnothing}$ &
\end{tabular}
\caption{\label{fig:two_loop_sectors}%
The eleven sectors of the two-loop ladder.  Propagators are numbered as
in \cref{fig:ladders}.}
\end{figure}

As a more involved example of our approach, here we consider the two-loop
ladder.
The top sector of the two-loop ladder contains two critical points whose
coordinates involve a square root.
This example illustrates the benefits of the companion matrix approach: we can
evaluate the sum of their contributions using traces of companion matrices,
keeping the calculation rational.
The integral family, with seven propagators and two ISPs, its regulated
Lee--Pomeransky representation and the normalisation of the differential forms
are defined in appendix~\labelcref{app:two_loop_ladder}.

From~\cref{eq:critical_points_expanded}, we find the eleven non-vanishing
sectors of~\cref{fig:two_loop_sectors}, whose dimensions $\nu_\Gamma$ add up to
$12$: two in the top sector and one in each subsector. Graph symmetries reduce
the number of independent master integrals to eight.
Following Ref.~\cite{Henn:2013pwa}, and keeping symmetry-related sectors
distinct, we choose the basis of differential forms
\begin{equation}
\label{eq:two_loop_basis}
\resizebox{0.9\textwidth}{!}{$\displaystyle
\begin{array}{lll}
e_{\set{2,3,5,7}}=\bigbrc{\varphi_{2\,0\,0\,1\,0\,2\,0}}\, ,
&
e_{\set{1,3,6,7}}=\bigbrc{\varphi_{0\,2\,0\,1\,2\,0\,0}}\, ,
&
e_{\set{1,2,5,6}}=\bigbrc{y\,\varphi_{0\,0\,1\,2\,0\,0\,2}}\, ,
\\[4pt]
e_{\set{3,4,7}}=\bigbrc{\varphi_{2\,2\,0\,0\,1\,1\,0}}\, ,
&
e_{\set{2,3,6}}=\bigbrc{\varepsilon\,\varphi_{1\,0\,0\,2\,1\,0\,1}}\, ,
&
e_{\set{1,5,7}}=\bigbrc{\varepsilon\,\varphi_{0\,1\,1\,2\,0\,1\,0}}\, ,
\\[4pt]
e_{\set{2,5}}=\bigbrc{(1+y)\,\varepsilon^2\,\varphi_{1\,0\,1\,1\,0\,1\,1}}\, ,
&
e_{\set{1,6}}=\bigbrc{(1+y)\,\varepsilon^2\,\varphi_{0\,1\,1\,1\,1\,0\,1}}\, ,
&
e_{\set{2,6}}=\bigbrc{y\,\varepsilon\,\varphi_{1\,0\,1\,2\,1\,0\,1}}\, ,
\\[4pt]
e_{\set{1,5}}=\bigbrc{y\,\varepsilon\,\varphi_{0\,1\,1\,2\,0\,1\,1}}\, ,
&
\multicolumn{2}{l}{
e_{\varnothing}=
\bigbrc{y\,\varepsilon^2\,\varphi_{1\,1\,1\,1\,1\,1\,1},
\varepsilon^2\,\varphi^{\set{8},(1)}_{1\,1\,1\,1\,1\,1\,1}}}\, .
\end{array}
$}
\end{equation}
We label the elements of each set by $e_{\Gamma,k}$ in the displayed order and take the
corresponding $\delta$-rescaled dual forms $\vartheta_{\Gamma,k}$ according
to~\cref{eq:basis_normalisation}.

With the basis fixed, let us now discuss how a pairing over the two top-sector
critical points is evaluated with companion matrices.
Specifically, we shall consider the leading pairing between the two top-sector
forms, an off-diagonal entry of $\mathbf C^{(0)}$,
\begin{equation}
    \vev{e_{\varnothing,1}\mid \vartheta_{\varnothing,2}}_{(0)}
    \>,
    \qquad
    \Gamma_a = \Gamma_b = \varnothing
    \>.
\end{equation}
Only the top sector contributes, so this pairing only involves the sum over the
top sector critical points, and~\cref{eq:lo_pairing} reduces to
\begin{equation}
\label{eq:two_loop_top_pairing}
    \vev{e_{\varnothing,1}\mid \vartheta_{\varnothing,2}}_{(0)}
    =
    \sum_{z^\star\in\operatorname{Crit}_\varnothing\brk{W}}
    \left(\frac{\psi_{\varnothing,1}\,\psi_{\varnothing,2}}{\det\Phi_{\overline{\varnothing}}} \right)
    \, ,
\end{equation}
where $\Phi_{\overline{\varnothing}}$ is the full unregulated Hessian.
For generic kinematics, the top sector has two critical points,
\begin{equation}
\label{eq:two_loop_critical_points}
    z^\star_{\varnothing,\pm}
    = \left(-\frac{2}{3},-\frac{2}{3}, \frac{-3\pm\sqrt{9+8y}}{3y}, \frac{-3\pm\sqrt{9+8y}}{3y},
        -\frac{2}{3},-\frac{2}{3}, \frac{-3\pm\sqrt{9+8y}}{3y}\right)
    \>,
\end{equation}
with coordinates ordered as $(z_1,\ldots,z_7)$.
Their contributions can be summed without manipulating the square root, by working
in the quotient ring introduced in~\cref{ssec:sector_sums}.
The relevant quotient ring $Q_\varnothing$ has dimension $\nu_\varnothing=2$,
and we use a monomial basis $\brc{z_7,1}$ and the multiplication matrices
$M\sbrk{f}$ defined in~\cref{eq:companion_matrix}.
In this basis, multiplication by $\cG$ is represented by
\begin{equation}
M\sbrk{\cG}=
\begin{bmatrix}
\dfrac{4(27+30y+4y^2)}{27y^2} &
-\dfrac{16(9+8y)}{81y^2}\\[4pt]
-\dfrac{2(9+8y)}{9y} &
\dfrac{8(3+2y)}{27y}
\end{bmatrix}.
\end{equation}
Since $M\sbrk{fg}=M\sbrk{f}\,M\sbrk{g}$, this matrix also gives the powers of $\cG$ needed below.
The denominators of $\psi_{\varnothing,1}$ and $\psi_{\varnothing,2}$ are $\cG^2$ and
$\cG^3$, respectively, the latter receiving one additional power from the
insertion factor $\cR_{\set{8}}^{(1)}$ associated with the ISP $D_8$.
Since companion matrices represent polynomials, we reorganise the summand
of~\cref{eq:two_loop_top_pairing} into the form
\begin{equation}
    \frac{\psi_{\varnothing,1}\,\psi_{\varnothing,2}}{\det\Phi_{\overline{\varnothing}}}
    =
    \frac{(\cG^{2}\,\psi_{\varnothing,1})(\cG^{3}\,\psi_{\varnothing,2})}{\det\Phi_{\overline{\varnothing}}\,\cG^{5}}
    \>,
\end{equation}
which is manifestly a ratio of polynomials.
The companion matrices of the two numerators are
\begin{equation}
M\sbrk{\cG^{2}\,\psi_{\varnothing,1}}=
\begin{bmatrix}
9y & 0\\
0 & 9y
\end{bmatrix},
\qquad
M\sbrk{\cG^{3}\,\psi_{\varnothing,2}}=
\begin{bmatrix}
\dfrac{4(9+8y)}{3y} & -\dfrac{16(3+2y)}{9y}\\[4pt]
-2(3+2y) & \dfrac{8}{3}
\end{bmatrix},
\end{equation}
while the denominator $\det\Phi_{\overline{\varnothing}}\,\cG^5$ is represented by
\begin{equation}
M\sbrk{\det\Phi_{\overline{\varnothing}} \,\cG^5}=
\begin{bmatrix}
-2(9+10y) & 8+\dfrac{32y}{3}\\[4pt]
3y(3+4y) & 4y
\end{bmatrix}.
\end{equation}
Therefore, the summand of~\cref{eq:two_loop_top_pairing} is represented in the quotient ring 
by the matrix
\begin{equation}
M\sbrk{\cG^{2}\,\psi_{\varnothing,1}} \cdot
M\sbrk{\cG^{3}\,\psi_{\varnothing,2}} \cdot
M\sbrk{\det\Phi_{\overline{\varnothing}} \,\cG^5}^{-1}
=
\begin{bmatrix}
-\dfrac{6(9+4y)}{9+8y} & \dfrac{24}{9+8y}\\[4pt]
\dfrac{27y}{9+8y} & -\dfrac{24y}{9+8y}
\end{bmatrix}\,.
\end{equation}
By~\cref{eq:trace_formula}, the sum over the critical points is the trace of this matrix,
\begin{equation}
\vev{e_{\varnothing,1}\mid \vartheta_{\varnothing,2}}_{(0)}
=
\Tr\Bigbrk{\,
M\sbrk{\cG^{2}\,\psi_{\varnothing,1}} \cdot
M\sbrk{\cG^{3}\,\psi_{\varnothing,2}} \cdot
M\sbrk{\det\Phi_{\overline{\varnothing}} \,\cG^5}^{-1}
\,}
=-6\, .
\end{equation}
While the contributions of the individual critical points involve square roots,
their sum is rational and is obtained here without explicitly solving the
critical point equations.

All other pairings are evaluated analogously.
Inserting the resulting matrices $\mathbf C$ and $\mathbf D$, which are provided
in the ancillary files, into~\cref{eq:Omegalin_unregulated}, we find that the
subleading contribution vanishes, $\mathbf\Omega^{(0)}=\mathbf 0$, while
$\mathbf\Omega^{(1)}$ has the $\dd\log$ form of~\cref{eq:dlog_form}. With the
basis~\labelcref{eq:two_loop_basis} ordered row by row, the matrices $\mathbf
A_0$ and $\mathbf A_1$ are
\begin{align}
\scalebox{0.97}{$\displaystyle
\mathbf A_{0} =
\begin{bmatrix}
\mzero & \mzero & \mzero & \mzero & \mzero & \mzero & \mzero & \mzero & \mzero & \mzero & \mzero & \mzero \\
\mzero & \mzero & \mzero & \mzero & \mzero & \mzero & \mzero & \mzero & \mzero & \mzero & \mzero & \mzero \\
\mzero & \mzero & -2 & \mzero & \mzero & \mzero & \mzero & \mzero & \mzero & \mzero & \mzero & \mzero \\
\mzero & \mzero & \mzero & \mzero & \mzero & \mzero & \mzero & \mzero & \mzero & \mzero & \mzero & \mzero \\
\mzero & \mzero & \mzero & \mzero & \mzero & \mzero & \mzero & \mzero & \mzero & \mzero & \mzero & \mzero \\
\mzero & \mzero & \mzero & \mzero & \mzero & \mzero & \mzero & \mzero & \mzero & \mzero & \mzero & \mzero \\
\frac{1}{2} & \mzero & -\frac{1}{2} & \mzero & \mzero & \mzero & -2 & \mzero & \mzero & \mzero & \mzero & \mzero \\
\mzero & \frac{1}{2} & -\frac{1}{2} & \mzero & \mzero & \mzero & \mzero & -2 & \mzero & \mzero & \mzero & \mzero \\
\mzero & \mzero & -\frac{3}{2} & \mzero & \mzero & \mzero & \mzero & \mzero & -2 & \mzero & \mzero & \mzero \\
\mzero & \mzero & -\frac{3}{2} & \mzero & \mzero & \mzero & \mzero & \mzero & \mzero & -2 & \mzero & \mzero \\
-\frac{3}{2} & -\frac{3}{2} & -3 & \mzero & \mzero & \mzero & 6 & 6 & -2 & -2 & -2 & \mzero \\
-\frac{3}{2} & -\frac{3}{2} & -\frac{9}{2} & -1 & 3 & \mzero & 9 & 9 & -2 & -2 & -1 & 1
\end{bmatrix}
$}
\quad 
\scalebox{0.97}{$\displaystyle
\mathbf A_{1} =
\begin{bmatrix}
\mzero & \mzero & \mzero & \mzero & \mzero & \mzero & \mzero & \mzero & \mzero & \mzero & \mzero & \mzero \\
\mzero & \mzero & \mzero & \mzero & \mzero & \mzero & \mzero & \mzero & \mzero & \mzero & \mzero & \mzero \\
\mzero & \mzero & \mzero & \mzero & \mzero & \mzero & \mzero & \mzero & \mzero & \mzero & \mzero & \mzero \\
\mzero & \mzero & \mzero & \mzero & \mzero & \mzero & \mzero & \mzero & \mzero & \mzero & \mzero & \mzero \\
\mzero & \mzero & \mzero & \mzero & \mzero & \mzero & \mzero & \mzero & \mzero & \mzero & \mzero & \mzero \\
\mzero & \mzero & \mzero & \mzero & \mzero & \mzero & \mzero & \mzero & \mzero & \mzero & \mzero & \mzero \\
\mzero & \mzero & \mzero & \mzero & \mzero & \mzero & 2 & \mzero & \mzero & \mzero & \mzero & \mzero \\
\mzero & \mzero & \mzero & \mzero & \mzero & \mzero & \mzero & 2 & \mzero & \mzero & \mzero & \mzero \\
\mzero & \mzero & \frac{3}{2} & \mzero & -3 & \mzero & \mzero & \mzero & 1 & \mzero & \mzero & \mzero \\
\mzero & \mzero & \frac{3}{2} & \mzero & \mzero & -3 & \mzero & \mzero & \mzero & 1 & \mzero & \mzero \\
3 & 3 & 3 & -2 & \mzero & 6 & -6 & -6 & 2 & 2 & 2 & -2 \\
\frac{3}{2} & \frac{3}{2} & \frac{9}{2} & -1 & -3 & \mzero & -9 & -9 & 2 & 2 & 1 & -1
\end{bmatrix}
$}
\end{align}

\subsection{Three- and Four-Loop Ladders}
\label{ssec:three_loop_ladder}
\label{ssec:four_loop_ladder}
\begin{figure}[t]
\centering
\setlength{\tabcolsep}{1pt}
\begin{tabular}{@{}ccccc@{}}
\sectorfig{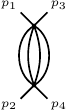} & \sectorfig{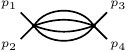} & \sectorfig{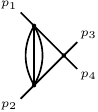} & \sectorfig{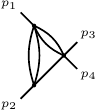} & \sectorfig{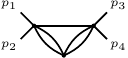} \\[17pt]
\sectorfig{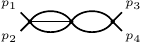} & \sectorfig{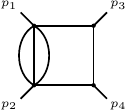} & \sectorfig{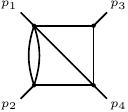} & \sectorfig{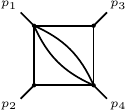} & \sectorfig{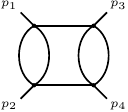} \\[17pt]
\sectorfig{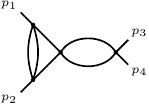} & \sectorfig{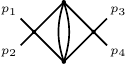} & \sectorfig{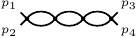} & \sectorfig{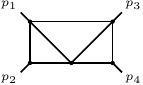} & \sectorfig{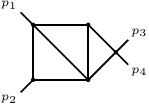} \\[17pt]
\sectorfig{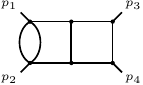} & \sectorfig{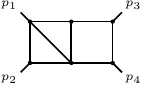} & \sectorfig{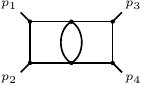} & \sectorfig{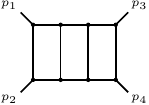} &  \\[17pt]
\end{tabular}
\caption{\label{fig:three_loop_sectors}%
The $19$ inequivalent sector topologies of the three-loop ladder,
after imposing graph symmetries.}
\end{figure}

Finally, in order to stress test our approach, we consider the three- and
four-loop ladder families. While the canonical differential equation for the three-loop ladder
is well-known~\cite{Henn:2013tua}, to our knowledge the one we provide for the four-loop ladder is new.\footnote{
    The $u$-channel cut of this family has been studied in Ref.~\cite{Bargiela:2024hgt}, while
    Ref.~\cite{Henn:2023tbo} has studied the maximal cut of the four-loop ladder with one off-shell leg.
}
As at one and two loops, our construction follows the procedure outlined in
\cref{ssec:sector_sums}, so we record only the data specific to each family.

The three-loop ladder family has ten propagators and five ISPs. We give its
regulated Lee--Pomeransky representation and the normalisation of the
differential forms are defined in appendix~\labelcref{app:three_loop_ladder}.
From~\cref{eq:critical_points_expanded}, we find $42$ non-vanishing sectors,
whose dimensions $\nu_\Gamma$ add up to $54$ before identifying graph
symmetries. After these identifications, the family has the $26$ independent
master integrals of Ref.~\cite{Henn:2013tua} in $19$ inequivalent sectors, as
shown in \cref{fig:three_loop_sectors}.
We use a $54$-element basis, obtained by carrying the $26$ canonical forms of
Ref.~\cite{Henn:2013tua} over to all sectors related to theirs by graph symmetries.
In the $54$-element basis, we obtain a $54\times54$
connection matrix. After identifying graph symmetries, it reduces to the
$26\times26$ canonical differential equation for the master integrals of
Ref.~\cite{Henn:2013tua}. As expected we find , which has the $\dd\log$ form
of~\cref{eq:dlog_form} and the subleading contribution vanishes,
$\mathbf\Omega^{(0)}=\mathbf 0$.

\begin{figure}[!p]
\centering
\setlength{\tabcolsep}{1pt}
\begin{tabular}{@{}ccccc@{}}
\sectorfig{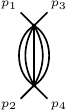} & \sectorfig{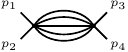} & \sectorfig{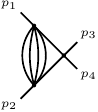} & \sectorfig{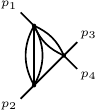} & \sectorfig{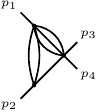} \\[17pt]
\sectorfig{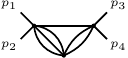} & \sectorfig{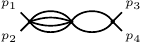} & \sectorfig{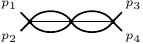} & \sectorfig{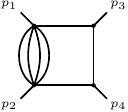} & \sectorfig{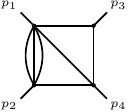} \\[17pt]
\sectorfig{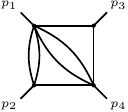} & \sectorfig{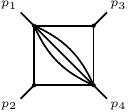} & \sectorfig{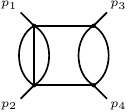} & \sectorfig{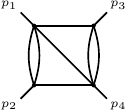} & \sectorfig{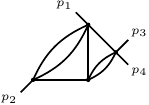} \\[17pt]
\sectorfig{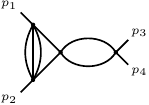} & \sectorfig{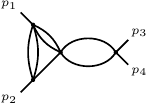} & \sectorfig{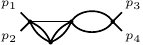} & \sectorfig{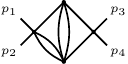} & \sectorfig{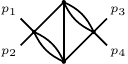} \\[17pt]
\sectorfig{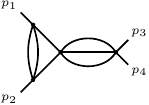} & \sectorfig{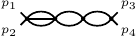} & \sectorfig{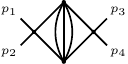} & \sectorfig{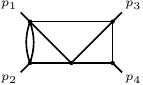} & \sectorfig{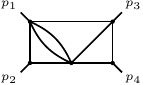} \\[17pt]
\sectorfig{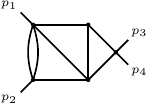} & \sectorfig{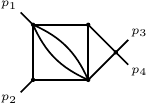} & \sectorfig{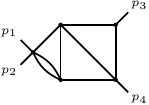} & \sectorfig{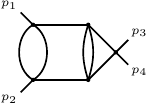} & \sectorfig{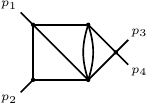} \\[17pt]
\sectorfig{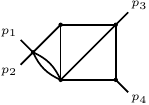} & \sectorfig{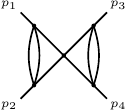} & \sectorfig{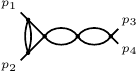} & \sectorfig{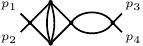} & \sectorfig{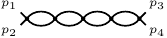} \\[17pt]
\sectorfig{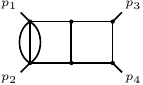} & \sectorfig{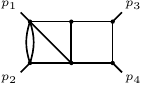} & \sectorfig{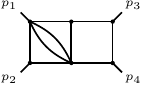} & \sectorfig{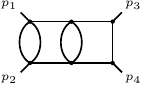} & \sectorfig{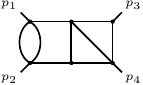} \\[17pt]
\sectorfig{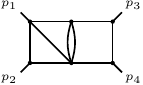} & \sectorfig{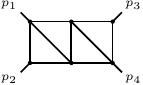} & \sectorfig{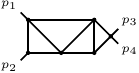} & \sectorfig{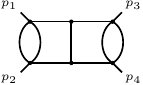} & \sectorfig{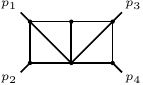} \\[17pt]
\sectorfig{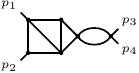} & \sectorfig{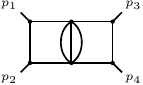} & \sectorfig{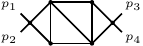} & \sectorfig{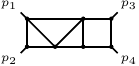} & \sectorfig{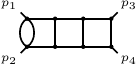} \\[17pt]
\sectorfig{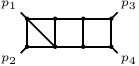} & \sectorfig{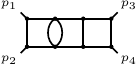} & \sectorfig{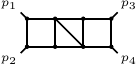} & \sectorfig{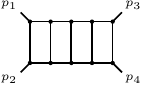} & \\[17pt]
\end{tabular}
\caption{\label{fig:four_loop_sectors}%
The $54$ inequivalent sector topologies of the four-loop ladder after imposing graph symmetries.}
\end{figure}

At four loops, we construct the canonical differential equation for the
four-loop ladder. This family serves as the most stringent test of our approach,
as it lies at the frontier of current IBP technology.
The integral family, with thirteen propagators and nine ISPs, its regulated
Lee--Pomeransky representation and the normalisation of the differential forms
are defined in appendix~\labelcref{app:four_loop_ladder}.
From~\cref{eq:critical_points_expanded}, we find $159$ non-vanishing sectors,
whose dimensions $\nu_\Gamma$ add up to $246$ before identifying graph
symmetries. After these identifications, the family has $88$ independent master
integrals, in $54$ inequivalent sectors shown in~\cref{fig:four_loop_sectors}.

Since no canonical basis is available in the literature at four loops, we construct one bottom-up, starting from the sectors
with the fewest propagators and following the sector hierarchy.
For each sector, we first bring the homogeneous part of the differential equation into
$\varepsilon$-factorised form, and then we choose the basis of forms so that the inhomogeneous
part is $\varepsilon$-factorised as well.
We find it particularly useful to read each diagram as a gluing of lower-loop sub-diagrams
and to choose the basis of forms for the sub-diagrams to be pure, using our knowledge of the
lower-loop cases from Refs.~\cite{Henn:2013pwa,Henn:2013tua}. In this way, we find a basis in
which pure integrals do not require mixing between topologies.
Extended to all $159$ non-vanishing sectors as at three loops, this basis has
$246$ elements. In it, we obtain a $246\times246$ connection matrix. After
identifying graph symmetries, it reduces to the $88\times88$ canonical
differential equation, which is $\varepsilon$-factorised and has the $\dd\log$
form of~\cref{eq:dlog_form}.
As at lower loops, the subleading contribution vanishes, $\mathbf\Omega^{(0)}=\mathbf 0$.

Our construction passes this stress test comfortably. With our proof-of-concept
implementation of~\cref{ssec:sector_sums}, we computed the $246\times246$
connection matrix in a few minutes on a laptop. By contrast, computing the same
matrix using IBP reductions with \texttt{Blade}~\cite{Guan:2024byi} required two
days on a compute node with $1$\,TB of RAM.

%% file: Sections/conclusions.tex
\section{Summary and Outlook}
\label{sec:conclusions}
In this work, we have developed a method for constructing differential equations
for Feynman integrals directly from multivariate intersection numbers, avoiding
both the recursive fibration of the intersection pairings and the generation of
integration-by-parts identities.
Since the connection matrix is rational in $\varepsilon$, it is
fixed by finitely many orders of the large-$\varepsilon$ expansion of the
intersection pairing, whose coefficients are Saito's higher residue pairings~\cite{Saito:1983,Mizera:2019vvs},
localised at the critical points of the potential.
For differential systems at most linear in $\varepsilon$, such as canonical
ones, the leading and first subleading pairings are sufficient to fully
reconstruct the matrices.
For Feynman integrals, these pairings apply only in the presence of analytic
regulators $\delta$ on the coordinate hyperplanes. In this work, we studied the
pairings in the vanishing-regulator limit $\delta\to0$.

Assuming that the regulated critical points are appropriately isolated,
non-degenerate, and analytic in a common regulator $\delta$, we have shown that,
as $\delta\to0$, they organise into sectors labelled by a set of pinched propagators.
Their limits are critical points of the unregulated sector problems, whose
dimensions add up to the number of master integrals before identifying graph
symmetries~\cite{Lee:2013hzt,Bitoun:2017nre}.
The pairings themselves are suppressed as $\delta\to0$ by a power of $\delta$
set by the pinch set of the dual form. Following Ref.~\cite{Fontana:2023amt}, we
compensate for this by rescaling each dual form by the inverse power, which
leaves the connection matrix unaffected.
With these rescaled dual forms, we have derived closed sectorwise formulae for
the leading and first subleading pairings, given
in~\cref{eq:lo_pairing,eq:subleading_pairing_final}.
The pairing between a form and a dual form receives contributions only from the sectors interpolating between their pinch sets.
All the ingredients are unregulated quantities evaluated on the critical points
of the sector ideals. The analytic regulator disappears entirely from the final
formulae. The leading-order intersection matrices are finite and block
triangular with respect to inclusion of pinch sets.

Combined with standard tools of computational commutative algebra, these results provide a practical algorithm.
The sums over the critical points of a sector are evaluated as traces of companion matrices in the quotient ring of the sector ideal. Hence, the critical points, which are generically algebraic, are never determined explicitly.
The construction of the differential system is thus reduced to multivariate polynomial division and linear algebra over the field of kinematic variables, and is amenable to finite-field reconstruction.

As an application, we have derived the canonical differential equations for the
massless planar four-point ladder families up to four loops, finding agreement
with independent integration-by-parts computations.
At four loops, where we constructed a canonical basis sector by sector, the
resulting $88\times88$ canonical system is, to our knowledge, new.
With a proof-of-concept implementation, it was obtained in a few minutes on a
laptop, whereas the corresponding IBP reduction required two days on a high-memory compute node.

Several directions deserve further investigation:
\begin{itemize}
\item \emph{Higher orders in $\varepsilon^{-1}$.}
Extending our analysis beyond the first subleading pairing would give access to differential
systems with a more general dependence on $\varepsilon$.
This requires obtaining the higher-order terms in the $\varepsilon^{-1}$
expansion of the intersection pairing, and studying their behaviour in the
$\delta\to0$ limit.
Moreover, for a given basis, a priori bounds on the $\varepsilon$-degree of the
connection matrix are needed to determine how many orders in $\varepsilon^{-1}$
must be computed.

\item \emph{Non-generic critical points.}
Our analysis assumes that the critical points are analytic in the regulator and
that their limits are isolated and non-degenerate.
These assumptions can fail (see e.g.~\cite{Page:2025gso,Crisanti:2026rbc}), and already for the democratic choice of regulators
the number of critical points can differ from the number of master integrals.
In these cases, the sectorwise organisation of the limit has to be revisited.

\item \emph{Connection to relative twisted cohomology.}
The precise relation between the $\delta \to 0$ limit of the regulated intersection pairing and
the intersection pairing of relative twisted
cohomology~\cite{Caron-Huot:2021iev,Caron-Huot:2021xqj} remains to be established.
A derivation working directly in relative twisted cohomology would provide a
clarifying perspective on our assumptions.

\item \emph{Canonical bases.}
Our construction takes as input a basis for which the differential system is at
most linear in $\varepsilon$.
Our pairing formulae could guide the systematic rotation of a given basis to an
$\varepsilon$-factorised one, removing the need for such a basis as input.

\item \emph{Other representations.}
It would be interesting to apply the approach in other representations such
as the Baikov representation, in which the regulated hyperplanes correspond to
cut rather than pinched propagators.
\end{itemize}
We expect the direct, fibration-free construction developed here to be a useful
tool for more demanding families, with internal masses, non-planar topologies, or non-polylogarithmic solutions.

%% file: Sections/subleading.tex
\section{%
    \texorpdfstring
        {Details of Derivation of the Subleading Pairing as $\delta \to 0$}
        {Details of Derivation of the Subleading Pairing as delta -> 0}
}
\label{app:subleading_derivation}
In this appendix, we supply further details on the derivation of the subleading
pairing of~\cref{ssec:subleading_order_pairing}. Specifically, we discuss the cancellation
of the double poles on the diagonal of the pinched block of $M_{ab}$ and the
derivation of the finite
terms~\labelcref{eq:T_cut_block,eq:T_mixed_block,eq:T_pinch_block}.

To proceed, we need a number of formulae, which we first collect: an
organisation of $M_{ab}$ into regular and pinch pieces, the derivatives of the
pinch factors, and the derivatives of $\log\det\bigbrk{\, \Phi_\delta \,}$ at
$z\brk{\delta}$.
Separating the regular and pinch factors coming from~\cref{eq:phi_g_decomp}, we write
\begin{align}
    \label{eq:Mab_split}
    M_{ab}^{ij} = M_{ab}^{ij}\brk{\psi}
    + M_{ab}^{ij}\brk{\psi, \Delta}
    + \psi_a \psi_b \> M_{ab}^{ij}\brk{\Delta}
    \>,
\end{align}
where the representation~\labelcref{eq:symmetric_Mab} gives
\begin{align}
    \label{eq:Mab_psi}
    M_{ab}^{ij}\brk{\psi}
    &\simeq
    \psi_a \> \partial_i \partial_j \psi_b
    -
    \psi_b \> \partial_i \partial_j \psi_a
    \>,
    \\
    \label{eq:Mab_psi_Delta}
    M_{ab}^{ij}\brk{\psi, \Delta}
    &\simeq
    \psi_a \> \brk{\partial_i \psi_b}
    \>
    \partial_j \log \Bigbrk{\frac{\Delta_b^2}{\det\bigbrk{\, \Phi_\delta \,}}}
    -
    \psi_b \> \brk{\partial_i \psi_a}
    \>
    \partial_j \log \Bigbrk{\frac{\Delta_a^2}{\det\bigbrk{\, \Phi_\delta \,}}}
    \>,
    \\
    \label{eq:Mab_Delta}
    M_{ab}^{ij}\brk{\Delta}
    &\simeq
    \frac{\partial_i \partial_j \Delta_b}{\Delta_b}
    -
    \frac{\partial_i \partial_j \Delta_a}{\Delta_a}
    - \Bigbrk{\partial_j \log\Bigbrk{\frac{\Delta_b}{\Delta_a}}}
    \>
    \partial_i \log \det\bigbrk{\, \Phi_\delta \,}
    \>,
\end{align}
which follows from~\cref{eq:symmetric_Mab} by the product rule and use\footnote{%
Since $A_i B_j - A_j B_i$ is antisymmetric, $A_i B_j \simeq A_j B_i$.}
of the equivalence relation $\simeq$.

Derivatives of the pinch factors are
\begin{equation}
    \label{eq:Delta_derivatives}
    \partial_i \log \Delta_x
    =
    -
    \frac{\delta_{i \Gamma_x}}{z_i}
    \>,
    \quad
    \frac{\partial_i \partial_j \Delta_x}{\Delta_x}
    =
    \frac{1 + \delta_{ij}}{z_i z_j}\delta_{i \Gamma_x} \delta_{j \Gamma_x}
    \>,
    \quad \text{with $x = a, b$}
    \>.
\end{equation}
Here $\delta_{i\Gamma_x}$ equals one for $i \in \Gamma_x$ and zero otherwise, so these derivatives have poles at $z\brk{\delta}$ for pinched indices.
The matrix $M_{ab}$ also involves the derivatives $\partial_k \log
\det\bigbrk{\, \Phi_\delta \,}$, which can have poles as well.
Since the regulator term in~\labelcref{eq:def_regulated_Hessian} is diagonal,
its derivative $\partial_k$ contributes only $2\,\delta/z_k^3$ in the entry
$\brk{k, k}$, and Jacobi's formula gives
\begin{equation}
    \label{eq:jacobi}
    \partial_k \log \det\bigbrk{\,
        \Phi_\delta
    \,}
    =
    \Tr\bigbrk{\, \Phi_\delta^{-1} \cdot \partial_k \Phi \,}
    +
    \frac{2 \, \delta}{z_k^3} \brk{\Phi_\delta^{-1}}_{kk}
    \>.
\end{equation}
For cut directions, the second term in~\cref{eq:jacobi} is suppressed by
$\delta$, and so, employing~\labelcref{eq:invhessian_expanded}, one finds
\begin{equation}
\label{eq:logdet_cut}
    \partial_{\bar{i}} \log \det\bigbrk{\, \Phi_\delta \,}
    \Bigr\rvert_{z = z\brk{\delta}}
    =
    \partial_{\bar{i}} \log \det\bigbrk{\, \Phi_{\overline{\Gamma}} \,}
    \Bigr\rvert_{z = z^\star}
    +
    \cO\brk{\delta}
    \>,
    \quad
    \text{for $\bar{i} \in \overline{\Gamma}$}
    \>.
\end{equation}
For pinched directions, the prefactor $2\,\delta/z_i^3$ of the second term in~\cref{eq:jacobi} scales as $\delta^{-2}$.
Combining it with the inverse Hessian~\labelcref{eq:invhessian_expanded} and the critical point equations~\labelcref{eq:critical_points_componentwise}, which give $\delta/z_i = -\omega_i$ at $z\brk{\delta}$, gives
\begin{equation}
    \label{eq:logdet_pinch}
    \partial_{i} \log \det\bigbrk{\, \Phi_\delta \,}
    \bigg\rvert_{z = z\brk{\delta}}
    =
    -\frac{2}{z_i\brk{\delta}}
    +
    \biggl(\,
    \frac{2 \, S_{\Gamma ,\, ii}}{\omega_{i}}
    +
    \partial_{i} \log \det\bigbrk{\, \Phi_{\overline{\Gamma}} \,}
    \,\biggr)\bigg\rvert_{z = z^\star}
    +
    \cO\brk{\delta}
    \>,
    \quad
    \text{for $i \in \Gamma$}
    \>.
\end{equation}

\paragraph{Double-pole cancellation}
At $z\brk{\delta}$, the regular factors and their derivatives are finite, and so are the cut derivatives $\partial_{\bar{i}}$ of the pinch factors and of $\log\det\bigbrk{\, \Phi_\delta \,}$.
Poles in $\delta$ arise only from pinched derivatives $\partial_i$, $i \in
\Gamma$, each of which produces at most a factor $1/z_i = \cO\brk{\delta^{-1}}$.
In~\cref{eq:Mab_split}, the piece $M_{ab}\brk{\psi}$ is therefore finite,
$M_{ab}\brk{\psi, \Delta}$ has at most simple poles, and only
$M_{ab}\brk{\Delta}$ can have double poles, in its pinched block.
In the power counting of~\cref{ssec:subleading_order_pairing}, the double poles of $M_{ab}\brk{\Delta}$ are harmless off the diagonal, where the pinched block of $\Phi_\delta^{-1}$ is of order $\delta^2$, but on the diagonal they meet a single power of $\delta$. There, they cancel. They come from $\psi_a \psi_b \, M_{ab}\brk{\Delta}$ alone, and by~\cref{eq:Delta_derivatives,eq:logdet_pinch} the coefficient of $\psi_a \psi_b / z_i^2$ in $M_{ab}^{ii}$ is
\begin{equation}
    \label{eq:double_pole_cancel}
    2 \, \delta_{i \Gamma_b}
    -
    2 \, \delta_{i \Gamma_a}
    -
    2 \, \bigbrk{\delta_{i \Gamma_b} - \delta_{i \Gamma_a}}
    =
    0
    \>,
\end{equation}
where the three terms come from the three terms of~\cref{eq:Mab_Delta}.
The remaining terms of $M_{ab}^{ii}$ have at most a simple pole, and their product with $\brk{\Phi_\delta^{-1}}_{ii} = \cO\brk{\delta}$ is finite.

\paragraph{Finite terms}
To derive the finite terms~\labelcref{eq:T_cut_block,eq:T_mixed_block,eq:T_pinch_block}, we now consider a critical point of a sector $\Gamma$ in the range~\labelcref{eq:contributing_sectors_hierarchical}.
Since $\Gamma_a \subseteq \Gamma \subseteq \Gamma_b$, every pinched index belongs to $\Gamma_b$ and no cut index belongs to $\Gamma_a$.
In particular, $\partial_{\bar{i}} \Delta_a = 0$, and for $i \in \Gamma$ the pole $-2/z_i$ of $2 \, \partial_i \log \Delta_b$ cancels the pole of $-\partial_i \log\det\bigbrk{\, \Phi_\delta \,}$ in~\cref{eq:logdet_pinch}.
Combining~\cref{eq:logdet_pinch} and~\cref{eq:Delta_derivatives}, we thus get
\begin{equation}
    \label{eq:DelbBlockRel}
    \partial_i \log \Bigbrk{\frac{\Delta_b^2}{\det\bigbrk{\, \Phi_\delta \,}}}
    \bigg\rvert_{z = z\brk{\delta}}
    =
    \biggl(\,
    - \frac{2 \, S_{\Gamma ,\, ii}}{\omega_i}
    - \partial_i \log \det\bigbrk{\, \Phi_{\overline{\Gamma}} \,}
    \,\biggr)\bigg\rvert_{z = z^\star}
    + \cO\brk{\delta}
    \>,
    \quad
    \text{for $i \in \Gamma$}
    \>,
\end{equation}
and the two ratios obey the exact identity
\begin{equation}
    \partial_i \log \Bigbrk{\frac{\Delta_a^2}{\det\bigbrk{\, \Phi_\delta \,}}}
    =
    - 2 \> \partial_i \log \Bigbrk{\frac{\Delta_b}{\Delta_a}}
    + \partial_i \log \Bigbrk{\frac{\Delta_b^2}{\det\bigbrk{\, \Phi_\delta \,}}}
    \>,
    \label{eq:DelaBlockRel}
\end{equation}
which isolates the possible pole in the second term
of~\cref{eq:Mab_psi_Delta}.
By~\cref{eq:Delta_derivatives}, the first term on the right-hand side
of~\cref{eq:DelaBlockRel} is $2 \, \delta_{i \Gamma_b \setminus \Gamma_a}
/ z_i$ for $i \in \Gamma$, while the second is finite
by~\cref{eq:DelbBlockRel}.
This is the simple pole that enters the mixed and pinched blocks below.
Each part of $T_{\delta, ab}$ in~\cref{eq:TraceDecomp} contracts one block of $M_{ab}$ with the corresponding block of the inverse Hessian, whose order in $\delta$ we read off from~\cref{eq:invhessian_expanded}.
We treat the three blocks in turn.

We begin with the cut block, $\bar{i}, \bar{j} \in \overline{\Gamma}$.
The cut block of $\Phi_\delta^{-1}$ is finite, so we need the cut block of $M_{ab}$ through order $\delta^0$.
The only simplification arises from $\partial_{\bar{i}}\Delta_a = 0$.
Inserting~\cref{eq:Delta_derivatives,eq:logdet_cut} and using $\simeq$ to move the indicators $\delta_{\bar{j}\Gamma_b}$ to the index $\bar{i}$, the three pieces of~\cref{eq:Mab_split} become
\begin{equation}
    \label{eq:cut_pieces}
\begin{aligned}
    M_{ab}^{\bar{i}\bar{j}}\brk{\psi}
    \bigg\rvert_{z = z\brk{\delta}}
    &=
    \biggbrk{\,
        \psi_a \> \partial_{\bar{i}} \partial_{\bar{j}} \psi_b
        - \psi_b \> \partial_{\bar{i}} \partial_{\bar{j}} \psi_a
    \,}\bigg\rvert_{z = z^\star}
    + \cO\brk{\delta}
    \>,
    \\
    M_{ab}^{\bar{i}\bar{j}}\brk{\psi, \Delta}
    \bigg\rvert_{z = z\brk{\delta}}
    &\simeq
    \biggl(\,
        - \brk{
            \psi_a \> \partial_{\bar{i}} \psi_b
            - \psi_b \> \partial_{\bar{i}} \psi_a
        }
        \> \partial_{\bar{j}} \log\det\bigbrk{\, \Phi_{\overline{\Gamma}} \,}
    - \frac{2 \, \delta_{\bar{i}\Gamma_b}}{z_{\bar{i}}} \> \psi_a \> \partial_{\bar{j}} \psi_b
    \,\biggr)\bigg\rvert_{z = z^\star}
    + \cO\brk{\delta}
    \>,
    \\
    \psi_a \psi_b \> M_{ab}^{\bar{i}\bar{j}}\brk{\Delta}
    \bigg\rvert_{z = z\brk{\delta}}
    &\simeq
        \biggbrk{\,
            \psi_a \psi_b \>
            \frac{1 + \delta_{\bar{i}\bar{j}}}{z_{\bar{i}} z_{\bar{j}}} \>
            \delta_{\bar{i}\Gamma_b} \delta_{\bar{j}\Gamma_b}
            + \frac{\delta_{\bar{i}\Gamma_b}}{z_{\bar{i}}} \> \psi_a \psi_b \> \partial_{\bar{j}} \log\det\bigbrk{\, \Phi_{\overline{\Gamma}} \,}
        \,}\bigg\rvert_{z = z^\star}
    + \cO\brk{\delta}
    \>.
\end{aligned}
\end{equation}
The terms without derivatives of the pinch factors make up the matrix $\midtilde{M}_{ab}$ of~\cref{eq:def_Mtilde}, and altogether
\begin{align}
    \label{eq:Mab_cut_block}
    M_{ab}^{\bar{i}\bar{j}}
    \bigg\rvert_{z = z\brk{\delta}}
    &\simeq
        \biggl(\,
        \midtilde{M}_{ab}^{\bar{i}\bar{j}}
        +
        \psi_a \psi_b \>
        \frac{1 + \delta_{\bar{i}\bar{j}}}{z_{\bar{i}}z_{\bar{j}}} \>
        \delta_{\bar{i}\Gamma_b} \delta_{\bar{j}\Gamma_b}
    \\
    &\quad-
        \frac{\delta_{\bar{i}\Gamma_b}\psi_a}{z_{\bar{i}}}
        \Bigbrk{
            2 \> \partial_{\bar{j}}\psi_b
            -
            \psi_b \> \partial_{\bar{j}}\log\det\bigbrk{\, \Phi_{\overline{\Gamma}} \,}
        }
        \,\biggr)\bigg\rvert_{z = z^\star}
        + \cO\brk{\delta}
        \nonumber
        \>,
        \qquad \text{for $\bar{i} ,\, \bar{j} \in \overline{\Gamma}$}
        \>\text{.}
\end{align}
Contracted with $\Phi_{\overline{\Gamma}}^{-1}$ in the cut part~\labelcref{eq:T_cut_def}, this gives $T_{ab}^{\overline{\Gamma}}$ in~\cref{eq:T_cut_block}.

For the mixed blocks, $i \in \Gamma$ and $\bar{j} \in \overline{\Gamma}$, the mixed blocks of the inverse Hessian are of order $\delta$, so we need the transposed sum of the mixed blocks of $M_{ab}$ through its simple pole.
Simple poles can only arise from the pinch factors, and $M_{ab}\brk{\psi}$ can be dropped. On the other hand,
\begin{equation}
    M_{ab}^{i\bar{j}}\brk{\Delta} + M_{ab}^{\bar{j}i}\brk{\Delta}
    \simeq
    2 \, \frac{\partial_i \partial_{\bar{j}} \Delta_b}{\Delta_b}
    - \partial_i \log\Bigbrk{\frac{\Delta_b}{\Delta_a}}
    \> \partial_{\bar{j}} \log \det\bigbrk{\, \Phi_\delta \,}
    - \partial_i \log \det\bigbrk{\, \Phi_\delta \,}
    \> \partial_{\bar{j}} \log \Delta_b
    \>.
\end{equation}
Here, the first and last terms cancel each other to leading order in $\delta$, leaving only the middle.
The other contribution comes from~\cref{eq:Mab_psi_Delta}.
There, a pole can only arise from the second term in the entry $\brk{\bar{j},
i}$, where the derivative is pinched. Altogether one finds
\begin{align}
    \Bigl(M_{ab}^{i\bar{j}} + M_{ab}^{\bar{j}i} \Bigr)\bigg\rvert_{z = z\brk{\delta}}
    &\simeq
    \biggl(\,
    - \psi_a \psi_b \> \partial_i \log\Bigbrk{\frac{\Delta_b}{\Delta_a}}
    \> \partial_{\bar{j}} \log \det\bigbrk{\, \Phi_\delta \,}
    \nonumber
    \\
    &\qquad
    - \partial_i \log\Bigbrk{\frac{\Delta_a^2}{\det\bigbrk{\, \Phi_\delta \,}}}
    \> \psi_b \> \partial_{\bar{j}} \psi_a
    \,\biggr)\bigg\rvert_{z = z\brk{\delta}}
    + \cO\brk{1}
    \nonumber
    \\
    \label{eq:Mab_mixed_block}
    &\simeq
    -\frac{\delta_{i \Gamma_b \setminus \Gamma_a}}{z_i\brk{\delta}}
    \biggl(\,
    \psi_b
    \Bigbrk{
        2 \> \partial_{\bar{j}} \psi_a
        -
        \psi_a \> \partial_{\bar{j}} \log\det\bigbrk{\, \Phi_{\overline{\Gamma}} \,}
    }
    \,\biggr)\bigg\rvert_{z = z^\star}
    + \cO\brk{1}
    \>,
\end{align}
where we apply~\cref{eq:DelaBlockRel} and~\cref{eq:logdet_cut}.
In the mixed part~\labelcref{eq:T_mixed_def}, this simple pole meets the
mixed block of the inverse Hessian in~\cref{eq:invhessian_expanded}, and
with $\delta/z_i = -\omega_i$ it gives $T_{ab}^{\Gamma\overline{\Gamma}}$
in~\cref{eq:T_mixed_block}.

Finally, the pinched block of the inverse Hessian is of order $\delta$ on the diagonal and $\delta^2$ off it, so for $i, j \in \Gamma$ we need $M_{ab}^{ij}$ through its double and simple poles.
By application of~\cref{eq:DelbBlockRel}, we see that only the second piece of~\cref{eq:Mab_psi_Delta} contributes, and that we can write
\begin{equation}
\begin{aligned}
    M_{ab}^{ij}
    \bigg\rvert_{z = z\brk{\delta}}
    &\simeq
    \biggl(\,
    \psi_a \psi_b \biggl(
    \frac{\partial_i \partial_j \Delta_b}{\Delta_b}
    - \frac{\partial_i \partial_j \Delta_a}{\Delta_a}
    - \partial_i \log \det\bigbrk{\, \Phi_\delta \,}
    \> \partial_j \log\Bigbrk{\frac{\Delta_b}{\Delta_a}}
    \biggr)
    \\
    &\quad
    - \psi_b \> \partial_i \psi_a
    \> \partial_j \log\Bigbrk{\frac{\Delta_a^2}{\det\bigbrk{\, \Phi_\delta \,}}}
    \,\biggr)\bigg\rvert_{z = z\brk{\delta}}
    + \cO\brk{1}
    \>.
\end{aligned}
\end{equation}
Here, the first two terms contribute pure double poles, the third both double and simple poles, and the fourth only a simple pole.
Inserting expressions in terms of Kronecker $\delta$ and the expansion of the
derivative of $\log \det\bigbrk{\, \Phi_\delta \,}$, symmetrising the third term
and moving the simple poles to the index $i$ with $\simeq$, one finds
\begin{align}
    \label{eq:Mab_pinch_block}
    M_{ab}^{ij}
    \bigg\rvert_{z = z\brk{\delta}}
    &\simeq
    -
    \biggl(\,
    \frac{\psi_a \psi_b \> \brk{1 - \delta_{ij}}}{z_i z_j}
    \delta_{i \Gamma_b \setminus \Gamma_a} \delta_{j \Gamma_b \setminus \Gamma_a}
    \,\biggr)\bigg\rvert_{z = z\brk{\delta}}
    \\
    &\quad +
    \frac{\delta_{i \Gamma_b \setminus \Gamma_a}}{z_i\brk{\delta}}
    \biggl(\,
    \psi_b
    \biggbrk{
        \psi_a \Bigbrk{
            \frac{2 \, S_{\Gamma ,\, jj}}{\omega_j}
            + \partial_j \log \det \bigbrk{\, \Phi_{\overline{\Gamma}} \,}
        }
        -2 \> \partial_j\psi_a
    }
    \,\biggr)\bigg\rvert_{z = z^\star}
    +\cO\brk{1}
    \>.
    \nonumber
\end{align}
The double pole has zero diagonal, in agreement with~\cref{eq:double_pole_cancel}.
In the pinched part~\labelcref{eq:T_pinch_def}, the simple pole on the diagonal meets the factor $-\delta \, D_\Gamma^{-1}$ and gives the terms $i = j$ of the first sum in~\cref{eq:T_pinch_block}, together with its second sum.
Off the diagonal, the double pole meets the factor $-\delta^2 \, D_\Gamma^{-1} \cdot S_\Gamma \cdot D_\Gamma^{-1}$, and $\delta^2/\brk{z_i z_j} = \omega_i \omega_j$ gives the terms $i \neq j$ of the first sum.

All other contributions to the three parts are of order $\delta$.
Every remaining quantity is continuous at $z^\star$, so the limit $\delta \to 0$ evaluates~\cref{eq:T_cut_block,eq:T_mixed_block,eq:T_pinch_block} at $z^\star$.
Together with the limit of the residue measure, this gives the pairing formula~\labelcref{eq:subleading_pairing_final}.

%% file: Sections/ladder_definitions.tex
\section{%
    \texorpdfstring
        {Definitions of the Ladder Integral Families}
        {Definitions of the Ladder Integral Families}
}
\label{app:ladder_definitions}
In this appendix, we collect the definitions of the two-, three- and four-loop ladder integral families studied in~\cref{sec:applications}: their propagators and irreducible scalar products, their regulated Lee--Pomeransky representations, and the normalisation of the corresponding differential forms.
The kinematics and the variable $y$ are those of~\cref{eq:mandelstam,eq:y_variable}.

\subsection{Two-Loop Ladder}
\label{app:two_loop_ladder}
The two-loop ladder integral family shown in~\cref{fig:ladders} is defined by
\begin{equation}
    I_{\alpha_1 \alpha_2 \alpha_3 \alpha_4 \alpha_5 \alpha_6 \alpha_7 \alpha_8 \alpha_9} = \int \left( \prod_{i=1}^2\frac{\dd^D \ell_i}{\im \pi^{D/2}}\right) \frac{D_8^{-\alpha_8}D_9^{-\alpha_9}}{D_1^{\alpha_1}D_2^{\alpha_2}D_3^{\alpha_3}D_4^{\alpha_4}D_5^{\alpha_5}D_6^{\alpha_6}D_7^{\alpha_7}}\, ,
\end{equation}
where the denominators are defined as
\begin{equation}
\begin{alignedat}{4}
D_1 &= -\ell_1^2, \qquad &
D_2 &= -\ell_2^2, \qquad &
D_3 &= -(\ell_1+p_1)^2, \qquad &
D_4 &= -(\ell_1-\ell_2)^2, \\[2pt]
D_5 &= -(\ell_1+p_1+p_2)^2, \qquad &
D_6 &= -(\ell_2+p_1+p_2)^2, \qquad &
D_7 &= -(\ell_2-p_3)^2, \qquad &
D_8 &= -(\ell_1-p_3)^2, \\[2pt]
D_9 &= -(\ell_2+p_1)^2.
\end{alignedat}
\end{equation}
For $\alpha_8=\alpha_9=0$, we abbreviate $I_{\alpha_1\ldots \alpha_7 0 0}$ as $I_{\alpha_1\ldots \alpha_7}$,
with the same convention for the regulated integrals.
The regulated Lee--Pomeransky representation~\labelcref{eq:leepom_reg} then reads
\begin{equation}
    I_{\alpha_1 \alpha_2 \alpha_3 \alpha_4 \alpha_5 \alpha_6 \alpha_7}^{\brk{\delta}} = \frac{\Gamma\brk{2-\varepsilon}}{\Gamma\brk{6-3\varepsilon-\vert\boldsymbol{\alpha}\vert-7\varepsilon\delta}} \int \cG^{\varepsilon-2}\prod_{i=1}^7\frac{\dd z_i \, z_i^{\alpha_i-1+\varepsilon\delta}}{\Gamma\brk{\alpha_i+\varepsilon\delta}} \, .
\end{equation}
We rewrite this family as a twisted period integral family, normalising the forms as in
\cref{eq:rational_normalisation}, with $L=2$, $c=1$ and $p=4$:
\begin{equation}
\frac{\Gamma\brk{1-3\varepsilon}}{\varepsilon^4\,\Gamma\brk{2-\varepsilon}}
I_{\alpha_1 \alpha_2 \alpha_3 \alpha_4 \alpha_5 \alpha_6 \alpha_7}^{\brk{\delta}}
= \int_{\mathbb{R}_+^7}e^{\varepsilon W_\delta}\,
\varphi_{\alpha_1 \alpha_2 \alpha_3 \alpha_4 \alpha_5 \alpha_6 \alpha_7} \, ,
\qquad
W_\delta=\log\brk{\cG}+\delta\sum_{i=1}^7\log\brk{z_i}
\, ,
\end{equation}
where
\begin{align}
\label{eq:DoubleBoxFormDefinition}
\varphi_{\alpha_1 \alpha_2 \alpha_3 \alpha_4 \alpha_5 \alpha_6 \alpha_7}
&=
\frac{\Gamma\brk{1-3\varepsilon}}
{\varepsilon^4\,
\Gamma\brk{6-\sum_{i=1}^7 \alpha_i-7\varepsilon\delta-3\varepsilon}}
\frac{1}{\cG^2}
\prod_{i=1}^7
\frac{\dd z_i\,z_i^{\alpha_i-1}}{\Gamma\brk{\alpha_i+\varepsilon\delta}}
\, .
\end{align}
As explained in \cref{ssec:lee_pomeransky_representation}, the ISPs $D_8$ and $D_9$ are represented by a rational insertion factor obtained by taking derivatives of the augmented twist $(\cG^{\mathrm{aug}})^{-D/2}$ with respect to unregulated auxiliary
parameters and then setting these parameters to zero. 
The explicit form of this factor is given in~\cref{eq:isp_numerator_factor}.
Differential forms are written as
\begin{align}
\label{eq:DoubleBoxISPFormDefinition}
\varphi^{A,\boldsymbol{\beta}}_{\alpha_1 \alpha_2 \alpha_3 \alpha_4 \alpha_5 \alpha_6 \alpha_7}
&=
\cR_A^{\boldsymbol{\beta}}\,
\frac{\Gamma\brk{1-3\varepsilon}}
{\varepsilon^4\,
\Gamma\brk{6-\sum_{i=1}^7 \alpha_i+\vert\boldsymbol{\beta}\vert-7\varepsilon\delta-3\varepsilon}}
\frac{1}{\cG^2}
\prod_{i=1}^7
\frac{\dd z_i\,z_i^{\alpha_i-1}}{\Gamma\brk{\alpha_i+\varepsilon\delta}}
\, ,
\end{align}
where  $A\subseteq\set{8,9}$, $\boldsymbol{\beta}=(\beta_j)_{j\in A}$, and $\vert\boldsymbol{\beta}\vert=\sum_{j\in A}\beta_j$.
For a single power of $D_8$ the rational insertion factor is
$\cR_{\set{8}}^{(1)}=(2-\varepsilon)(\partial_8\cG^{\mathrm{aug}})\Bigr\rvert_{z_8=0}/\cG$.

\subsection{Three-Loop Ladder}
\label{app:three_loop_ladder}
The three-loop ladder integral family is depicted in~\cref{fig:ladders} and is given by
\begin{equation}
    I_{\alpha_1 \ldots \alpha_{15}} = \int \left( \prod_{i=1}^3\frac{\dd^D \ell_i}{\im \pi^{D/2}}\right) \frac{\prod_{i=11}^{15}D_i^{-\alpha_i}}{\prod_{i=1}^{10}D_i^{\alpha_i}}\, ,
\end{equation}
where the denominators are defined as
\begin{align}
    \label{eq:dens_l3l}
\begin{alignedat}{4}
D_1    &= -\ell_1^2 \,,\, &
D_2    &= -\ell_2^2 \,,\, &
D_3    &= -\ell_3^2 \,,\, &
D_4    &= -(\ell_1+p_1)^2 \,, \\
D_5    &= -(\ell_1-\ell_2)^2 \,,\, &
D_6    &= -(\ell_2-\ell_3)^2 \,,\, &
D_7    &= -(\ell_1+p_1+p_2)^2 \,, &
D_8    &= -(\ell_2+p_1+p_2)^2 \,,\, \\
D_9    &= -(\ell_3+p_1+p_2)^2 \,,\, &
D_{10} &= -(\ell_3-p_3)^2 \,,\, &
D_{11} &= -(\ell_1-p_3)^2 \,,\, &
D_{12} &= -(\ell_2-p_3)^2 \,, \\
D_{13} &= -(\ell_2+p_1)^2 \,,\, &
D_{14} &= -(\ell_3+p_1)^2 \,,\, &
D_{15} &= -(\ell_1-\ell_3)^2 \,.
\end{alignedat}
\end{align}
The family has ten propagators, $D_1,\ldots,D_{10}$, 
while $D_{11},\ldots,D_{15}$ are irreducible
scalar products.
With the five ISP indices set to zero and omitted, the regulated Lee--Pomeransky
representation~\labelcref{eq:leepom_reg}, with $L=3$ and $n=10$, reads
\begin{equation}
\label{eq:threeloop_LP}
    I_{\alpha_1 \ldots \alpha_{10}}^{\brk{\delta}}
    =
    \frac{\Gamma\brk{2-\varepsilon}}{\Gamma\brk{8-4\varepsilon-\vert\boldsymbol{\alpha}\vert-10\varepsilon\delta}}
    \int \cG^{\varepsilon-2}\prod_{i=1}^{10}\frac{\dd z_i \, z_i^{\alpha_i-1+\varepsilon\delta}}{\Gamma\brk{\alpha_i+\varepsilon\delta}}
    \, ,
\end{equation}
where $\cG$ is the Lee--Pomeransky polynomial of the three-loop ladder, in the ten Feynman
parameters $z_1,\ldots,z_{10}$.
We rewrite this family as a twisted period integral family, normalising the forms as
in~\cref{eq:rational_normalisation}, with $L=3$, $c=2$ and $p=7$:
\begin{equation}
\label{eq:threeloop_tpi}
\frac{\Gamma\brk{2-4\varepsilon}}{\varepsilon^7\,\Gamma\brk{2-\varepsilon}}
\, I_{\alpha_1 \ldots \alpha_{10}}^{\brk{\delta}}
= \int_{\mathbb{R}_+^{10}}e^{\varepsilon W_\delta}\,
\varphi_{\alpha_1 \ldots \alpha_{10}} \, ,
\qquad
W_\delta=\log\brk{\cG}+\delta\sum_{i=1}^{10}\log\brk{z_i}
\, ,
\end{equation}
where
\begin{align}
\label{eq:ThreeLoopFormDefinition}
\varphi_{\alpha_1 \ldots \alpha_{10}}
&=
\frac{\Gamma\brk{2-4\varepsilon}}
{\varepsilon^7\,
\Gamma\brk{8-\sum_{i=1}^{10} \alpha_i-10\varepsilon\delta-4\varepsilon}}
\frac{1}{\cG^2}
\prod_{i=1}^{10}
\frac{\dd z_i\,z_i^{\alpha_i-1}}{\Gamma\brk{\alpha_i+\varepsilon\delta}}
\, .
\end{align}
Auxiliary ISP insertions are included through~\cref{eq:isp_numerator_factor}.
Negative indices among the ten propagators are reduced to simple poles, as
in~\cref{eq:form_pole_structure}.

\subsection{Four-Loop Ladder}
\label{app:four_loop_ladder}
The four-loop ladder integral family shown in~\cref{fig:ladders} is defined by
\begin{equation}
    I_{\alpha_1 \ldots \alpha_{22}} = \int \left( \prod_{i=1}^4 \frac{\dd^D \ell_i}{\im \pi^{D/2}}\right) \frac{\prod_{i=14}^{22}D_i^{-\alpha_i}}{\prod_{i=1}^{13}D_i^{\alpha_i}}\, ,
\end{equation}
where the denominators are defined as
\begin{align}
    \label{eq:dens_l4l}
\begin{alignedat}{4}
D_1    &= -\ell_1^2 \,,\, &
D_2    &= -\ell_2^2 \,,\, &
D_3    &= -\ell_3^2 \,,\, &
D_4    &= -\ell_4^2 \,, \\
D_5    &= -(\ell_1+p_1)^2 \,,\, &
D_6    &= -(\ell_1-\ell_2)^2 \,,\, &
D_7    &= -(\ell_2-\ell_3)^2 \,,\, &
D_8    &= -(\ell_4-\ell_3)^2 \,, \\
D_9    &= -(\ell_1+p_1+p_2)^2 \,,\, &
D_{10} &= -(\ell_2+p_1+p_2)^2 \,,\, &
D_{11} &= -(\ell_3+p_1+p_2)^2 \,,\, &
D_{12} &= -(\ell_4+p_1+p_2)^2 \,, \\
D_{13} &= -(\ell_4-p_3)^2 \,,\, &
D_{14} &= -(\ell_1-p_3)^2 \,,\, &
D_{15} &= -(\ell_2-p_3)^2 \,,\, &
D_{16} &= -(\ell_3-p_3)^2 \,, \\
D_{17} &= -(\ell_2+p_1)^2 \,,\, &
D_{18} &= -(\ell_3+p_1)^2 \,,\, &
D_{19} &= -(\ell_4+p_1)^2 \,,\, &
D_{20} &= -(\ell_1-\ell_3)^2 \,, \\
D_{21} &= -(\ell_1-\ell_4)^2 \,,\, &
D_{22} &= -(\ell_2-\ell_4)^2 \,.
\end{alignedat}
\end{align}

The family has thirteen propagators, $D_1,\ldots,D_{13}$, while $D_{14},\ldots,D_{22}$ are
irreducible scalar products.
With the nine ISP indices set to zero and omitted, the regulated Lee--Pomeransky
representation~\labelcref{eq:leepom_reg}, with $L=4$ and $n=13$, reads
\begin{equation}
\label{eq:fourloop_LP}
    I_{\alpha_1 \ldots \alpha_{13}}^{\brk{\delta}}
    =
    \frac{\Gamma\brk{2-\varepsilon}}{\Gamma\brk{10-5\varepsilon-\vert\boldsymbol{\alpha}\vert-13\varepsilon\delta}}
    \int \cG^{\varepsilon-2}\prod_{i=1}^{13}\frac{\dd z_i \, z_i^{\alpha_i-1+\varepsilon\delta}}{\Gamma\brk{\alpha_i+\varepsilon\delta}}
    \, ,
\end{equation}
where $\cG$ is the Lee--Pomeransky polynomial of the four-loop ladder, in the thirteen Feynman
parameters $z_1,\ldots,z_{13}$.
We rewrite this family as a twisted period integral family, normalising the forms as
in~\cref{eq:rational_normalisation}, with $L=4$, $c=3$ and $p=10$:
\begin{equation}
\label{eq:fourloop_tpi}
\frac{\Gamma\brk{3-5\varepsilon}}{\varepsilon^{10}\,\Gamma\brk{2-\varepsilon}}
\, I_{\alpha_1 \ldots \alpha_{13}}^{\brk{\delta}}
= \int_{\mathbb{R}_+^{13}}e^{\varepsilon W_\delta}\,
\varphi_{\alpha_1 \ldots \alpha_{13}} \, ,
\qquad
W_\delta=\log\brk{\cG}+\delta\sum_{i=1}^{13}\log\brk{z_i}
\, ,
\end{equation}
where
\begin{align}
\label{eq:FourLoopFormDefinition}
\varphi_{\alpha_1 \ldots \alpha_{13}}
&=
\frac{\Gamma\brk{3-5\varepsilon}}
{\varepsilon^{10}\,
\Gamma\brk{10-\sum_{i=1}^{13} \alpha_i-13\varepsilon\delta-5\varepsilon}}
\frac{1}{\cG^2}
\prod_{i=1}^{13}
\frac{\dd z_i\,z_i^{\alpha_i-1}}{\Gamma\brk{\alpha_i+\varepsilon\delta}}
\, .
\end{align}
Auxiliary ISP insertions are treated with~\cref{eq:isp_numerator_factor},
and negative propagator indices are reduced to simple coordinate poles as in \cref{eq:form_pole_structure}.

%% file: main.bbl
\providecommand{\href}[2]{#2}\begingroup\raggedright\begin{thebibliography}{10}

\bibitem{Kotikov:1990kg}
A.V.~Kotikov, \emph{{Differential equations method: New technique for massive
  Feynman diagrams calculation}},
  \href{https://doi.org/10.1016/0370-2693(91)90413-K}{\emph{Phys. Lett. B}
  {\bfseries 254} (1991) 158}.

\bibitem{Kotikov:1991pm}
A.V.~Kotikov, \emph{{Differential equation method: The Calculation of $n$-point
  Feynman diagrams}},
  \href{https://doi.org/10.1016/0370-2693(91)90536-Y}{\emph{Phys. Lett. B}
  {\bfseries 267} (1991) 123}.

\bibitem{Remiddi:1997ny}
E.~Remiddi, \emph{{Differential equations for Feynman graph amplitudes}},
  \href{https://doi.org/10.1007/BF03185566}{\emph{Nuovo Cim. A} {\bfseries 110}
  (1997) 1435} [\href{https://arxiv.org/abs/hep-th/9711188}{{\ttfamily
  hep-th/9711188}}].

\bibitem{Gehrmann:1999as}
T.~Gehrmann and E.~Remiddi, \emph{{Differential equations for two-loop
  four-point functions}},
  \href{https://doi.org/10.1016/S0550-3213(00)00223-6}{\emph{Nucl. Phys. B}
  {\bfseries 580} (2000) 485}
  [\href{https://arxiv.org/abs/hep-ph/9912329}{{\ttfamily hep-ph/9912329}}].

\bibitem{Argeri:2007up}
M.~Argeri and P.~Mastrolia, \emph{{Feynman Diagrams and Differential
  Equations}}, \href{https://doi.org/10.1142/S0217751X07037147}{\emph{Int. J.
  Mod. Phys. A} {\bfseries 22} (2007) 4375}
  [\href{https://arxiv.org/abs/0707.4037}{{\ttfamily 0707.4037}}].

\bibitem{Tkachov:1981wb}
F.V.~Tkachov, \emph{{A Theorem on Analytical Calculability of Four Loop
  Renormalization Group Functions}},
  \href{https://doi.org/10.1016/0370-2693(81)90288-4}{\emph{Phys. Lett. B}
  {\bfseries 100} (1981) 65}.

\bibitem{Chetyrkin:1981qh}
K.G.~Chetyrkin and F.V.~Tkachov, \emph{{Integration by Parts: The Algorithm to
  Calculate beta Functions in 4 Loops}},
  \href{https://doi.org/10.1016/0550-3213(81)90199-1}{\emph{Nucl. Phys.}
  {\bfseries B192} (1981) 159}.

\bibitem{Laporta:2000dsw}
S.~Laporta, \emph{{High-precision calculation of multiloop Feynman integrals by
  difference equations}},
  \href{https://doi.org/10.1142/S0217751X00002159}{\emph{Int. J. Mod. Phys. A}
  {\bfseries 15} (2000) 5087}
  [\href{https://arxiv.org/abs/hep-ph/0102033}{{\ttfamily hep-ph/0102033}}].

\bibitem{Lee:2013mka}
R.N.~Lee, \emph{{LiteRed 1.4: a powerful tool for reduction of multiloop
  integrals}}, \href{https://doi.org/10.1088/1742-6596/523/1/012059}{\emph{J.
  Phys. Conf. Ser.} {\bfseries 523} (2014) 012059}
  [\href{https://arxiv.org/abs/1310.1145}{{\ttfamily 1310.1145}}].

\bibitem{Peraro:2019svx}
T.~Peraro, \emph{{$\text{FiniteFlow}$: multivariate functional reconstruction
  using finite fields and dataflow graphs}},
  \href{https://doi.org/10.1007/JHEP07(2019)031}{\emph{JHEP} {\bfseries 07}
  (2019) 031} [\href{https://arxiv.org/abs/1905.08019}{{\ttfamily
  1905.08019}}].

\bibitem{Guan:2024byi}
X.~Guan, X.~Liu, Y.-Q.~Ma and W.-H.~Wu, \emph{{Blade: A package for
  block-triangular form improved Feynman integrals decomposition}},
  \href{https://doi.org/10.1016/j.cpc.2025.109538}{\emph{Comput. Phys. Commun.}
  {\bfseries 310} (2025) 109538}
  [\href{https://arxiv.org/abs/2405.14621}{{\ttfamily 2405.14621}}].

\bibitem{Lange:2025fba}
F.~Lange, J.~Usovitsch and Z.~Wu, \emph{{Kira 3: integral reduction with
  efficient seeding and optimized equation selection}},
  \href{https://doi.org/10.1016/j.cpc.2025.109999}{\emph{Comput. Phys. Commun.}
  {\bfseries 322} (2026) 109999}
  [\href{https://arxiv.org/abs/2505.20197}{{\ttfamily 2505.20197}}].

\bibitem{Smirnov:2019qkx}
A.V.~Smirnov and F.S.~Chukharev, \emph{{FIRE6: Feynman Integral REduction with
  modular arithmetic}},
  \href{https://doi.org/10.1016/j.cpc.2019.106877}{\emph{Comput. Phys. Commun.}
  {\bfseries 247} (2020) 106877}
  [\href{https://arxiv.org/abs/1901.07808}{{\ttfamily 1901.07808}}].

\bibitem{matsumoto1994}
K.~Matsumoto, \emph{{Quadratic Identities for Hypergeometric Series of Type
  $(k,l)$}}, \href{https://doi.org/10.2206/kyushujm.48.335}{\emph{Kyushu
  Journal of Mathematics} {\bfseries 48} (1994) 335}.

\bibitem{matsumoto1998}
K.~Matsumoto, \emph{Intersection numbers for logarithmic $k$-forms},
  {\emph{Osaka J. Math.} {\bfseries 35} (1998) 873}.

\bibitem{OST2003}
K.~Ohara, Y.~Sugiki and N.~Takayama, \emph{{Quadratic Relations for Generalized
  Hypergeometric Functions $_p F_{p-1}$}},
  \href{https://doi.org/10.1619/fesi.46.213}{\emph{Funkcialaj Ekvacioj}
  {\bfseries 46} (2003) 213}.

\bibitem{doi:10.1142/S0129167X13500948}
Y.~Goto, \emph{{Twisted Cycles and Twisted Period Relations for Lauricella's
  Hypergeometric Function $F_C$}},
  \href{https://doi.org/10.1142/S0129167X13500948}{\emph{International Journal
  of Mathematics} {\bfseries 24} (2013) 1350094}
  [\href{https://arxiv.org/abs/1308.5535}{{\ttfamily 1308.5535}}].

\bibitem{goto2015}
Y.~Goto and K.~Matsumoto, \emph{{The monodromy representation and twisted
  period relations for Appell’s hypergeometric function $F_{4}$}},
  \href{https://doi.org/10.1215/00277630-2873714}{\emph{Nagoya Math. J.}
  {\bfseries 217} (2015) 61}.

\bibitem{goto2015b}
Y.~Goto, \emph{{Twisted period relations for Lauricella's hypergeometric
  functions $F_{A}$}}, {\emph{Osaka J. Math.} {\bfseries 52} (2015) 861}.

\bibitem{Yoshiaki-GOTO2015203}
Y.~Goto, \emph{{Intersection Numbers and Twisted Period Relations for the
  Generalized Hypergeometric Function $_{m+1}F_{m}$}},
  \href{https://doi.org/10.2206/kyushujm.69.203}{\emph{Kyushu Journal of
  Mathematics} {\bfseries 69} (2015) 203}.

\bibitem{Mizera:2017rqa}
S.~Mizera, \emph{{Scattering Amplitudes from Intersection Theory}},
  \href{https://doi.org/10.1103/PhysRevLett.120.141602}{\emph{Phys. Rev. Lett.}
  {\bfseries 120} (2018) 141602}
  [\href{https://arxiv.org/abs/1711.00469}{{\ttfamily 1711.00469}}].

\bibitem{matsubaraheo2019algorithm}
S.-J.~Matsubara-Heo and N.~Takayama, \emph{{An Algorithm of Computing
  Cohomology Intersection Number of Hypergeometric Integrals}},
  \href{https://doi.org/10.1017/nmj.2021.2}{\emph{Nagoya Math. J.} {\bfseries
  246} (2022) 256} [\href{https://arxiv.org/abs/1904.01253}{{\ttfamily
  1904.01253}}].

\bibitem{Ohara98intersectionnumbers}
K.~Ohara, ``{Intersection numbers of twisted cohomology groups associated with
  Selberg-type integrals}.''
  \url{http://www.math.kobe-u.ac.jp/HOME/ohara/Math/980523.ps}, 1998.

\bibitem{https://doi.org/10.48550/arxiv.2006.07848}
Y.~Goto and S.-J.~Matsubara-Heo, \emph{{Homology and cohomology intersection
  numbers of GKZ systems}},
  \href{https://doi.org/10.1016/j.indag.2021.12.002}{\emph{Indag. Math.}
  {\bfseries 33} (2022) 546}
  [\href{https://arxiv.org/abs/2006.07848}{{\ttfamily 2006.07848}}].

\bibitem{https://doi.org/10.48550/arxiv.2008.03176}
S.-J.~Matsubara-Heo, \emph{{Computing cohomology intersection numbers of GKZ
  hypergeometric systems}},
  \href{https://doi.org/10.22323/1.383.0013}{\emph{PoS} {\bfseries MA2019}
  (2022) 013} [\href{https://arxiv.org/abs/2008.03176}{{\ttfamily
  2008.03176}}].

\bibitem{https://doi.org/10.48550/arxiv.2104.12584}
S.-J.~Matsubara-Heo, \emph{{Localization formulas of cohomology intersection
  numbers}}, \href{https://doi.org/10.2969/jmsj/87738773}{\emph{J. Math. Soc.
  Jap.} {\bfseries 75} (2023) 909}
  [\href{https://arxiv.org/abs/2104.12584}{{\ttfamily 2104.12584}}].

\bibitem{Mastrolia:2018uzb}
P.~Mastrolia and S.~Mizera, \emph{{Feynman Integrals and Intersection Theory}},
  \href{https://doi.org/10.1007/JHEP02(2019)139}{\emph{JHEP} {\bfseries 02}
  (2019) 139} [\href{https://arxiv.org/abs/1810.03818}{{\ttfamily
  1810.03818}}].

\bibitem{Frellesvig:2019kgj}
H.~Frellesvig, F.~Gasparotto, S.~Laporta, M.K.~Mandal, P.~Mastrolia,
  L.~Mattiazzi et~al., \emph{{Decomposition of Feynman Integrals on the Maximal
  Cut by Intersection Numbers}},
  \href{https://doi.org/10.1007/JHEP05(2019)153}{\emph{JHEP} {\bfseries 05}
  (2019) 153} [\href{https://arxiv.org/abs/1901.11510}{{\ttfamily
  1901.11510}}].

\bibitem{Frellesvig:2019uqt}
H.~Frellesvig, F.~Gasparotto, M.K.~Mandal, P.~Mastrolia, L.~Mattiazzi and
  S.~Mizera, \emph{{Vector Space of Feynman Integrals and Multivariate
  Intersection Numbers}},
  \href{https://doi.org/10.1103/PhysRevLett.123.201602}{\emph{Phys. Rev. Lett.}
  {\bfseries 123} (2019) 201602}
  [\href{https://arxiv.org/abs/1907.02000}{{\ttfamily 1907.02000}}].

\bibitem{Frellesvig:2020qot}
H.~Frellesvig, F.~Gasparotto, S.~Laporta, M.K.~Mandal, P.~Mastrolia,
  L.~Mattiazzi et~al., \emph{{Decomposition of Feynman Integrals by
  Multivariate Intersection Numbers}},
  \href{https://doi.org/10.1007/JHEP03(2021)027}{\emph{JHEP} {\bfseries 03}
  (2021) 027} [\href{https://arxiv.org/abs/2008.04823}{{\ttfamily
  2008.04823}}].

\bibitem{Weinzierl:2020xyy}
S.~Weinzierl, \emph{{On the computation of intersection numbers for twisted
  cocycles}}, \href{https://doi.org/10.1063/5.0054292}{\emph{J. Math. Phys.}
  {\bfseries 62} (2021) 072301}
  [\href{https://arxiv.org/abs/2002.01930}{{\ttfamily 2002.01930}}].

\bibitem{Duhr:2024rxe}
C.~Duhr, F.~Porkert, C.~Semper and S.F.~Stawinski, \emph{{Twisted Riemann
  bilinear relations and Feynman integrals}},
  \href{https://doi.org/10.1007/JHEP03(2025)019}{\emph{JHEP} {\bfseries 03}
  (2025) 019} [\href{https://arxiv.org/abs/2407.17175}{{\ttfamily
  2407.17175}}].

\bibitem{Duhr:2024xsy}
C.~Duhr, F.~Porkert, C.~Semper and S.F.~Stawinski, \emph{{Self-duality from
  twisted cohomology}},
  \href{https://doi.org/10.1007/JHEP03(2025)053}{\emph{JHEP} {\bfseries 03}
  (2025) 053} [\href{https://arxiv.org/abs/2408.04904}{{\ttfamily
  2408.04904}}].

\bibitem{Duhr:2025xyy}
C.~Duhr, S.~Maggio, F.~Porkert, C.~Semper, Y.~Sohnle and S.F.~Stawinski,
  \emph{{Canonical differential equations and intersection matrices}},
  \href{https://doi.org/10.1007/JHEP02(2026)211}{\emph{JHEP} {\bfseries 02}
  (2026) 211} [\href{https://arxiv.org/abs/2509.17787}{{\ttfamily
  2509.17787}}].

\bibitem{Bree:2026qww}
I.~Bree, F.~Gasparotto, S.~P{\"o}gel, X.~Wang, S.~Weinzierl and X.~Xu,
  \emph{{Intersection matrices associated to geometric-ordered bases of Feynman
  integrals}},  \href{https://arxiv.org/abs/2608.03646}{{\ttfamily
  2608.03646}}.

\bibitem{Duhr:2026elp}
C.~Duhr, S.~Maggio, C.~Semper and S.F.~Stawinski, \emph{{Discrete symmetries of
  Feynman integrals}},
  \href{https://doi.org/10.1007/JHEP10(2026)031}{\emph{JHEP} {\bfseries 10}
  (2026) 031} [\href{https://arxiv.org/abs/2604.08332}{{\ttfamily
  2604.08332}}].

\bibitem{Brunello:2023fef}
G.~Brunello, G.~Crisanti, M.~Giroux, P.~Mastrolia and S.~Smith, \emph{{Fourier
  calculus from intersection theory}},
  \href{https://doi.org/10.1103/PhysRevD.109.094047}{\emph{Phys. Rev. D}
  {\bfseries 109} (2024) 094047}
  [\href{https://arxiv.org/abs/2311.14432}{{\ttfamily 2311.14432}}].

\bibitem{Brunello:2024ibk}
G.~Brunello and S.~De~Angelis, \emph{{An improved framework for computing
  waveforms}}, \href{https://doi.org/10.1007/JHEP07(2024)062}{\emph{JHEP}
  {\bfseries 07} (2024) 062}
  [\href{https://arxiv.org/abs/2403.08009}{{\ttfamily 2403.08009}}].

\bibitem{Brunello:2025cot}
G.~Brunello, V.~Chestnov, G.~Crisanti, M.~Giroux and S.~Smith,
  \emph{{Gravitational waveforms from restriction theory and rapid-decay
  homology}}, \href{https://doi.org/10.1103/ynk9-ykw7}{\emph{Phys. Rev. D}
  {\bfseries 113} (2026) 085011}
  [\href{https://arxiv.org/abs/2510.26874}{{\ttfamily 2510.26874}}].

\bibitem{Cacciatori:2022mbi}
S.L.~Cacciatori and P.~Mastrolia, \emph{{Intersection Numbers in Quantum
  Mechanics and Field Theory}},
  \href{https://arxiv.org/abs/2211.03729}{{\ttfamily 2211.03729}}.

\bibitem{Weinzierl:2020nhw}
S.~Weinzierl, \emph{{Correlation functions on the lattice and twisted
  cocycles}}, \href{https://doi.org/10.1016/j.physletb.2020.135449}{\emph{Phys.
  Lett. B} {\bfseries 805} (2020) 135449}
  [\href{https://arxiv.org/abs/2003.05839}{{\ttfamily 2003.05839}}].

\bibitem{Gasparotto:2022mmp}
F.~Gasparotto, A.~Rapakoulias and S.~Weinzierl, \emph{{Nonperturbative
  computation of lattice correlation functions by differential equations}},
  \href{https://doi.org/10.1103/PhysRevD.107.014502}{\emph{Phys. Rev. D}
  {\bfseries 107} (2023) 014502}
  [\href{https://arxiv.org/abs/2210.16052}{{\ttfamily 2210.16052}}].

\bibitem{Gasparotto:2023roh}
F.~Gasparotto, S.~Weinzierl and X.~Xu, \emph{{Real time lattice correlation
  functions from differential equations}},
  \href{https://doi.org/10.1007/JHEP06(2023)128}{\emph{JHEP} {\bfseries 06}
  (2023) 128} [\href{https://arxiv.org/abs/2305.05447}{{\ttfamily
  2305.05447}}].

\bibitem{Arkani-Hamed:2023kig}
N.~Arkani-Hamed, D.~Baumann, A.~Hillman, A.~Joyce, H.~Lee and G.L.~Pimentel,
  \emph{{Differential equations for cosmological correlators}},
  \href{https://doi.org/10.1007/JHEP09(2025)009}{\emph{JHEP} {\bfseries 09}
  (2025) 009} [\href{https://arxiv.org/abs/2312.05303}{{\ttfamily
  2312.05303}}].

\bibitem{De:2023xue}
S.~De and A.~Pokraka, \emph{{Cosmology meets cohomology}},
  \href{https://doi.org/10.1007/JHEP03(2024)156}{\emph{JHEP} {\bfseries 03}
  (2024) 156} [\href{https://arxiv.org/abs/2308.03753}{{\ttfamily
  2308.03753}}].

\bibitem{Benincasa:2024ptf}
P.~Benincasa, G.~Brunello, M.K.~Mandal, P.~Mastrolia and F.~Vaz{\~a}o,
  \emph{{One-loop corrections to the Bunch-Davies wave function of the
  universe}}, \href{https://doi.org/10.1103/PhysRevD.111.085016}{\emph{Phys.
  Rev. D} {\bfseries 111} (2025) 085016}
  [\href{https://arxiv.org/abs/2408.16386}{{\ttfamily 2408.16386}}].

\bibitem{Fontana:2023amt}
G.~Fontana and T.~Peraro, \emph{{Reduction to master integrals via intersection
  numbers and polynomial expansions}},
  \href{https://doi.org/10.1007/JHEP08(2023)175}{\emph{JHEP} {\bfseries 08}
  (2023) 175} [\href{https://arxiv.org/abs/2304.14336}{{\ttfamily
  2304.14336}}].

\bibitem{Brunello:2023rpq}
G.~Brunello, V.~Chestnov, G.~Crisanti, H.~Frellesvig, M.K.~Mandal and
  P.~Mastrolia, \emph{{Intersection numbers, polynomial division and relative
  cohomology}}, \href{https://doi.org/10.1007/JHEP09(2024)015}{\emph{JHEP}
  {\bfseries 09} (2024) 015}
  [\href{https://arxiv.org/abs/2401.01897}{{\ttfamily 2401.01897}}].

\bibitem{Caron-Huot:2021iev}
S.~Caron-Huot and A.~Pokraka, \emph{{Duals of Feynman Integrals. Part II.
  Generalized unitarity}},
  \href{https://doi.org/10.1007/JHEP04(2022)078}{\emph{JHEP} {\bfseries 04}
  (2022) 078} [\href{https://arxiv.org/abs/2112.00055}{{\ttfamily
  2112.00055}}].

\bibitem{Caron-Huot:2021xqj}
S.~Caron-Huot and A.~Pokraka, \emph{{Duals of Feynman integrals. Part I.
  Differential equations}},
  \href{https://doi.org/10.1007/JHEP12(2021)045}{\emph{JHEP} {\bfseries 12}
  (2021) 045} [\href{https://arxiv.org/abs/2104.06898}{{\ttfamily
  2104.06898}}].

\bibitem{Brunello:2024tqf}
G.~Brunello, V.~Chestnov and P.~Mastrolia, \emph{{Intersection numbers from
  companion tensor algebra}},
  \href{https://doi.org/10.1007/JHEP07(2025)045}{\emph{JHEP} {\bfseries 07}
  (2025) 045} [\href{https://arxiv.org/abs/2408.16668}{{\ttfamily
  2408.16668}}].

\bibitem{Lu:2024dsb}
M.~Lu, Z.~Wang and L.L.~Yang, \emph{{Intersection theory, relative cohomology
  and the Feynman parametrization}},
  \href{https://doi.org/10.1007/JHEP05(2025)158}{\emph{JHEP} {\bfseries 05}
  (2025) 158} [\href{https://arxiv.org/abs/2411.05226}{{\ttfamily
  2411.05226}}].

\bibitem{Huang:2026xnq}
L.-H.~Huang, Y.-Q.~Ma, Z.~Wang and L.L.~Yang, \emph{{Feynman integral reduction
  with intersection theory made simple}},
  \href{https://arxiv.org/abs/2604.05025}{{\ttfamily 2604.05025}}.

\bibitem{Chestnov:2022xsy}
V.~Chestnov, H.~Frellesvig, F.~Gasparotto, M.K.~Mandal and P.~Mastrolia,
  \emph{{Intersection numbers from higher-order partial differential
  equations}}, \href{https://doi.org/10.1007/JHEP06(2023)131}{\emph{JHEP}
  {\bfseries 06} (2023) 131}
  [\href{https://arxiv.org/abs/2209.01997}{{\ttfamily 2209.01997}}].

\bibitem{Chestnov:2022alh}
V.~Chestnov, F.~Gasparotto, M.K.~Mandal, P.~Mastrolia, S.J.~Matsubara-Heo,
  H.J.~Munch et~al., \emph{{Macaulay matrix for Feynman integrals: linear
  relations and intersection numbers}},
  \href{https://doi.org/10.1007/JHEP09(2022)187}{\emph{JHEP} {\bfseries 09}
  (2022) 187} [\href{https://arxiv.org/abs/2204.12983}{{\ttfamily
  2204.12983}}].

\bibitem{Saito:1983}
K.~Saito, \emph{The higher residue pairings {$K_F^{(k)}$} for a family of
  hypersurface singular points},  in \emph{Singularities, Part 2 (Arcata,
  Calif., 1981)}, P.~Orlik, ed., vol.~40 of \emph{Proceedings of Symposia in
  Pure Mathematics}, (Providence, RI), pp.~441--463, American Mathematical
  Society (1983), \href{https://doi.org/10.1090/pspum/040.2/713270}{DOI}.

\bibitem{Mizera:2019vvs}
S.~Mizera and A.~Pokraka, \emph{{From Infinity to Four Dimensions: Higher
  Residue Pairings and Feynman Integrals}},
  \href{https://doi.org/10.1007/JHEP02(2020)159}{\emph{JHEP} {\bfseries 02}
  (2020) 159} [\href{https://arxiv.org/abs/1910.11852}{{\ttfamily
  1910.11852}}].

\bibitem{Lee:2014ioa}
R.N.~Lee, \emph{{Reducing differential equations for multiloop master
  integrals}}, \href{https://doi.org/10.1007/JHEP04(2015)108}{\emph{JHEP}
  {\bfseries 04} (2015) 108} [\href{https://arxiv.org/abs/1411.0911}{{\ttfamily
  1411.0911}}].

\bibitem{Henn:2014qga}
J.M.~Henn, \emph{{Lectures on differential equations for Feynman integrals}},
  \href{https://doi.org/10.1088/1751-8113/48/15/153001}{\emph{J. Phys. A}
  {\bfseries 48} (2015) 153001}
  [\href{https://arxiv.org/abs/1412.2296}{{\ttfamily 1412.2296}}].

\bibitem{e-collaboration:2025frv}
{\scshape {\ensuremath{\varepsilon}}-collaboration} collaboration,
  \emph{{Geometric Bookkeeping Guide to Feynman Integral Reduction and
  {\ensuremath{\epsilon}}-Factorized Differential Equations}},
  \href{https://doi.org/10.1103/pyt8-d7rt}{\emph{Phys. Rev. Lett.} {\bfseries
  136} (2026) 241602} [\href{https://arxiv.org/abs/2506.09124}{{\ttfamily
  2506.09124}}].

\bibitem{Henn:2013pwa}
J.M.~Henn, \emph{{Multiloop integrals in dimensional regularization made
  simple}}, \href{https://doi.org/10.1103/PhysRevLett.110.251601}{\emph{Phys.
  Rev. Lett.} {\bfseries 110} (2013) 251601}
  [\href{https://arxiv.org/abs/1304.1806}{{\ttfamily 1304.1806}}].

\bibitem{Lee:2013hzt}
R.N.~Lee and A.A.~Pomeransky, \emph{{Critical points and number of master
  integrals}}, \href{https://doi.org/10.1007/JHEP11(2013)165}{\emph{JHEP}
  {\bfseries 11} (2013) 165} [\href{https://arxiv.org/abs/1308.6676}{{\ttfamily
  1308.6676}}].

\bibitem{Baikov:1996iu}
P.A.~Baikov, \emph{{Explicit solutions of the multiloop integral recurrence
  relations and its application}},
  \href{https://doi.org/10.1016/S0168-9002(97)00126-5}{\emph{Nucl. Instrum.
  Meth. A} {\bfseries 389} (1997) 347}
  [\href{https://arxiv.org/abs/hep-ph/9611449}{{\ttfamily hep-ph/9611449}}].

\bibitem{Frellesvig:2017aai}
H.~Frellesvig and C.G.~Papadopoulos, \emph{{Cuts of Feynman Integrals in Baikov
  representation}}, \href{https://doi.org/10.1007/JHEP04(2017)083}{\emph{JHEP}
  {\bfseries 04} (2017) 083}
  [\href{https://arxiv.org/abs/1701.07356}{{\ttfamily 1701.07356}}].

\bibitem{Frellesvig:2024ymq}
H.~Frellesvig, \emph{{The loop-by-loop Baikov representation {\textemdash}
  Strategies and implementation}},
  \href{https://doi.org/10.1007/JHEP04(2025)111}{\emph{JHEP} {\bfseries 04}
  (2025) 111} [\href{https://arxiv.org/abs/2412.01804}{{\ttfamily
  2412.01804}}].

\bibitem{Matsumoto:2018igm}
K.~Matsumoto, \emph{{Relative twisted homology and cohomology groups associated
  with Lauricella's $F_D$}},
  \href{https://doi.org/10.1619/fesi.67.105}{\emph{Funkcial. Ekvac.} {\bfseries
  67} (2024) 105} [\href{https://arxiv.org/abs/1804.00366}{{\ttfamily
  1804.00366}}].

\bibitem{Bitoun:2017nre}
T.~Bitoun, C.~Bogner, R.P.~Klausen and E.~Panzer, \emph{{Feynman integral
  relations from parametric annihilators}},
  \href{https://doi.org/10.1007/s11005-018-1114-8}{\emph{Lett. Math. Phys.}
  {\bfseries 109} (2019) 497}
  [\href{https://arxiv.org/abs/1712.09215}{{\ttfamily 1712.09215}}].

\bibitem{Henn:2013tua}
J.M.~Henn, A.V.~Smirnov and V.A.~Smirnov, \emph{{Analytic results for planar
  three-loop four-point integrals from a Knizhnik-Zamolodchikov equation}},
  \href{https://doi.org/10.1007/JHEP07(2013)128}{\emph{JHEP} {\bfseries 07}
  (2013) 128} [\href{https://arxiv.org/abs/1306.2799}{{\ttfamily 1306.2799}}].

\bibitem{GelfandShilov1964}
I.M.~Gelfand and G.E.~Shilov, \emph{Generalized Functions, Vol. 1: Properties
  and Operations}, Academic Press, New York (1964).

\bibitem{Lee:2014tja}
R.N.~Lee, \emph{{Modern techniques of multiloop calculations}},  in \emph{{49th
  Rencontres de Moriond on QCD and High Energy Interactions}}, pp.~297--300,
  2014 [\href{https://arxiv.org/abs/1405.5616}{{\ttfamily 1405.5616}}].

\bibitem{Chen:2023eqx}
W.~Chen, \emph{{A representation transformation of parametric Feynman
  integrals}},
  \href{https://doi.org/10.1016/j.physletb.2025.139340}{\emph{Phys. Lett. B}
  {\bfseries 862} (2025) 139340}
  [\href{https://arxiv.org/abs/2303.12427}{{\ttfamily 2303.12427}}].

\bibitem{Agostini:2022cgv}
D.~Agostini, C.~Fevola, A.-L.~Sattelberger and S.~Telen, \emph{{Vector spaces
  of generalized Euler integrals}},
  \href{https://doi.org/10.4310/CNTP.2024.v18.n2.a2}{\emph{Commun. Num. Theor.
  Phys.} {\bfseries 18} (2024) 327}
  [\href{https://arxiv.org/abs/2208.08967}{{\ttfamily 2208.08967}}].

\bibitem{Aomoto:1975}
K.~Aomoto, \emph{{On vanishing of cohomology attached to certain many valued
  meromorphic functions}},
  \href{https://doi.org/10.2969/jmsj/02720248}{\emph{Journal of the
  Mathematical Society of Japan} {\bfseries 27} (1975) 248 }.

\bibitem{KitaNoumi:1982}
M.~Kita and M.~Noumi, \emph{{On the structure of cohomology groups attached to
  the integral of certain many-valued analytic functions}},
  \href{https://doi.org/10.3792/pjaa.58.97}{\emph{Proceedings of the Japan
  Academy, Series A, Mathematical Sciences} {\bfseries 58} (1982) 97 }.

\bibitem{Cho:1997}
K.~Cho, \emph{{A generalization of Kita and Noumi's vanishing theorems of
  cohomology groups of local system}}, {\emph{Nagoya Mathematical Journal}
  {\bfseries 147} (1997) 63 }.

\bibitem{AomotoKitaKohnoIohara}
K.~Aomoto, M.~Kita, T.~Kohno and K.~Iohara, \emph{Theory of Hypergeometric
  Functions}, Springer Monographs in Mathematics, Springer Japan (2011).

\bibitem{huh2013maximum}
J.~Huh, \emph{{The maximum likelihood degree of a very affine variety}},
  \href{https://doi.org/10.1112/S0010437X13007057}{\emph{Compositio Math.}
  {\bfseries 149} (2013) 1245}
  [\href{https://arxiv.org/abs/1207.0553}{{\ttfamily 1207.0553}}].

\bibitem{Crisanti:2026rbc}
G.~Crisanti, H.~Frellesvig, A.~Pokraka and S.~Smith, \emph{{Magic Relations and
  Critical Varieties of Feynman Integrals}},
  \href{https://arxiv.org/abs/2605.29789}{{\ttfamily 2605.29789}}.

\bibitem{Wu:2026xea}
Z.~Wu and Y.~Xu, \emph{{On the spanning cuts consistency problem in the IBP
  reductions of Feynman integrals}},
  \href{https://arxiv.org/abs/2606.02485}{{\ttfamily 2606.02485}}.

\bibitem{Crisanti:2026cuf}
G.~Crisanti, L.~Lippstreu, A.J.~McLeod and M.~Polackova, \emph{{DiscKosky:
  Efficient Genealogical Constraints from Sector-By-Sector Critical Point
  Counting}},  \href{https://arxiv.org/abs/2610.40772}{{\ttfamily 2610.40772}}.

\bibitem{Page:2025gso}
B.~Page and Q.~Song, \emph{{Critical points and syzygies for Feynman
  integrals}}, \href{https://doi.org/10.1007/JHEP02(2026)004}{\emph{JHEP}
  {\bfseries 02} (2026) 004}
  [\href{https://arxiv.org/abs/2509.17681}{{\ttfamily 2509.17681}}].

\bibitem{DGPS}
W.~Decker, G.-M.~Greuel, G.~Pfister and H.~Sch\"onemann, ``{\sc Singular}
  {4-4-0} --- {A} computer algebra system for polynomial computations.''
  \url{http://www.singular.uni-kl.de}, 2024.

\bibitem{msolve}
J.~Berthomieu, C.~Eder and M.~{Safey El Din}, \emph{{msolve: A Library for
  Solving Polynomial Systems}},  in \emph{{2021 International Symposium on
  Symbolic and Algebraic Computation}}, 46th International Symposium on
  Symbolic and Algebraic Computation, (Saint Petersburg, Russia), pp.~51--58,
  {ACM}, July, 2021, \href{https://doi.org/10.1145/3452143.3465545}{DOI}.

\bibitem{CoxLittleOShea2005}
D.A.~Cox, J.~Little and D.~O'Shea, \emph{Using Algebraic Geometry}, vol.~185 of
  \emph{Graduate Texts in Mathematics}, Springer, New York, 2~ed. (2005),
  \href{https://doi.org/10.1007/b138611}{10.1007/b138611}.

\bibitem{Chestnov:2025svg}
V.~Chestnov and G.~Crisanti, \emph{{Sampling Polynomial Rational Remainders
  with SP$\mathbb{Q}$R: A new Package for Polynomial Division and
  Elimination}},  \href{https://arxiv.org/abs/2511.14875}{{\ttfamily
  2511.14875}}.

\bibitem{vonManteuffel:2014ixa}
A.~von Manteuffel and R.M.~Schabinger, \emph{{A novel approach to integration
  by parts reduction}},
  \href{https://doi.org/10.1016/j.physletb.2015.03.029}{\emph{Phys. Lett. B}
  {\bfseries 744} (2015) 101}
  [\href{https://arxiv.org/abs/1406.4513}{{\ttfamily 1406.4513}}].

\bibitem{Peraro:2016wsq}
T.~Peraro, \emph{{Scattering amplitudes over finite fields and multivariate
  functional reconstruction}},
  \href{https://doi.org/10.1007/JHEP12(2016)030}{\emph{JHEP} {\bfseries 12}
  (2016) 030} [\href{https://arxiv.org/abs/1608.01902}{{\ttfamily
  1608.01902}}].

\bibitem{Bargiela:2024hgt}
P.~Bargiela, \emph{{Integrated unitarity for scattering amplitudes and the
  four-loop ladder Feynman integral}},
  \href{https://doi.org/10.1007/JHEP05(2025)004}{\emph{JHEP} {\bfseries 05}
  (2025) 004} [\href{https://arxiv.org/abs/2403.18047}{{\ttfamily
  2403.18047}}].

\bibitem{Henn:2023tbo}
J.~Henn, E.~Pratt, A.-L.~Sattelberger and S.~Zoia, \emph{{D-module techniques
  for solving differential equations in the context of Feynman integrals}},
  \href{https://doi.org/10.1007/s11005-024-01835-7}{\emph{Lett. Math. Phys.}
  {\bfseries 114} (2024) 87}
  [\href{https://arxiv.org/abs/2303.11105}{{\ttfamily 2303.11105}}].

\end{thebibliography}\endgroup
